\documentclass[twocolumn, twocolappendix, astrosymb, times]{aastex631}

\usepackage{xspace}
\usepackage{amsmath}
\usepackage{txfonts}
\usepackage{multirow}

\newcommand{\methanol}{CH$_3$OH\xspace}
\newcommand{\acetaldehyde}{CH$_3$CHO\xspace}
\newcommand{\methylformate}{CH$_3$OCHO\xspace}
\newcommand{\dimethylether}{CH$_3$OCH$_3$\xspace}
\newcommand{\acetone}{CH$_3$COCH$_3$\xspace}
\newcommand{\ethyleneoxide}{$c$-C$_2$H$_4$O\xspace}
\newcommand{\propenal}{$t$-C$_2$H$_3$CHO\xspace}
\newcommand{\propanal}{$s$-C$_2$H$_5$CHO\xspace}

\shorttitle{V883 Ori Band 3}
\shortauthors{Yamato et al.}
\graphicspath{{./}{figures/}}

\begin{document}

\title{Chemistry of Complex Organic Molecules in the V883 Ori Disk Revealed by ALMA Band 3 Observations}

\author[0000-0003-4099-6941]{Yoshihide Yamato}
\affiliation{Department of Astronomy, Graduate School of Science, The University of Tokyo, 7-3-1 Hongo, Bunkyo-ku, Tokyo 113-0033, Japan}

\author[0000-0003-2493-912X]{Shota Notsu}
\affiliation{Department of Astronomy, Graduate School of Science, The University of Tokyo, 7-3-1 Hongo, Bunkyo-ku, Tokyo 113-0033, Japan}
\affiliation{Star and Planet Formation Laboratory, RIKEN Cluster for Pioneering Research, 2-1 Hirosawa, Wako, Saitama 351-0198, Japan}
\affiliation{Department of Earth and Planetary Science, Graduate School of Science, The University of Tokyo, 7-3-1 Hongo, Bunkyo-ku, Tokyo 113-0033, Japan}

\author[0000-0003-3283-6884]{Yuri Aikawa}
\affiliation{Department of Astronomy, Graduate School of Science, The University of Tokyo, 7-3-1 Hongo, Bunkyo-ku, Tokyo 113-0033, Japan}

\author[0000-0003-3655-5270]{Yuki Okoda}
\affiliation{Star and Planet Formation Laboratory, RIKEN Cluster for Pioneering Research, 2-1 Hirosawa, Wako, Saitama 351-0198, Japan}

\author[0000-0002-7058-7682]{Hideko Nomura}
\affiliation{Division of Science, National Astronomical Observatory of Japan, 2-21-1 Osawa, Mitaka, Tokyo 181-8588, Japan}

\author[0000-0002-3297-4497]{Nami Sakai}
\affiliation{Star and Planet Formation Laboratory, RIKEN Cluster for Pioneering Research, 2-1 Hirosawa, Wako, Saitama 351-0198, Japan}



\begin{abstract}
Complex organic molecules (COMs) in protoplanetary disks are key to understanding the origin of volatiles in comets in our solar system, yet the chemistry of COMs in protoplanetary disks remains poorly understood.
Here we present Atacama Large Millimeter/submillimeter Array (ALMA) Band 3 observations of the disk around the young outbursting star V883 Ori, where the COMs sublimate from ices and are thus observable thanks to the warm condition of the disk. We have robustly identified ten oxygen-bearing COMs including $^{13}$C-isotopologues in the disk-integrated spectra. The radial distributions of the COM emission, revealed by the detailed analyses of the line profiles, show the inner emission cavity, similar to the previous observations in Band 6 and Band 7. We found that the COMs abundance ratios with respect to methanol are significantly higher than those in the warm protostellar envelopes of IRAS~16293-2422 and similar to the ratios in the solar system comet 67P/Churyumov-Gerasimenko, suggesting the efficient (re-)formation of COMs in protoplanetary disks. We also constrained the $^{12}$C/$^{13}$C and D/H ratios of COMs in protoplanetary disks for the first time. The $^{12}$C/$^{13}$C ratios of acetaldehyde, methyl formate, and dimethyl ether are consistently lower ($\sim$\,20--30) than the canonical ratio in the interstellar medium ($\sim$\,69), indicating the efficient $^{13}$C-fractionation of CO. The D/H ratios of methyl formate are slightly lower than the values in IRAS 16293-2422, possibly pointing to the destruction and reformation of COMs in disks. We also discuss the implications for nitrogen and sulfur chemistry in protoplanetary disks.  

\end{abstract}



\section{Introduction} \label{sec:intro}
Our Solar System formed from its protoplanetary disk, or the proto-solar disk, a byproduct of the Sun's formation via the gravitational collapse of the parent molecular cloud. Observations have found many volatile ices in solar system comets, which should contain information about the chemical composition of the solar nebula, suggesting that the proto-solar disk was rich in volatiles such as water and organics. Organic molecules are of particular interest because they could be the precursors of the prebiotic molecules that may have given rise to life on Earth. \citep{Ceccarelli2023}.

The origin of the cometary water and organics has been extensively discussed but remains controversial. The most commonly discussed possibility is that the cometary ices are, at least in part, inherited from the interstellar medium (ISM) through the formation of the Solar System. 
For example, the higher D$_2$O/HDO ratio than the HDO/H$_2$O ratio in both the warm inner envelopes of protostars \citep{Coutens2014, Jensen2021} and a solar system comet \citep{Altwegg2017}
strongly suggests the inheritance of water from the ISM rather than the chemical reset (e.g., destruction and reformation).
Organic molecules have also been detected in both protostellar envelopes and solar system comets \citep[e.g.,][]{Jorgensen2016, Rubin2019}. In the warm inner envelopes heated by the central protostar, organic molecules are thermally desorbed from the grain surfaces at a typical sublimation temperature of $\sim100$\,K, allowing us to observe them in the gas phase. Among these molecules, carbon-bearing molecules with six or more atoms are empirically referred to \textrm{as} complex organic molecules (COMs). The composition of COMs in protostellar envelopes is compared with that in comets, from which the possible inheritance of COMs from the prestellar and protostellar phases is discussed \citep[e.g.,][]{Drozdovskaya2019}.

Observations of COMs in protoplanetary disks are rather sparse compared to those in protostellar envelopes, since COMs are mainly locked into ices due to the colder nature ($<$\,100~K) of typical disks around T Tauri stars.  Relatively simple organic molecules such as \methanol, CH$_3$CN, and HCOOH, which are non-thermally desorbed and/or formed via the gas-phase chemistry in the outer cold region, are detected in a handful of disks \citep{Walsh2016, Oberg2015, Bergner2018, Loomis2018, Ilee2021, Favre2018}, but these observations do not directly trace the bulk composition of COMs hidden in the ice mantles. Recently, high-sensitivity observations with Atacama Large Millimeter/submillimeter Array (ALMA) have just begun to detect the emission of thermally desorbed COMs in warm disks around Herbig Ae/Be stars \citep{vanderMarel2021, Booth2021_CH3OH, Brunken2022, Booth2023_HD169142}, but the detections are still limited to a few numbers of disks and molecular species. Since protoplanetary disks could be potential sites for the chemical reset from the ISM \citep[e.g.,][]{Walsh2014, Furuya2014, Eistrup2016}, constraining the composition and distribution of COMs in disks is crucial to fully understand the chemical evolution during star and planet formation.

Recent observations of the disk around the young outbursting star, V883 Ori, have opened a new window to COMs in protoplanetary disks. V883 Ori is a low-mass ($M_\star = 1.3\,M_\odot$; \citealt{Cieza2016}) FU Orionis type object located in the L1641 cluster in the Orion molecular clouds \citep[$d\approx400$\,pc;][]{Strom1993} with a well-developed Keplerian-rotating disk \citep{Cieza2016}, which shows a large-amplitude outburst in the optical ($\Delta m_\mathrm{V} > 4$\,mag) with an increased luminosity of $\sim200\,L_\odot$ \citep{Furlan2016}. This high luminosity warms the disk and shifts the sublimation front (or snowline) of water and COMs \citep{Tobin2023}, allowing for observing thermally desorbed molecules in the disk. In addition, because the duration of the outburst ($\sim10^2$\,yr) is shorter than the timescale of the typical gas-phase chemical reaction (several $10^4$\,yr; e.g., \citealt{Nomura2009}), it is possible to observe the fresh sublimates without significant gas-phase chemical changes and probe the composition of disk ices almost directly. 
\citet{vantHoff2018} detected \methanol emission for the first time in this disk with high-resolution ($\sim0\farcs14$) ALMA Band 7 observations, followed by the additional detection of several COMs such as \methylformate, \acetaldehyde, and CH$_3$CN in ALMA Band 7 at $\sim0\farcs1$ resolution \citep{Lee2019}. These COM emissions all show the velocity structure consistent with the Keplerian rotation traced by the C$^{18}$O emission \citep{Cieza2016, vantHoff2018}, proving that the COM emission originates from the rotationally supported disk. The COM emission spreads over 40--120 au radii with an inner emission cavity ($\lesssim40$\,au radius), similar to the spatial distribution of the water emission \citep{Tobin2023}. \citet{Lee2019} compared the abundances of a few COMs with respect to \methanol, the simplest COM, with those in the warm envelopes of the Class 0 protostar IRAS~16293-2422~B \citep[hereafter IRAS~16293B; e.g.,][]{Jorgensen2016} and the solar system comet 67P/Churyumov-Gerasimenko \citep[hereafter 67P/C-G; e.g.,][]{Altwegg2019}. They found that the abundances of COMs in the V883 Ori disk are generally higher than those in IRAS~16293B (except for CH$_3$CN) and similar to those in 67P/C-G, suggesting potential chemical evolution in protoplanetary disks.

However, there are several limitations to these observations. First, in ALMA Band 7, i.e., the sub-mm wavelengths, the intense dust continuum emission obscures the molecular line emission in the innermost region ($\lesssim40$\,au radius), resulting in an apparent inner cavity in the molecular line emission \citep{vantHoff2018, Lee2019, Tobin2023}. This necessitates observations at longer wavelengths, where the dust continuum emission is generally expected to be fainter, to investigate the chemical composition there, i.e., the comet- and terrestrial-planet-forming zone.
Second, the limited number of unambiguously detected species \citep{Lee2019} makes it difficult to derive a general and concrete picture of chemical evolution in protoplanetary disks. Third, previous observations did not measure the isotopic ratios (e.g., $^{12}$C/$^{13}$C and D/H) of COMs \citep[except for the D/H ratio of \methanol;][]{Lee2019}, the most sensitive chemical fingerprint for tracing the formation conditions and thermal history. 

In this paper, we present the new observations of COMs toward the V883 Ori disk in ALMA Band 3. Observational and imaging details are described in Section \ref{sec:observation}, followed by a detailed analysis of the disk-integrated spectra in Section \ref{sec:analysis_result} where we report the first (tentative) detection of several species including $^{13}$C- and D-isotopologues. We discuss the implications of the present observations for the physical and chemical structure of the V883~Ori disk, as well as for the chemical composition and isotopic chemistry of COMs in protoplanetary disks in Section \ref{sec:discussion}. We finally summarize the present work in Section \ref{sec:summary}.


\section{Observations} \label{sec:observation}

\subsection{ALMA Band 3 Observations}
V883 Ori was observed in Band 3 during ALMA Cycle 8 (project code: 2021.1.00357.S, PI: S. Notsu). The observations were carried out in a total of four executions, one with a compact antenna configuration and the other three with an extended antenna configuration. The observation dates, number of antennas, on-source integration time, precipitable water vapor (PWV), baseline coverage, angular resolution, maximum recoverable scales (MRS), and calibrator information are listed in Table \ref{tab:observations}. 

The correlator setup included eleven spectral windows (SPWs) in Frequency Division Mode (FDM), one of which was dedicated to continuum acquisition with a wide bandwidth of 937.5\,MHz for an accurate determination of the continuum level. The frequency resolution of the continuum SPW was 0.488\,MHz, resulting in a velocity resolution of $\sim$1.5\,km \,s$^{-1}$. The continuum SPW included many spectral lines of COMs. Other ten SPWs targeted specific spectral lines with narrower bandwidths of 58.59\,MHz or 117.19\,MHz, which also covered many COMs lines. The frequency resolution of these SPWs was 0.141\,MHz, resulting in a velocity resolution of $\sim$0.4--0.5\,km\,s$^{-1}$. 
The detailed properties of the SPWs are summarized in Table \ref{tab:cube_properties}. We note that two SPWs (8 and 9) were partially overlapped, and only the wider SPW 9 is used for analysis. 

\subsection{Calibration and Imaging}
Initial calibrations were performed by the ALMA staff using the standard ALMA calibration pipeline version 2021.2.0.128. Subsequent self-calibration and imaging were performed using the Common Astronomy Software Applications \citep[CASA;][]{CASA} version 6.2.1.7. The pipeline-calibrated data were first imaged without any deconvolution (i.e., dirty imaging) or continuum subtraction to accurately specify the line-free channels by visually inspecting the dirty image cubes. These line-free channels are averaged to obtain the continuum visibilities. \textrm{The continuum emission peak on the image from each execution block was aligned to a common direction, $\alpha\mathrm{(ICRS)}= 05^\mathrm{h}38^\mathrm{m}18\fs101$, $\delta\mathrm{(ICRS)} = -07^\mathrm{d}02^\mathrm{m}26\fs00$,
which was used as the disk center in the following analysis.} Self-calibration was then performed using the continuum visibilities with the \texttt{gaincal} and \texttt{applycal} tasks. A round of each phase-only and amplitude self-calibration was first performed on the data with a compact antenna configuration. The extended antenna configuration data were then concatenated and self-calibrated together. Two rounds of phase-only self-calibration and one round of amplitude self-calibration were performed on the combined data. The solutions were then applied to the spectral line visibilities.

\textrm{The combined dust continuum image at 3.3\,mm (or 92\,GHz) was made using the \texttt{tclean} task with the Briggs weighting (\texttt{robust} $=0.5$) and is shown in Figure \ref{fig:moment_zero_gallery}. The beam size and the RMS noise level were $0\farcs39\times0\farcs27$ (P.A. $= -74\arcdeg$) and 27\,\textmu Jy\,beam$^{-1}$, respectively. We performed a Gaussian fit to the continuum image using the \texttt{imfit} task and obtained a deconvolved size of $0\farcs23\times0\farcs21$ (P.A. $= 106\arcdeg$), indicating that the dust continuum disk is marginally spatially resolved. The brightness temperature at the disk center was $\sim65$\,K, which was lower than that of the higher-resolution Band 6 image \citep{Cieza2016} mainly due to the beam dilution effect.}

As for the line image cubes, each of the SPWs was imaged using the \texttt{tclean} task \citep{Hogbom1974} with Briggs weighting (\texttt{robust} $=$ 0.5). The \texttt{tclean} task was run in parallel using \texttt{mpicasa} implementation. As a specific weighting scheme, \texttt{briggsbwtaper} was used with independent weight densities for each channel (i.e., \texttt{perchanweightdensity = True}) to achieve more uniform sensitivity across channels. For each SPW, a common restoring beam across all channels was used. A few channels at the edge of the SPW were removed during the imaging process due to the large deviation in beam size caused by a slight difference in the frequency coverage between the data with compact and extended antenna configurations\footnote{See \url{https://casadocs.readthedocs.io/en/stable/notebooks/memo-series.html\#Correcting-bad-common-beam}.}. All SPWs were cleaned down to 3\,$\times$\,RMS with the native channel widths. An automatic masking algorithm with \texttt{automultithresh} was also used to generate the CLEAN mask. The typical beam size and the RMS noise level were $\sim0\farcs3$--$0\farcs4$ and 0.6--1\,mJy\,beam$^{-1}$, respectively. \textrm{The beam sizes were comparable to the emitting region sizes of COM emission in Band 6 and Band 7 \citep{Lee2019, Tobin2023}.} The RMS noise level was measured on the images without primary beam correction. The beam sizes and RMS noise levels for each SPW are listed in Table \ref{tab:cube_properties}. \textrm{We adopt a 10\% uncertainty as an absolute flux calibration accuracy\footnote{See Chapter 10.2.6 of ALMA Cycle 8 2021 Technical Handbook (\url{https://almascience.nao.ac.jp/documents-and-tools/cycle8/alma-technical-handbook})}.}  Finally, the continuum component is subtracted from these image cubes using the \texttt{imcontsub} task to produce the spectral-line-only image cubes. The continuum subtraction on the image plane instead of the visibility plane (e.g., with the \texttt{uvcontsub} task) mitigates as much as possible the Jorsater \& van Moorsel (JvM) effect \citep{JvM, Czekala2021} on the line emission, which is critical for the flux scale (see Appendix \ref{appendix:JvM_effect}). Throughout this paper, we use the image cubes that were continuum-subtracted on the image plane, unless otherwise noted.


\begin{deluxetable*}{ccccccccc}
\label{tab:observations}
\tablecaption{Observational Details}
\tablehead{\colhead{Date} & \colhead{\# of Ant.} & \colhead{On-source Int.} & \colhead{PWV}  & \colhead{Baseline} & \colhead{Ang. Res.} & \colhead{MRS} & \colhead{Bandpass/Amplitude Cal.} & \colhead{Phase Cal.} \\
\colhead{} & \colhead{} & \colhead{(min)} & \colhead{(mm)} & \colhead{(m)} & \colhead{($\arcsec$)} & \colhead{($\arcsec$)} & \colhead{} & \colhead{}}
\startdata
2021 Nov. 21 & 44 & 43 & 2.2 & 41--3396 & 0.3 & 5.2 & J0423-0120 & J0541-0541 \\
2021 Nov. 21 & 44 & 43 & 2.0 & 41--3396 & 0.3 & 5.2 & J0538-4405 & J0541-0541 \\
2021 Nov. 22 & 43 & 43 & 3.8 & 41--3396 & 0.3 & 4.7 & J0423-0120 & J0541-0541 \\
2022 Jan. 20 & 41 & 47 & 3.6 & 14--740  & 1.5 & 17.4 & J0423-0120 & J0541-0541
\enddata
\end{deluxetable*}

\begin{deluxetable*}{cCCCCCCCC}
\label{tab:cube_properties}
\tablecaption{Properties of Image Cubes}
\tablehead{\colhead{SPW} & \colhead{Cent. Freq.} & \colhead{\# of Chan.} & \colhead{Bandwidth} & \multicolumn{2}{c}{Channel Width}  & \colhead{Vel. Res.} & \colhead{Beam Size (P.A.)} & \colhead{RMS} \\
\colhead{} & \colhead{(GHz)} & \colhead{} & \colhead{(MHz)} & \colhead{(MHz)} & \colhead{(km\,s$^{-1}$)} & \colhead{(km\,s$^{-1}$)} & \colhead{} & \colhead{(mJy\,beam$^{-1}$})}
\startdata
0 & 85.160960 & 478 & 58.229 & 0.122 & 0.43 & 0.51 & 0$\farcs$42$\times$0$\farcs$31\,(-75$\arcdeg$) & 1.1 \\
1 & 85.530202 & 478 & 58.229 & 0.122 & 0.43 & 0.50 & 0$\farcs$42$\times$0$\farcs$31\, (-75$\arcdeg$) & 1.0 \\
2 & 85.851713 & 478 & 58.229 & 0.122 & 0.43 & 0.50 & 0$\farcs$42$\times$0$\farcs$31\, (-75$\arcdeg$) & 0.99 \\
3 & 85.924957 & 478 & 58.229 & 0.122 & 0.43 & 0.50 & 0$\farcs$41$\times$0$\farcs$31\, (-75$\arcdeg$) & 0.99 \\
4 & 86.669481 & 478 & 58.229 & 0.122 & 0.42 & 0.50 & 0$\farcs$41$\times$0$\farcs$31\, (-75$\arcdeg$) & 0.94 \\
5 & 86.752980 & 958 & 116.824 & 0.122 & 0.42 & 0.50 & 0$\farcs$41$\times$0$\farcs$31\, (-75$\arcdeg$) & 0.94 \\
6 & 87.327244 & 478 & 58.229 & 0.122 & 0.42 & 0.49 & 0$\farcs$41$\times$0$\farcs$31\, (-75$\arcdeg$) & 0.98 \\
7 & 96.490854 & 478 & 58.229 & 0.122 & 0.38 & 0.45 & 0$\farcs$37$\times$0$\farcs$28\, (-75$\arcdeg$) & 1.0 \\
8 & 96.743210 & 478 & 58.229 & 0.122 & 0.38 & 0.45 & 0$\farcs$37$\times$0$\farcs$27\, (-75$\arcdeg$) & 0.99 \\
9 & 96.754197 & 958 & 116.824 & 0.122 & 0.38 & 0.45 & 0$\farcs$37$\times$0$\farcs$27\, (-75$\arcdeg$) & 0.99 \\
10 & 97.979676 & 3839 & 937.036 & 0.244 & 0.75 & 1.5 & 0$\farcs$36$\times$0$\farcs$26\, (-73$\arcdeg$) & 0.59
\enddata
\end{deluxetable*}

\section{Data Analysis and Result}\label{sec:analysis_result}
\subsection{Spectrum Extraction and Line Identification}\label{subsec:line_identification}
Figure \ref{fig:spectra_gallery} shows the spectra integrated over 1\farcs2 aperture and corrected for the line broadening due to the Keplerian rotation of the disk. The original disk-integrated spectra are shown in Figures \ref{fig:disk-integrated_spectra0}--\ref{fig:disk-integrated_spectra3} in Appendix \ref{appendix:spectra}. More zoom-in spectra with the Keplerian rotation correction are also shown in Figures \ref{fig:disk-integrated_spectra_fit0}--\ref{fig:disk-integrated_spectra_fit3} in Appendix \ref{appendix:spectra}. To obtain the spectra corrected for the Keplerian line broadening, we use the \texttt{integrated\_spectrum()} function implemented in the Python package \texttt{GoFish} \citep{GoFish}. This method first deprojects the disk and then aligns the Dopper-shifted spectra at each position within the disk to a common velocity (the systemic velocity of the source) to recover the single peak spectra for each transition, facilitating the identification of the blended transitions. In this procedure, we assume a position angle (32\arcdeg) and inclination angle (38.3\arcdeg) of the disk, a central stellar mass of $1.29\,M_\odot$, and a distance of 400\,pc, based on previous works \citep{Cieza2016, Tobin2023}. We also assume that the emission originates from the midplane of the disk, and do not consider the vertical extent of the emission. The uncertainties of the spectra are calculated per channel basis within the \texttt{integrated\_spectrum()} function taking into account the spectral decorrelation \citep{Yen2016}.

As shown in Figure \ref{fig:spectra_gallery}, numerous spectral features are detected in the disk-integrated spectra. Since many of the COM transitions exist in narrow frequency ranges and overlap between different molecular species, visual identification of the lines is challenging. Alternatively, we fitted a synthetic spectrum to the observed spectra to identify each of these features. The details of the synthetic spectral model are described in Appendix \ref{appendix:spectral_model}. Briefly, the model used a common emitting region size, a common excitation temperature, and a common line width under the local thermodynamic equilibrium (LTE) conditions. We varied the emitting region size, excitation temperature, line width, and the column density of each species to fit the observed spectra, and test if they could be reproduced with the model. To construct the model, the spectroscopic data of the molecular lines are taken from the Jet Propulsion Spectroscopy \citep[JPL;][]{JPL} and the Cologne Database for Molecular Spectroscopy \citep[CDMS;][]{CDMS1, CDMS2, CDMS3}, as detailed in Appendix \ref{appendix:spectroscopic_data}. Only the transitions with an Einstein A coefficient for spontaneous emission of $\geq10^{-8}$\,s$^{-1}$ and an upper state energy of $\leq1000$\,K, which are detectable under the physical condition of the V883 Ori disk, are considered in the model.

Following \citet{Jorgensen2020}, we set two criteria for the identification of molecular species. First, the synthetic spectrum of multiple (more than one) transitions of a given species accounts for the observed spectra including at least one detected transitions that is not blended with transitions of other species. Second, no undetected transitions are over-predicted by the synthetic spectrum. Species that fulfill the second criterion but with no unblended transitions detected are considered as tentative identifications. Based on these criteria, we robustly identified 10 species and their isotopologues: \methanol, \acetaldehyde, \methylformate, \dimethylether, \ethyleneoxide, \acetone, \propenal, \propanal, $^{13}$CH$_3$OCHO, and CH$_3$O$^{13}$CHO. We also tentatively identified 6 species and isotopologues: CH$_3^{13}$CHO, $^{13}$CH$_3$OCH$_3$, CH$_2$DOH, CH$_2$DCHO, CH$_3$CDO, and C$_2$H$_3$CN. In addition, one transition of each sulfur-bearing molecule (OCS and SO$_2$) was detected. The detected transitions are listed in Table \ref{tab:transitions} in Appendix \ref{appendix:transitions}. In addition to the disk-integrated spectra for the full frequency range in Figures \ref{fig:disk-integrated_spectra0}--\ref{fig:disk-integrated_spectra_fit3}, we also present the zoom-in spectra of $^{13}$CH$_3$OCHO and CH$_3$O$^{13}$CHO in Figure \ref{fig:13C-methylformate_spectra} in Appendix \ref{appendix:spectra} for the purpose of visually clarity to confirm the detection of these species.


\begin{figure*}
\epsscale{1.15}
\plotone{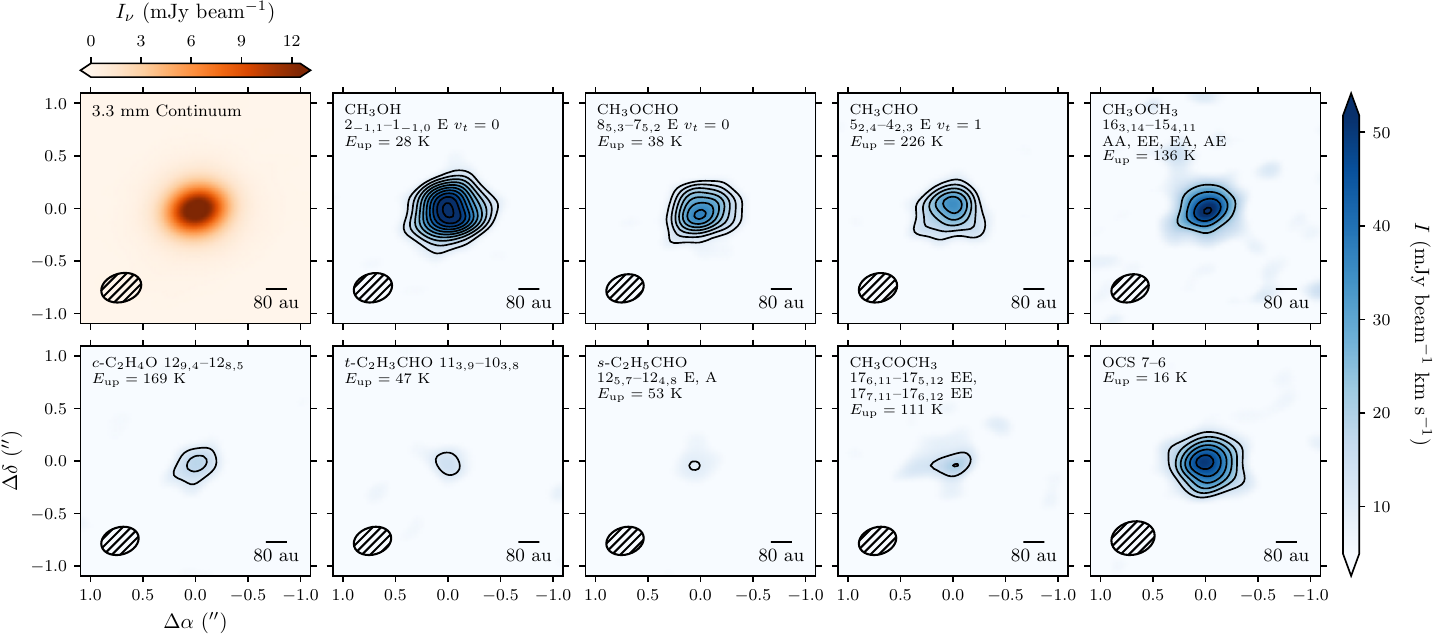}
\caption{The 3.3 mm dust continuum image (left top) and the velocity-integrated intensity maps of the molecular line emission (others) in the V883 Ori disk. The velocity range for integration are $\pm3.5$ km s$^{-1}$ with respect to the source systemic velocity $v_\mathrm{sys}=4.25$ km s$^{-1}$ \citep{Tobin2023} including \dimethylether and \acetone, where multiple blended transitions are integrated together. The molecular species, transitions, and upper state energy levels are indicated in the upper-left corner of each panel. The black contours start from 5$\sigma$ with steps of 2.5$\sigma$, where $\sigma$ are the noise level of each map measured on the emission-free region, spanning $\sigma = 1.8$--4.4 mJy beam$^{-1}$ km s$^{-1}$ depending on the maps. The synthesized beam and a scale bar of 80 au are shown in the lower left and right corner of each panel, respectively.}
\label{fig:moment_zero_gallery}
\end{figure*}

\begin{figure*}
\epsscale{1.15}
\plotone{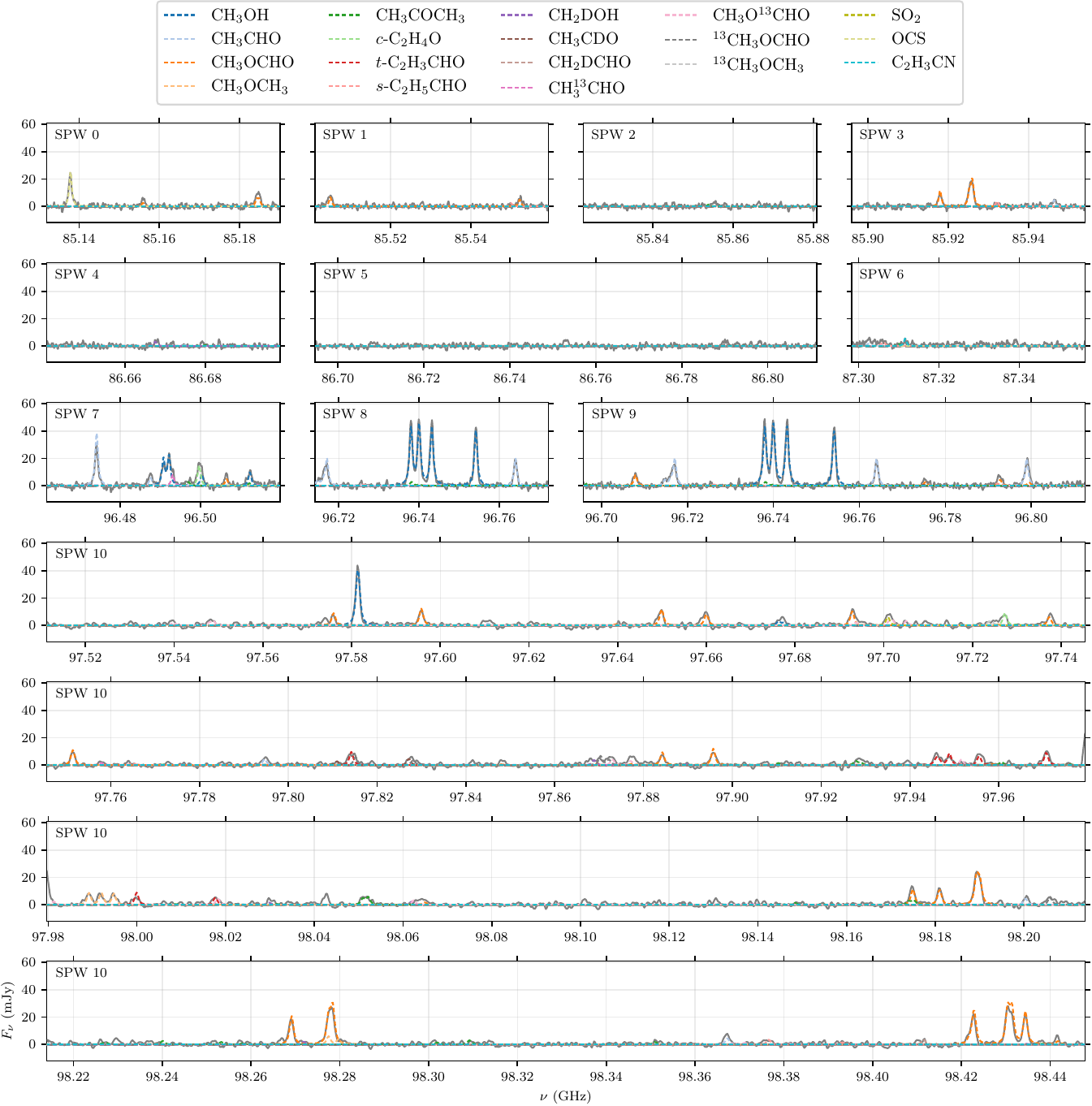}
\caption{Spectra corrected for Keplerian rotation toward the V883 Ori disk (gray). The model spectra for each species are shown in colored lines. }
\label{fig:spectra_gallery}
\end{figure*}





\subsection{Spatial Distributions}
\subsubsection{Velocity-integrated Intensity Maps}
Figure \ref{fig:moment_zero_gallery} shows the velocity-integrated intensity map (zeroth moment map) of the selected transitions without significant blending with other species. These maps are created using \texttt{bettermoments} \citep{bettermoments} by integrating over the velocity range of $\pm3.5$ km s$^{-1}$ with respect to the source systemic velocity ($v_\mathrm{sys} = 4.25$ km s$^{-1}$; \citealt{Tobin2023}) without any masking. For \dimethylether and \acetone, multiple blended transitions of each species are integrated together. For these molecules, the integration range is the combination of the default integration range ($\pm3.5$ km s$^{-1}$) for the multiple blended transitions. The molecular line emission is all confined to the disk region ($\sim0\farcs3$ or $\sim120$ au radius) and is marginally spatially resolved. While the spatial extent of the emission is consistent with previous ALMA observations in Band 6 and Band 7 \citep{vantHoff2018, Lee2019, Tobin2023} considering the difference in the beam size, the central emission cavities seen in the Band 6/7 data are not detected in these maps at the current spatial resolution of $\sim0\farcs3$.

\subsubsection{Line Profile Analysis}\label{subsubsec:line_profile_analysis}
\begin{figure*}
\epsscale{1.15}
\plotone{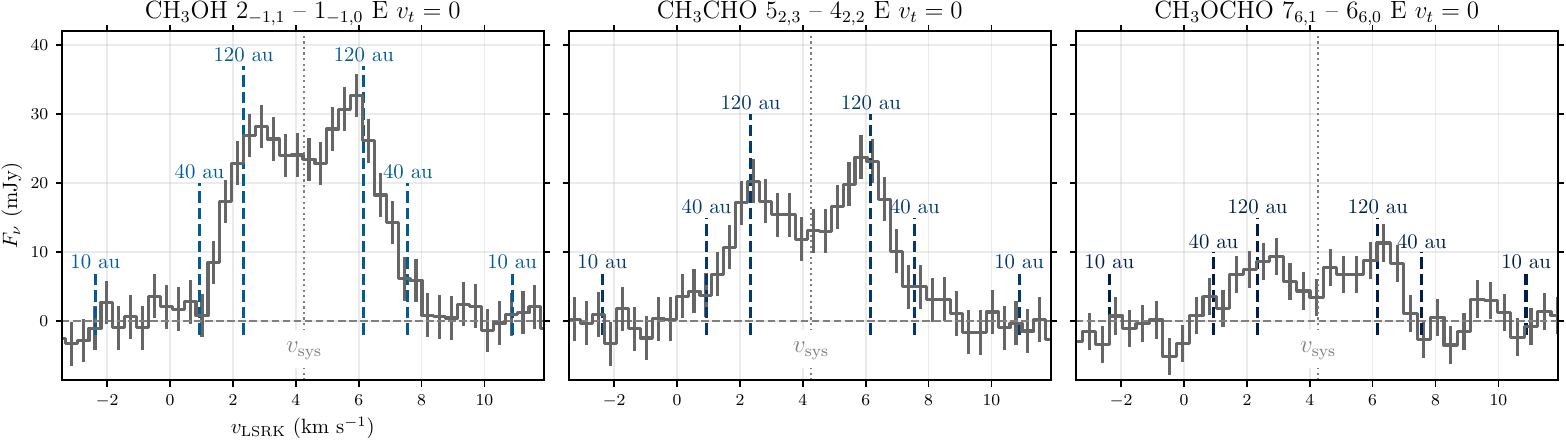}
\caption{Line profiles of \methanol $2_{-1,1}$ -- $1_{-1,0}$ E $v_t=0$, \acetaldehyde $5_{2,3}$ -- $4_{2,2}$ E $v_t=0$, and \methylformate $7_{6,1}$ -- $6_{6,0}$ E $v_t=0$. The vertical gray  dotted line marks the systemic velocity (4.25 km s$^{-1}$). The vertical blue dashed lines indicate the corresponding disk radii at each velocity channel based on the maximum velocity of Keplerian rotation in Equation (\ref{eq:Keplerian_rotation}). The horizontal gray dashed line indicates the zero-flux level. There are no significant emission at the velocity channels which corresponds to inside $\sim$ 40\,au radius.}
\label{fig:line_profile}
\end{figure*}

As the disk emission is only marginally spatially resolved, it is difficult to directly infer the distribution of the emission in the inner region, particularly the emission cavity seen in the sub-mm line observations in Bands 6 and 7 \citep{vantHoff2018, Lee2019, Tobin2023}. Alternatively, we can indirectly probe the emission in the inner region, which is not spatially resolved in the velocity-integrated intensity maps (Figure \ref{fig:moment_zero_gallery}), from the line profile by assuming gas kinematics. 
For the line emission of gas in a Keplerian-rotating disk, the high-velocity components with respect to the systemic velocity, or the line wings, purely contain information about the emission from the inner region. More specifically, the maximum velocity of Keplerian rotation as a function of disk radius ($v_\mathrm{max}(r)$) can be written as 
\begin{equation}\label{eq:Keplerian_rotation}
    v_\mathrm{max}(r) = \sqrt{\frac{GM_\star}{r}}\sin i,
\end{equation}
where $G$ is the gravitational constant, $M_\star$ is the central stellar mass, and $i$ is the inclination of the disk. The V883 Ori disk has been revealed to be a Keplerian-rotating disk by the higher-resolution observations in Bands 6 and 7 \citep{Cieza2016, Lee2019}. To confirm this in the present Band 3 data, we examined the line profiles of three bright transitions of \methanol, \acetaldehyde, and \methylformate (Figure \ref{fig:line_profile}). These line profiles are all double-peaked, consistent with the Keplerian rotation. We then compared these profiles with the maximum velocity of Keplerian rotation at certain radii (Equation \ref{eq:Keplerian_rotation}) in Figure \ref{fig:line_profile}. As seen in the line wings of each line profile, no significant emission is detected in the velocity channels corresponding to $\lesssim$ 40 au radius. The channel maps (Figures \ref{fig:channelmap_CH3OH}, \ref{fig:channelmap_CH3CHO}, and \ref{fig:channelmap_CH3OCHO}) also show no significant emission at $\lesssim$ 40 au radius.

\begin{figure}
\plotone{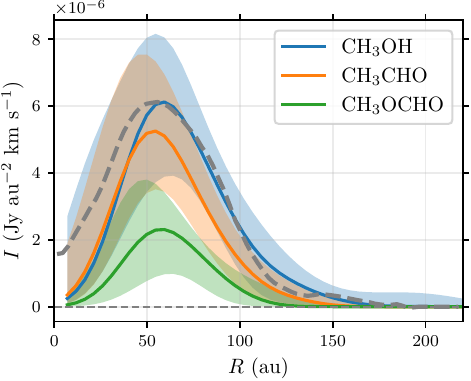}
\caption{Reconstructed radial intensity profiles of \methanol (blue), \acetaldehyde (orange), and \methylformate (green) transitions. The horizontal gray dashed line indicates the zero-intensity level. The gray dashed curve shows the radial intensity profile of \methanol transition in Band 6 created by deprojection and azimuthal averaging on the higher-resolution ($\sim$ 0\farcs1) velocity-integrated intensity maps \citep{Tobin2023}. The Band 6 intensity profile is normalized by its radial peak being matched to the radial peak of the Band 3 \methanol profile.}
\label{fig:radial_profile_from_line_profile}
\end{figure}

To quantitatively assess the contribution of the emission at each radius, we used a forward modeling approach to reconstruct the radial intensity profiles from the line profiles. Using the relation in Equation (\ref{eq:Keplerian_rotation}), the radial intensity profile (as a function of radius) can be constructed from the line profile (as a function of velocity) as demonstrated in \citet{Bosman2021}. We used an approach similar to that of \citet{Bosman2021} to construct the radial intensity profiles of the aforementioned three bright transitions from their line profiles. Details of the modeling can be found in Appendix \ref{appendix:line_profile_analysis_method}. Figure \ref{fig:radial_profile_from_line_profile} shows the reconstructed radial intensity profiles of these three transitions. All three transitions show the depression in the innermost region with a radial peak at $\sim$ 50--60 au, which is consistent with the higher-resolution Band 6 profile \citep{Tobin2023}. This suggests that the bulk of the observed Band 3 emission can be explained by emission from the 40--80 au region, and that the inner emission cavity likely exists even in the Band 3 emission as well. 

\subsection{Column Density Retrieval}\label{subsec:spectral_fit}
To derive the disk-averaged column density of each species, we first employ a simultaneous fit of an LTE spectral model to the extracted spectra, taking into account the spectral blending and line optical depth. The same spectral model is used as for line identification (see Appendix \ref{appendix:spectral_model} for details), where all (tentatively) identified species as in Section \ref{subsec:line_identification} are considered. We excluded the optically thick ($\tau > 0.2$) transitions from the fit (except for OCS which only have one detected transition) to minimize the optical depth effect; the selection is based on the optical depth at the line center in the best-fit model for line identification (Section \ref{subsec:line_identification}). Specifically, we excluded the transitions of abundant species (e.g., \methanol, \methylformate, and \acetaldehyde) with a large Einstein A coefficient and a relatively low upper state energy. The excluded transitions are indicated in Table \ref{tab:transitions} in Appendix \ref{appendix:transitions}. The optical depth threshold of 0.2 is determined to minimize the optical depth effect while leaving a sufficient number of transitions available for the fit. When higher optical depth threshold of $\tau > 0.4$ is adopted, the resulting column densities are lower by a factor of $\sim2$ or less. Although such difference does not change the conclusion of this paper, we employ the threshold value of 0.2 to discuss the result carefully.

\begin{figure*}
\epsscale{1.15}
\plotone{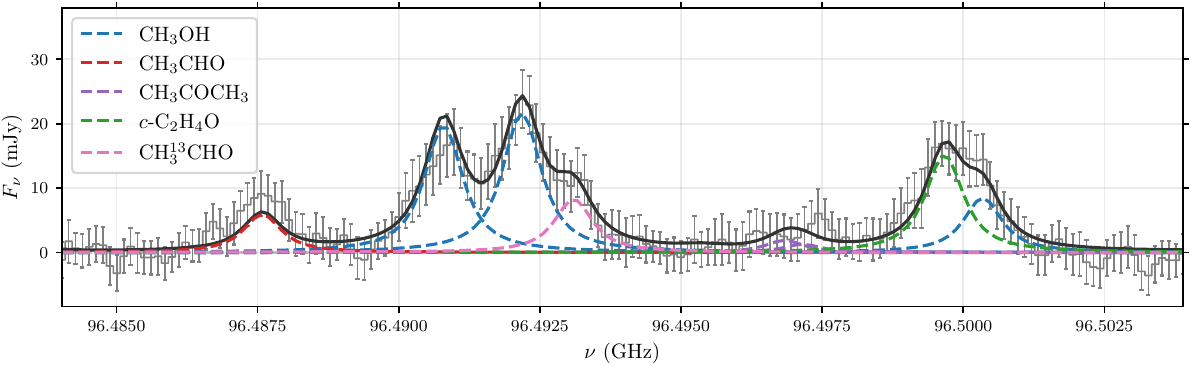}
\caption{Demonstration of the simultaneous spectral fit for a selected frequency range in SPW 7. The observed spectra (gray) is well reproduced by the model (black) composed of multiple blended transitions of different species, \methanol, \acetaldehyde, \acetone, \ethyleneoxide, and CH$_3^{13}$CHO (dashed colored lines).}
\label{fig:spectral_fit_demo}
\end{figure*}

Since the emission is only marginally spatially resolved, we assumed that the emitting region in Band 3 is the same as that of \methanol in Band 6 \citep{Tobin2023}. The inner and outer radius of the emitting area are fixed to 0\farcs1 and 0\farcs3, respectively. 
In addition to the (tentatively) identified transitions in Section \ref{subsec:line_identification}, we included several undetected species in the spectral fit: $^{13}$CH$_3$CHO, CH$_2$DOCHO, and CH$_3$OCDO. In total, we consider 24 free parameters: excitation temperature ($T_\mathrm{ex}$), line width ($\Delta V_\mathrm{FWHM}$), column density of 21 species, and an additional parameter $\gamma$ for line broadening (see Appendix \ref{appendix:spectral_model}). \textrm{For the justification of the common excitation temperature and line width, we measured the line width of each isolated transitions with Gaussian fits, and found no systematic variations depending on the molecular species nor the upper state energies.}
We used the affine-invariant Markov Chain Monte Carlo (MCMC) algorithm implemented in the \texttt{emcee} Python package \citep{emcee} to explore the parameter spaces. We run 100 walkers for 15000 steps, and the initial 12000 steps are discarded as burn-in. Figure \ref{fig:spectral_fit_demo} demonstrates the unique ability of the simultaneous spectral fit to fully exploit the transitions for column density derivation even with a severe spectral blending. 

The results of the fits are summarized in Table \ref{tab:specfit_result}. The derived excitation temperature is $106.7_{-3.8}^{+4.3}$ K, similar to the typical sublimation temperature of COMs ($\sim100$\,K), indicating the thermal sublimation of molecules. The column densities of COMs are $\sim$10$^{15}$--10$^{18}$ cm$^{-2}$, broadly in agreement with previous estimates using Band 7 data \citep{Lee2019}. \textrm{The uncertainties of the derived quantities are 16th and 84th percentiles of the posterior distributions, where only the statistical uncertainty is considered. While the column density itself could have a larger uncertainty due to the absolute flux calibration uncertainty of 2.5\% (1$\sigma$), the column density ratios, from which we discuss the chemistry of COMs in the following sections, are not affected by the absolute flux calibration uncertainty.}

\begin{deluxetable}{lccc}
\tablecaption{Column Density of Molecules}
\tablehead{\colhead{Molecule} & \colhead{$N$ (cm$^{-2}$)}}
\startdata
CH$_3$OH & $3.9_{-0.6}^{+0.7} \times 10^{18}$ \\ 
CH$_3$CHO & $6.5_{-0.5}^{+0.5} \times 10^{17}$ \\ 
CH$_3$OCHO & $7.4_{-0.3}^{+0.3} \times 10^{17}$ \\ 
CH$_3$OCH$_3$ & $5.7_{-0.5}^{+0.5} \times 10^{17}$ \\ 
CH$_3$COCH$_3$ & $4.4_{-0.5}^{+0.5} \times 10^{16}$ \\ 
$c$-C$_2$H$_4$O & $3.5_{-0.4}^{+0.5} \times 10^{16}$ \\ 
$t$-C$_2$H$_3$CHO & $5.0_{-0.3}^{+0.3} \times 10^{15}$ \\ 
$s$-C$_2$H$_5$CHO & $1.8_{-0.3}^{+0.3} \times 10^{16}$ \\ 
CH$_2$DOH$^\ast$ & $2.2_{-0.4}^{+0.4} \times 10^{16}$ \\ 
CH$_3$CDO$^\ast$ & $5.5_{-1.3}^{+1.3} \times 10^{15}$ \\ 
CH$_2$DCHO$^\ast$ & $2.8_{-0.8}^{+0.8} \times 10^{16}$ \\ 
CH$_3^{13}$CHO$^\ast$ & $2.3_{-0.8}^{+0.8} \times 10^{16}$ \\ 
$^{13}$CH$_3$CHO & $< 1.5 \times 10^{17}$ \\ 
CH$_3$O$^{13}$CHO & $3.2_{-0.3}^{+0.4} \times 10^{16}$ \\ 
$^{13}$CH$_3$OCHO & $3.2_{-0.3}^{+0.3} \times 10^{16}$ \\ 
CH$_2$DOCHO & $< 1.2 \times 10^{16}$ \\ 
CH$_3$OCDO & $< 1.4 \times 10^{16}$ \\ 
$^{13}$CH$_3$OCH$_3$$^\ast$ & $2.3_{-0.6}^{+0.6} \times 10^{16}$ \\ 
SO$_2$ & $3.0_{-0.7}^{+0.7} \times 10^{16}$ \\ 
OCS & $5.5_{-1.3}^{+2.0} \times 10^{16}$ \\ 
C$_2$H$_3$CN$^\ast$ & $2.2_{-0.9}^{+0.9} \times 10^{15}$ \\ 
\enddata
\tablenotetext{\dagger}{The excitation temperature and line width, which are assumed to be common for all molecules, are estimated to be $106.8_{-3.9}^{+4.2}$\,K and $0.71_{-0.05}^{+0.07}$\,km\,s$^{-1}$, respectively.}
\tablenotetext{\ddagger}{Uncertainties are the 16th and 84th percentile of the posterior distributions, where only the statistical uncertainty is included. The upper limits are determined based on the 99.7th percentile of the posterior distributions.}
\tablenotetext{\ast}{\textrm{Based on the tentatively detected transitions and therefore should be considered as upper limits.}}
\label{tab:specfit_result}
\end{deluxetable}

\section{Discussion} \label{sec:discussion}
\subsection{Comparison of the spatial distributions with the sub-mm observations}
The spatial distributions of the COM emission in Band 3, where the dust continuum emission would be fainter than in the sub-mm regime, are essential information to infer the origin of the inner cavity seen in the sub-mm observations. If the central region where the emission cavity is seen in Band 6/7 is filled with COM emission in Band 3, it would be evidence that the COM emission is reduced in the sub-mm observations due to the intense dust continuum emission. 

We found that the reconstructed radial intensity profiles (Figure \ref{fig:radial_profile_from_line_profile}) show the inner depression of the emission while the spatial distributions of the COM emission in Band 3 are consistently centrally-peaked in the velocity-integrated intensity maps (Figure \ref{fig:moment_zero_gallery}). This indicates that the inner emission cavity is smeared out by the beam on the velocity-integrated intensity maps. We confirmed that the inner emission cavity of a \methanol line in Band 6\footnote{Taken from the Harvard Dataverse repository (\url{https://dataverse.harvard.edu/dataset.xhtml?persistentId=doi:10.7910/DVN/MDQJEU)})} \citep{Tobin2023} is smeared out when the image is smoothed to the same spatial resolution as that of the presented Band 3 data.
Although the uncertainties of the reconstructed radial intensity profiles are quite large, the profiles all have radial peaks at $\sim$ 50--60\,au, which is consistent with the higher-resolution Band 6 profile (Figure \ref{fig:radial_profile_from_line_profile}). Therefore, it is likely that the emission observed in Band 3 mainly traces the components outside the emission cavity ($\gtrsim 40\,$ au, i.e., the same component as that traced by the sub-mm observations). 

We note, however, that there are some additional systematic uncertainties in the reconstructed radial intensity profiles (Figure \ref{fig:radial_profile_from_line_profile}). In reality, the molecular line emission could originate from the disk surface \citep[or the warm molecular layer,][]{Aikawa2002, Law2021, Law2022, Law2023, PanequeCarreno2023}, where the Keplerian velocity is slightly smaller than that in the midplane. 
This invalidates the assumption made in the reconstruction of the radial intensity profiles that the emission comes entirely from the disk midplane. Therefore, while the depression of the emission in the innermost region is robust, the actual emission cavity sizes in Band 3 emission are rather uncertain. In addition, transitions with different optical depths or upper state energies, which could trace the different disk heights, may have different emission distributions. 

\subsection{The physical and chemical structure of the innermost region of the V883 Ori disk}\label{subsec:physicochemical_structure}
Our line profile analyses suggest that the COM line emission in Band 3 also shows the inner emission cavity at $\lesssim$ 40 au radius. Here we discuss the possible origins of the emission cavity and the associated physical and chemical structures of the V883 Ori disk.

The first possible explanation for the emission cavity of the COM emission in Band 3 is that the dust continuum emission is still too intense even in Band 3 to detect the line emission in the innermost disk. In the sub-mm wavelengths, the dust continuum emission at $\lesssim$ 40 au radius has been shown to be optically thick ($\tau \gtrsim 2$) by the intra-band analysis of the ALMA Band 6 observations with a spatial resolution of $\sim0\farcs04$ \citep{Cieza2016}. Since the dust opacity is generally smaller in lower frequencies, the dust continuum emission is expected to be more optically thin in the longer wavelengths. However, the factors that affect the line intensity are not only the dust optical depth, but also the temperature of the dust emitting layer. We consider a simple slab model for the vertical structure of a disk. If the line-emitting layer is well separated from the dust-emitting layer (near the midplane) and is closer to the observer \citep[see e.g.,][]{Bosman2021}, the observed line intensity after the continuum subtraction is expressed as 
\begin{align}\label{equation:line_intensity_separated}
    I_\mathrm{L-C} &= B_\nu(T_\mathrm{gas})(1 - e^{-\tau_\mathrm{line}}) + I_{\nu, \mathrm{dust}}e^{-\tau_\mathrm{line}} - I_{\nu, \mathrm{dust}} \notag \\
    &= (B_\nu(T_\mathrm{gas}) - I_{\nu, \mathrm{dust}})(1 - e^{-\tau_\mathrm{line}}),
\end{align}
where $B_\nu$ is the Planck function for blackbody radiation, $T_\mathrm{gas}$ is the gas temperature representing the temperature of the line emitting region, $\tau_\mathrm{line}$ is the optical depth of the line emission at the line center, and $I_{\nu, \mathrm{dust}} = \chi B_\nu(T_\mathrm{dust})(1 - e^{-\tau_\mathrm{dust}})$ is the intensity of the dust continuum emission. Here $\chi$ is the intensity reduction factor due to the scattering effect \citep[e.g.,][]{Bosman2021}, $T_\mathrm{dust}$ is the dust temperature, and $\tau_\mathrm{dust}$ is the optical depth of the dust continuum emission. Equation (\ref{equation:line_intensity_separated}) is an extension of the situation discussed in \citet{Bosman2021}, where the dust continuum emission is optically thick (i.e., $1 - e^{-\tau_\mathrm{dust}} \sim 1$) and the temperature of the line- and dust-emitting region is the same (i.e., $T_\mathrm{gas} = T_\mathrm{dust}$, vertically isothermal disk).

In archetypal disks where the temperature structure is mainly determined by the passive irradiation from the central star, the temperature in the line-emitting region ($T_\mathrm{gas}$) is usually much higher than the temperature in the disk midplane, from which the dust emission is mostly originating ($T_\mathrm{dust}$), allowing us to observe the line emission originating from the disk surface. However, in the case of the V883 Ori disk, where the disk temperature is expected to be elevated due to the accretion outburst, viscous heating may be efficient in the midplane, which could lead to a higher temperature in the disk midplane ($T_\mathrm{dust}$\textrm{; \citealt{Alarcon2023}}). Indeed, \citet{Lee2019} suggest that the intensity depression of the COM emission can be reproduced by the presence of the additional heating in the midplane. Although the intensity of the dust emission also depends on the scattering effect ($\chi$), this warmer temperature in the midplane makes it even more difficult to detect the line emission originating from its emitting layer closer to the observer, even if the optical depth of the dust continuum emission is not very high. 

It is also possible that the bulk of the line emission in the innermost region ($\lesssim$ 40 au radius) is well mixed with the dust emission and originates from the disk midplane. In this case, the observed line intensity after the continuum subtraction will completely disappear if the line opacity is negligible compared to the dust opacity (see \citealt{Bosman2021}, Equation (4)). 



Another possible explanation for the inner emission cavity is that the molecules at $\lesssim40$ au radius are destroyed by the strong UV/X-ray radiation from the central protostar. The UV radiation can destroy COMs via photodissociation \citep[e.g.,][]{Garrod2006, Oberg2009} in the region where the dust is optically thin. This UV effect will thus be effective only in low-density regions (e.g., disk surfaces) where the dust could be optically thin.
On the other hand, X-rays can penetrate into higher-density regions than UV radiation.
Recently, \citet{Notsu2021} modeled the effect of the X-ray radiation from the central protostar on the chemistry of the protostellar envelopes, and found that the gas-phase fractional abundances of \methanol and other COMs within their snowlines decrease as the X-ray luminosity of the central protostar increases. In the presence of the strong X-ray radiation, COMs are mainly destroyed by the X-ray-induced photodissociation \citep[e.g.,][]{Garrod2006, Taquet2016, Notsu2021}. 
In the model by \citet{Notsu2021}, \methanol is destroyed in $\sim\,10^3$\,yr when the X-ray ionization rate is $\sim10^{-13}$\,s$^{-1}$ in the infalling envelope. Here we use the ionization rate rather than the flux, since the former is the direct input parameter for the chemical reaction network models. While the destruction timescale may weakly depend on the gas density, which could be different between the inner \textrm{$\lesssim100$\,au radius region of the protostellar} envelopes and the $\lesssim40$\,au radius region of the V883~Ori disk, \methanol and other COMs could be destroyed in $\sim100$ yr if the X-ray ionization rate is $\gtrsim 10^{-12}$\,s$^{-1}$.
We note that \citet{Kuhn2019} reported that X-ray luminosities and plasma temperature of X-ray radiation in FU Ori type stars tend to be larger than in typical non-bursting YSOs, based on the comparison with a sample of low-mass stars in the Orion Nebula Cluster.
In addition, cosmic-rays accelerated near the central star (in strongly magnetized shocks along the outflow or in accretion shocks near the stellar surface) could penetrate into the higher-density disk midplane and cause the destruction of COMs via cosmic-ray-induced photodissociation \citep[e.g.,][]{Padovani2020, Cabedo2023}.


In summary, the inner cavity of the COM emission in Band 3 can be explained by the absorption of the molecular line emission by the bright dust continuum emission, and/or by the chemical destruction of the molecules. Our ALMA Band 3 observations suggest that sensitive observations at longer wavelengths, where the dust continuum emission should be fainter, with ALMA Band 1, Karl G. Jansky Very Large Array (VLA), and the future next generation Vary Large Array (ngVLA), are essential to directly probe the chemistry of the innermost region of the V883 Ori disk.

\subsection{Chemical abundance ratios of COMs}\label{subsec:abundance_ratio}
\begin{figure}
\epsscale{1.15}
\plotone{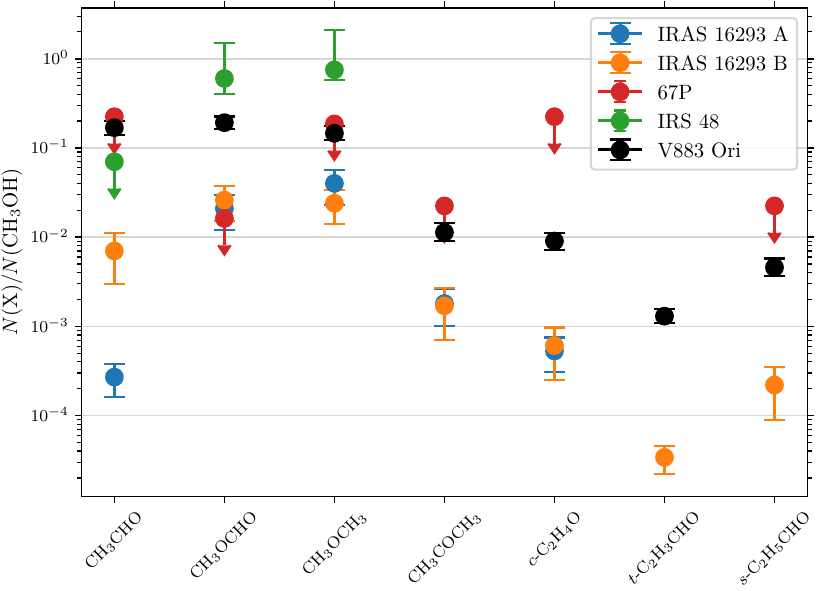}
\caption{Comparison of the COM abundance ratios with respect to \methanol among different evolutionary stages including the V883 Ori disk measured in the present work. The data point with a down arrow indicate an upper limit. We compiled the literature values for the protostellar envelopes of IRAS 16293-2422 A/B \citep{Lykke2017, Jorgensen2018, Manigand2020, Manigand2021}, the comet 67P/C-G \citep{Rubin2019, Schuhmann2019}, and the IRS 48 disk \citep{Brunken2022}. Note that while all the cometary values are limited to be upper limits because cometary measurement (mass spectra) cannot distinguish the isomeric molecules with the same mass (e.g., \acetaldehyde and \ethyleneoxide pair and \acetone and \propanal pair), the upper limit of \acetaldehyde in the IRS 48 disk is due to non-detection.}
\label{fig:ratio_comparison}
\end{figure}

The V883 Ori disk is a unique and ideal target for disk chemistry since we can observe fresh sublimates and probe the chemical composition of COMs which is hidden in typical colder disks. Here we discuss the chemical composition of the COMs detected in our observations and their implications for the chemical evolution during star and planet formation. 
Figure \ref{fig:ratio_comparison} compares the column density ratios of COMs with respect to \methanol in the V883 Ori disk with those in different evolutionary stages, including the warm inner envelopes of the Class 0 protobinary IRAS 16293-2422 A/B \citep[][see also \citealt{Drozdovskaya2019}]{Lykke2017, Jorgensen2018, Manigand2020, Manigand2021}, the protoplanetary disk around the Herbig Ae star Oph-IRS~48 \citep{Brunken2022}, and the solar system comet 67P/C-G \citep[][see also \citealt{Drozdovskaya2019}]{Rubin2019, Schuhmann2019}. IRAS 16293-2422 is a famous ``hot corino'' source, extensively studied in the ALMA Protostellar Interferometric Line Survey \citep[PILS;][]{Jorgensen2016}, where we can observe COM-rich warm gas sublimated from ices. The Oph-IRS~48 disk is the only disk so far to show emission of multiple COM species. A localized COM emission offset from the disk center has been observed, which is interpreted to be sublimated from the icy dust mantle stirred up to the warm disk surface \citep{vanderMarel2021, Brunken2022}. The comet 67P/C-G is also a well-known object whose chemical composition has recently been extensively studied by in-situ measurements in the Rosetta mission \citep[e.g.,][]{Altwegg2019}. We found higher abundance ratios of COMs in the V883 Ori disk compared to those in the warm envelopes of IRAS 16293-2422 by a factor of $\gtrsim$ 5--10, indicating that the abundance of chemically complex species is enhanced in protoplanetary disks. \textrm{We note that the COM abundance ratios in protostellar envelopes also show scatter depending on the source \citep{Belloche2020, Yang2021}.} The COM abundance ratios in the V883 Ori disk are similar to those in another disk Oph-IRS~48 and in comet 67P/C-G, except for one species, \methylformate, which shows a lower abundance ratios in the comet compared to the disks. These results suggest that the more complex species than \methanol efficiently form during the evolution from the protostellar envelope to the protoplanetary disks, and that the chemical composition of disks can be inherited to comets without drastic chemical reprocessing. 

In general, the formation of COMs are thought to occur mainly via reactions on dust grain surfaces \citep[][and references therein]{Herbst2009}. In the cold temperature region ($\lesssim 10$\,K), the \methanol ice forms via efficient hydrogenation of the CO ice \citep[e.g.,][]{Tielens1982, Watanabe2002, Watanabe2003}. Therefore, the \methanol ice forms in cold molecular clouds and could be delivered to protoplanetary disks. On the other hand, 
diffusion of heavier radicals (e.g., CH$_3$, CH$_2$OH, and CH$_3$O), formed via the UV/X-ray/cosmic-ray-induced photolysis and hydrogen abstraction of simpler molecules including \methanol, are efficient on the warmer ($\sim$ 30--50\,K) dust grain surfaces. More complex molecules such as \acetaldehyde, \methylformate, \dimethylether, and \acetone could thus be formed via radical-radical reactions.
For example, theoretical and experimental studies \citep[e.g.,][]{Garrod2006, Chuang2016} have shown that \acetaldehyde, \methylformate, \dimethylether, and \acetone are mainly formed via following reactions,
\begin{align}
    \mathrm{HCO} + \mathrm{CH_3} &\longrightarrow \mathrm{CH_3CHO}, \label{eq:r-r_reaction1} \\ 
    \mathrm{HCO} + \mathrm{CH_3O} &\longrightarrow \mathrm{CH_3OCHO}, \label{eq:r-r_reaction2} \\
    \mathrm{CH_3} + \mathrm{CH_3O} &\longrightarrow \mathrm{CH_3OCH_3}, \label{eq:r-r_reaction3} \\
    \mathrm{CH_3} + \mathrm{CH_3CO} &\longrightarrow \mathrm{CH_3COCH_3}. \label{eq:r-r_reaction4}
\end{align}
These reactions may contribute to the efficient formation of these species in protoplanetary disks \citep{Walsh2014, Furuya2014}, which may be the origin of abundant complex species in the V883 Ori disk (Figure \ref{fig:ratio_comparison}).



The present observations also detected several isomeric pairs, i.e., \acetaldehyde and \ethyleneoxide pair and \acetone and \propanal pair. The abundance ratios between isomers are key to constraining the formation pathways. We found a \acetaldehyde/\ethyleneoxide ratio of $9.2_{-1.1}^{+1.2}$ and a \acetone/\propanal ratio of $2.2_{-0.4}^{+0.4}$ in the V883 Ori disk. These ratios are consistent with the values in IRAS 16293B \citep{Lykke2017} within an order of magnitude, suggesting that the isomers are also similarly formed.

\subsection{Isotopic ratios of COMs}
The present observations also detected several isotopologues, including D and $^{13}$C, of some abundant molecules. Here, we discuss the implications of these detections and the $^{12}$C/$^{13}$C and D/H ratios for the isotopic chemistry in protoplanetary disks. 

\subsubsection{\texorpdfstring{$^{12}$C/$^{13}$C}{} ratio}\label{subsubsec:12C/13C}

\begin{figure*}
\plotone{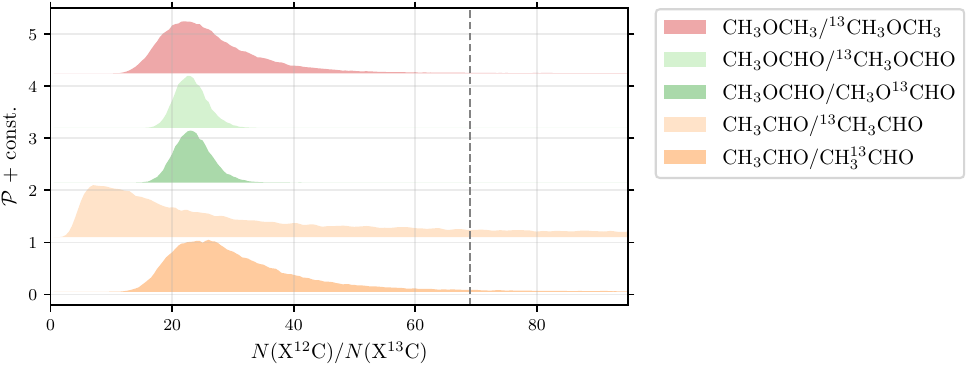}
\caption{Kernel density estimate (KDE) of the posterior probability density distributions of column density ratios of $^{12}$C- to $^{13}$C-isotopologues for \acetaldehyde, \methylformate, and \dimethylether. Two isomers of $^{13}$C-isotopologues are observed for \acetaldehyde and \methylformate. The probability density is normalized by the peak being unity and offset for visual clarity. The ISM value of 69 is marked by the vertical dashed line. \textrm{We note that the $^{12}$C/$^{13}$C ratios of \acetaldehyde and \dimethylether are based on the tentatively detected transitions of $^{13}$C-isotopologues, and therefore should be considered as lower limits.}}
\label{fig:12C13C}
\end{figure*}

We have identified $^{13}$C-methyl formate ($^{13}$CH$_3$OCHO and CH$_3$O$^{13}$CHO) and tentatively identified $^{13}$C isotopologues of acetaldehyde and dimethyl ether. While previous observations reported the detection of $^{13}$CH$_3$OH \citep{Lee2019} and $^{13}$CH$_3$CHO as a potential blended line with the HDO line \citep{Tobin2023}, measurements of $^{12}$C/$^{13}$C ratios in COMs have not been performed. Figure \ref{fig:12C13C} shows the probability density distributions of the $^{12}$C/$^{13}$C ratios of \acetaldehyde, \methylformate, and \dimethylether. The measured $^{12}$C/$^{13}$C ratios are also summarized in Table \ref{tab:12C13C_summary} together with the value in the warm envelopes of IRAS 16293-2422.  The measured $^{12}$C/$^{13}$C ratios are all $\sim$20--30\footnote{We note that the statistical correction has not been applied for $^{13}$\dimethylether, which apparently contains equivalent two carbon atoms in the CH$_3$- functional groups. However, these two carbon atoms are actually not equivalent based on the formation pathway of \dimethylether (Equation \ref{eq:r-r_reaction3}); the formation of \dimethylether (and therefore the inclusion of $^{13}$C into \dimethylether) happens via the reaction between non-equivalent reactants (CH$_3$ and CH$_3$O) unlike e.g., the equivalent hydrogen addition to CO which forms CH$_3$OH and its deuterated isotopologues. We here simply compare the $^{12}$C/$^{13}$C of \dimethylether without statistical correction to other species, although the actual correction factor depends on the $^{12}$C/$^{13}$C ratios of reactants (i.e., CH$_3$ and CH$_3$O), which can be varied among them.}, which is significantly lower than the elemental abundance ratio of $^{12}$C/$^{13}$C in the local ISM ($\sim$ 69; \citealt{Wilson1999}), although the $^{12}$C/$^{13}$C ratios of acetaldehyde and dimethyl ether have a probability at higher values as well due to the tentative detection or non-detection. Interestingly, a similarly low $^{12}$C/$^{13}$C ratio ($21\pm5$) of CO has been reported in the 70--110 au region of the protoplanetary disk around TW Hya by \citet{Yoshida2022_12CO13CO} \citep[see also][]{Zhang2017}. They suggest that the gas-phase isotope-exchange reaction, 
\begin{equation}\label{eq:12C13C_reaction}
    ^{13}\mathrm{C}^+ + ^{12}\mathrm{CO} \to {}^{12}\mathrm{C}^+ + {}^{13}\mathrm{CO} + \Delta E,
\end{equation}
($\Delta E \approx 35$\,K; \citealt{Langer1984, Furuya2011}), with the help of the high gas-phase C/O ratio ($> 1$), would lead to such a low $^{12}$CO/$^{13}$CO ratio in the warm molecular layer. The same mechanism would explain the observed low $^{12}$C/$^{13}$C ratio in COMs if the $^{13}$C-rich CO is incorporated into the ice on dust grains in the disk midplane via vertical mixing and freeze-out \citep{Furuya2022}, from which the COMs such as \acetaldehyde and \methylformate are synthesized via hydrogenation and radical-radical reactions (Equations (\ref{eq:r-r_reaction1})--(\ref{eq:r-r_reaction4})) on dust grain surfaces.

An alternative mechanism that could alter the $^{12}$CO/$^{13}$CO ratio is the difference in the binding energies between the two isotopologues. \citet{Smith2015} proposed that the slightly higher binding energy of $^{13}$CO than that of $^{12}$CO ($840\pm4$\,K and $835\pm5$\,K on $^{12}$CO ice; \citealt{Smith2021}) could explain the high $^{12}$CO/$^{13}$CO ratios in the gas phase ($\sim85$--165) observed toward several low-mass young stellar objects. The difference in binding energies leads to the sublimation of $^{12}$CO at a slightly lower temperature than $^{13}$CO, making the gas-phase CO $^{13}$C-poor (and CO ice $^{13}$C-rich). However, this fractionation mechanism should work only in a very narrow temperature range. Also, the difference in binding energy between $^{12}$CO and $^{13}$CO is rather uncertain. It is therefore speculative that this mechanism contributes to the $^{13}$C fractionation. In addition, the gas and dust dynamics in disks (e.g., vertical turbulent mixing) is a critical unknown factor for this mechanism to work for the $^{13}$C fractionation of CO and consequently of COMs \citep[e.g.,][]{Yoshida2022_12CO13CO}.

\textrm{The molecular $^{12}$C/$^{13}$C ratios have been measured in different evolutionary stages \citep{Nomura2023_PPVII}. In cold dense cores, single-dish observations have revealed that CO are enriched in $^{13}$C \citep{Agundez2019}, potentially indicating the isotope-exchange reaction (Equation \ref{eq:12C13C_reaction}) in the cold environment.} In the warm envelope around the Class 0 protostar IRAS 16293B, similar lower values of the $^{12}$C/$^{13}$C ratio ($\sim$ 30) have been observed in some COMs (CH$_2$(OH)CHO (glycolaldehyde), \dimethylether, and possibly \methylformate) \citep[see Table \ref{tab:12C13C_summary};][]{Jorgensen2016, Jorgensen2018}, but for IRAS 16293A no evidence for $^{13}$C fractionation in COMs has been observed \citep[Table \ref{tab:12C13C_summary};][]{Manigand2020}. CO (and thus COMs) can be enriched in $^{13}$C via the exchange reaction (Equation \ref{eq:12C13C_reaction}) only when C/O ratio is greater than unity (or more specifically, CO is not the dominant C reservoir). In a rotationally supported disk, the C/O ratio could exceed unity due to the decoupling of the ice-coated grains from the gas. The low $^{12}$C/$^{13}$C ratio of some COMs in IRAS 16293B may indicate that similar dust-gas decoupling already occurs in the protostellar envelope \citep[e.g.,][]{Koga2022}. 

We note that the elemental abundance ratios of C/O and C/H in the gas phase become lower with increasing ionization rate in the disk ($\gtrsim10^{-17}$ s$^{-1}$; \citealt{Eistrup2016, Schwarz2018, Notsu2020}), which may inhibit the increase of $^{13}$CO abundance via the exchange reaction (Equation \ref{eq:12C13C_reaction}) \citep{Woods2009}. Alternatively, it may be possible that the destruction of carbonaceous grains enhances the C/H abundance and C/O ratio in protostellar disks and envelopes \citep[e.g.,][]{Wei2019, vantHoff2020}, although it is unclear whether the destruction of carbonaceous grains efficiently occurs in the outer cold region where CO and COMs form ($\lesssim100$ K).

In the comet 67P/C-G, only the \methanol/$^{13}$CH$_3$OH ratio has been measured ($91\pm10$; \citealt{Altwegg2020_isotopes}, see Table \ref{tab:12C13C_summary}), showing no $^{13}$C fractionation from the solar value \citep[$\sim89$; e.g.,][]{Mumma2011}.
The $^{12}$C/$^{13}$C ratios of simpler molecular species such as CO, CO$_2$, and H$_2$CO have also been measured, but no $^{13}$C fractionation has been observed \citep{Hassig2017, Rubin2017} except for H$_2$CO ($40\pm14$; \citealt{Altwegg2020_isotopes}). This trend is in contrast to the observed $^{13}$C fractionation in COMs in the V883 Ori disk. Although the origin of this difference is unclear, a speculative explanation could be the formation environment of the comet 67P/C-G; it may be formed from the dust grains with non-$^{13}$C-rich ices, which can be produced from lower gas-phase C/O ratios ($\lesssim 1$). \citet{Yoshida2022_12CO13CO} reported a radial variation of the gas-phase $^{12}$CO/$^{13}$CO ratio in the disk around TW~Hya, where the outer region ($>130$ au) shows a higher $^{12}$CO/$^{13}$CO ratio ($>84$) than that in the inner region (70--110 au; $21\pm5$). If a similar radial variance in the $^{12}$CO/$^{13}$CO ratio existed in the proto-solar disk, the $^{13}$C fractionation in comets depends on the formation location within the disk.




\begin{deluxetable*}{lcccc}
\tablecaption{$^{12}$C/$^{13}$C Ratios of COMs}
\tablehead{\colhead{} & \colhead{V883 Ori (this work)$^\dagger$} & \colhead{IRAS~16293A\tablenotemark{a}} & \colhead{IRAS~16293B\tablenotemark{b}} & \colhead{67P/C-G\tablenotemark{c}}}
\startdata
Methanol & & & \\
\quad \methanol/$^{13}$CH$_3$OH & --- & $65\pm27$ & --- & $91\pm10$ \\
Acetaldehyde & & & \\
\quad \acetaldehyde/CH$_3^{13}$CHO & (29$_{-8}^{+16}$)$^{\ast}$ & --- & 67$^{\ddagger}$ & --- \\
\quad \acetaldehyde/$^{13}$CH$_3$CHO & $> 4.2$ & --- & 67$^{\ddagger}$ & ---\\
Methyl Formate & & & \\
\quad \methylformate/CH$_3$O$^{13}$CHO & 23$_{-3}^{+3}$ & $75\pm32$ & (41)$^{\ast}$ & --- \\
\quad \methylformate/$^{13}$CH$_3$OCHO & 23$_{-2}^{+3}$ & --- & --- & --- \\
Dimethyl Ether & & & \\
\quad \dimethylether/$^{13}$CH$_3$OCH$_3$ & ($25_{-6}^{+10}$)$^{\ast}$ & $86\pm18$ & 17 & --- \\
Glycolaldehyde & & & \\
\quad CH$_2$OHCHO/CH$_2$OH$^{13}$CHO & --- & --- & 27$^{\ddagger}$ & --- \\
\quad CH$_2$OHCHO/$^{13}$CH$_2$OHCHO & --- & --- & 27$^{\ddagger}$ & --- \\
\enddata
\tablenotetext{\dagger}{For the measurement in the V883 Ori disk, uncertainties are the 16th and 84th percentile of the posterior distributions. The 0.3rd percentile (corresponding to $3\sigma$) of the posterior distributions are adopted for the lower limit of \acetaldehyde/$^{13}$CH$_3$CHO.}
\tablenotetext{\ddagger}{Derived from the fit assuming the same column density for these isomeric pairs.}
\tablenotetext{\ast}{Tentative measurement using tentatively detected transitions.}
\tablenotetext{a}{\citet{Manigand2020}.}
\tablenotetext{b}{\citet{Jorgensen2016} and \citet{Jorgensen2018}.}
\tablenotetext{c}{\citet{Altwegg2020_isotopes}}
\label{tab:12C13C_summary}
\end{deluxetable*}

\subsubsection{D/H ratio}
Deuterium fractionation (an enhancement of D/H ratios in molecules) is also a key to understanding the formation environment and thermal history of planetary medium \citep[e.g.,][]{Ceccarelli2014, Nomura2023_PPVII}. The D/H ratios of COMs, in particular \methanol, has been measured in warm protostellar envelopes \citep[e.g.,][]{Jorgensen2018, Drozdovskaya2021}. Even multiply-deuterated species have been detected toward IRAS 16293-2422 protobinary \citep[e.g.,][]{Manigand2019, Richard2021, Drozdovskaya2022}, and the similarity of the D enrichment between the protostellar envelopes and the solar system comets suggests the inheritance of interstellar molecules to comets \citep{Drozdovskaya2021}. In protoplanetary disks, while the D/H ratios of several simpler molecules such as HCN, HCO$^{+}$, and N$_2$H$^{+}$ have been measured \citep[e.g.,][see also \citealt{Aikawa2022}]{Cataldi2021}, no measurements of COM deuteration exist so far except for the CH$_2$DOH/\methanol ratio in the V883 Ori disk \citep{Lee2019}.

The deuterium fractionation is initiated by the deuteration of $\mathrm{H_3^+}$ with the gas-phase ion-molecule reaction, 
\begin{equation}\label{eq:deuteration}
    \mathrm{H}_3^+ + \mathrm{HD} \rightarrow \mathrm{H_2D^+} + \mathrm{H_2} + \Delta E.
\end{equation}
This reaction is exothermic ($\Delta E \approx 230$\,K; e.g., \citealt{Millar1989}), and therefore at the low temperature ($\lesssim 30$\,K) the backward reaction is suppressed, enhancing the $\mathrm{H_2D^+}$ abundance in the gas phase. In dense cold regions such as prestellar cores, a freeze-out of CO, a main reactant with $\mathrm{H_2D^+}$, further enhances the fractionation \citep[e.g.,][]{Roberts2000}. Dissociative recombination of $\mathrm{H_2D^+}$ enhances the abundance of D atoms, which subsequently causes the deuteration of molecules formed on dust grain surfaces, including COMs. In the warmer conditions ($\gtrsim 30$\,K), the fractionation is more moderate, since the backward reaction of Equation (\ref{eq:deuteration}) is not severely suppressed. Molecular D/H ratio thus can be used to probe the formation environment of molecules including COMs. For example, some molecules showing relatively high D/H ratios ($\gtrsim$ several \%) are considered to be formed in cold prestellar cores even though those are observed at warm protostellar envelopes \citep[e.g.,][]{Jorgensen2018, Yamato2022}

We have observed multiple deuterated species of COMs, \methanol, \acetaldehyde, and \methylformate, for the first time in protoplanetary disks, while the deuterated methanol, CH$_2$DOH, has also been detected in the previous Band 7 observations \citep{Lee2019}. We also tentatively identified CH$_2$DOH in the present observations. For deuterated \acetaldehyde, two different isomers (CH$_2$DCHO and CH$_3$CDO) have been tentatively identified. Several transitions of two isomers of deuterated \methylformate (CH$_3$OCDO and CH$_2$DOCHO) are covered but not detected, from which we constrain the upper limit on their column densities. 

\textrm{We derived a CH$_2$DOH/CH$_3$OH ratio of $0.0057_{-0.0015}^{+0.0018}$, although this value should be considered as an upper limit due to the tentative identification of CH$_2$DOH. In addition, the available spectroscopic data of CH$_2$DOH transitions in the JPL database have large uncertainties, particularly in their intensities (see Appendix \ref{appendix:specdata_methanol}), making comparison and discussion more difficult. We therefore consider this measurement as a reference value. As for \acetaldehyde, the CH$_2$DCHO/\acetaldehyde and CH$_3$CDO/\acetaldehyde ratios are measured to be $0.0084_{-0.0019}^{+0.0021}$ and $0.043_{-0.012}^{+0.012}$, respectively, but again these values should be considered as upper limits due to tentative identifications. We also constrain the upper limit on both CH$_2$DOCHO/\methylformate and CH$_3$OCDO/\methylformate ratios to be $\lesssim0.02$ based on the non-detection of CH$_2$DOCHO and CH$_3$OCDO.}

\begin{figure*}
\plotone{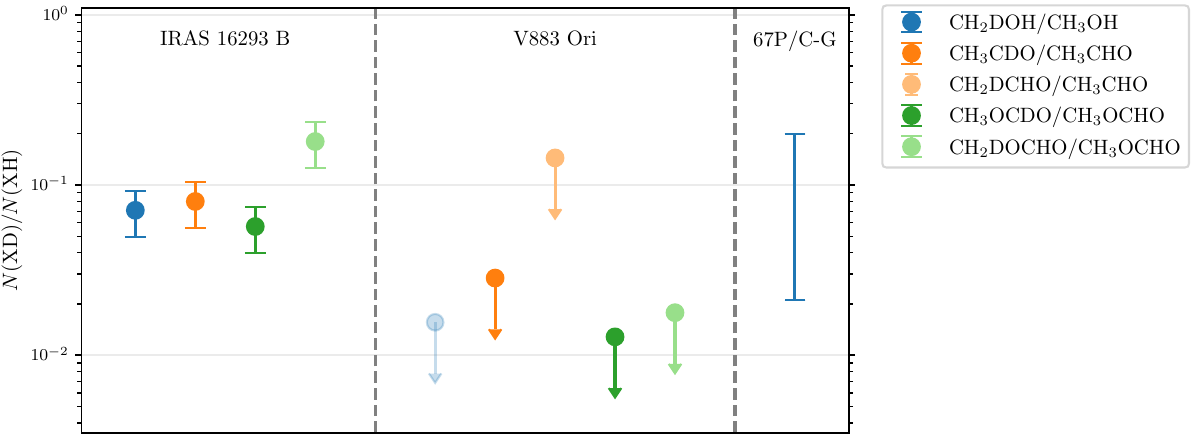}
\caption{\textrm{Comparison of the COM D/H ratios in the V883 Ori disk with those in IRAS 16293 B \citep{Jorgensen2018} and 67P/C-G \citep{Drozdovskaya2021} from literature. The different colors indicate the different molecular species of COMs as shown in the legend. Upper limits are shown by the downside arrows. The CH$_2$DOH/\methanol ratio in V883 Ori is shown by a semi-transparent circle to indicate its large uncertainty due to the large uncertainty in the spectroscopic data of CH$_2$DOH.}}
\label{fig:DH_comparison}
\end{figure*}

Figure \ref{fig:DH_comparison} compare the D/H column density ratios of COMs in the V883 Ori disk with those in IRAS 16293B and 67P/C-G. The upper limit of COM D/H ratios obtained in the V883 Ori disk are lower than the ratios in IRAS 16293B. 
As the D/H ratios are very sensitive to the formation temperature, the lower D/H ratio of COMs in the V883 Ori disk may be explained by the formation on the lukewarm ($\sim30$--50\,K) dust grain surfaces within the disk. On the other hand, the D/H ratio of \methanol (not the column density ratio, but the elemental D/H ratio after the statistical correction) in 67P/C-G has been measured to be 0.71--6.6~\% \citep{Drozdovskaya2021}, which reduced to a CH$_2$DOH/CH$_3$OH ratio (0.021--0.20), consistent with the values in IRAS 16293B, pointing to the prestellar origins of \methanol \citep{Drozdovskaya2021}. The D/H ratios of different COMs measured in the V883 Ori disk seem to be slightly lower than the reported CH$_2$DOH/CH$_3$OH ratio in 67P/C-G. This might suggest the different origins of \methanol in 67P/C-G and COMs in the V883 Ori disk; while COMs in the V883 Ori disk may have been (re-)formed on the lukewarm ($\sim$~30--50 K) dust grain surfaces in the inner region of the disk in its quiescent phase, the comet 67P/C-G may have been formed in the cold ($\sim$~10 K) outer region of the proto-solar disk, where molecules are directly inherited from the natal envelope without significant chemical reset. To draw a more concrete picture on the chemical evolution of COMs, it is essential to constrain the D/H ratios of various COMs in both disks and comets with more sensitive observations of multiple transitions and accurate spectroscopic data, in particular for CH$_2$DOH (see Appendix \ref{appendix:spectroscopic_data}). \textrm{Higher-resolution observations are also helpful to constrain the radial distributions of D/H ratios \citep[e.g.,][]{Cataldi2021}.}

\subsection{Deficiency of Nitrogen-bearing COMs}
While the number of detected oxygen-bearing COMs in the V883 Ori disk is comparable to that in protostellar envelopes, nitrogen-bearing COMs seem to be deficient in the V883 Ori disk compared to those in protostellar envelopes. In the present observations, 
only C$_2$H$_3$CN has been tentatively detected as a nitrogen-bearing COM. Our spectral setup covers the transitions of other nitrogen-bearing COMs, such as NH$_2$CHO and C$_2$H$_5$CN, but they are not detected. In Band 7 observations, CH$_3$CN is the only nitrogen-bearing COMs that has been detected \citep{Lee2019}. Even for CH$_3$CN, its column density is lower than those of oxygen-bearing COMs \citep{Lee2019}. The spectral setup of the present observations did not cover intense transitions of CH$_3$CN, and therefore it is not detected. \textrm{This apparent deficiency of nitrogen-bearing COMs may indicate that the nitrogen is locked into less volatile components, e.g., refractory carbonaceous grains \citep{vantHoff2020} and/or ammonium salt \citep{Poch2020, Altwegg2020}. This scenario is supported by the recent observations of protostellar envelopes which report a difference in the spatial distributions between nitrogen- and oxygen-bearing COMs in protostellar envelopes \citep{Csengeri2019, Lee2022, Nazari2023, Okoda2021, Okoda2022}. While the gaseous nitrogen-bearing COMs may exist in the inner hot regions of the V883 Ori disk, their emission may be beam-diluted and/or hidden by the intense dust continuum emission as discussed in Section \ref{subsec:physicochemical_structure} in the present observations.} To fully reveal the nitrogen content in the V883 Ori disk, high-resolution observations at longer wavelengths (ALMA Band 1, VLA, and ngVLA) are needed to avoid the dust absorption.   

\subsection{Implications for Sulfur Chemistry}
Sulfur-bearing molecules have been routinely observed in protostellar environments and solar system comets, and their potential chemical links have been discussed \citep[e.g.,][]{Drozdovskaya2018}. On the other hand, detection of gas-phase sulfur-bearing molecules are still sparse in Class II disks except for CS and its isotopologue \citep[][and references therein]{LeGal2021}. Recent observations have just began to report the (tentative) detection of sulfur-bearing molecules such as SO, SO$_2$, and SiS in a few warm disks, which might be originated from the thermal desorption and/or the accretion shock toward the protoplanets \citep{Booth2021, Booth2023, Law2023_SiS}. 

In the V883 Ori disk, \citet{Lee2019} reported the detection of CH$_3$SH and tentative detection of SO$_2$. The present observations have confirmed SO$_2$ and newly detected OCS.
Since the emission distributions are similar to that of other oxygen-bearing COMs in the V883 Ori disk (see Figure \ref{fig:moment_zero_gallery} for OCS), the SO$_2$ and OCS detected in the present observations likely originate from the thermal sublimation as with the other molecules. The SO$_2$/OCS ratios are measured to be $0.56_{-0.20}^{+0.21}$ in the V883 Ori disk. In molecular clouds, SO$_2$ and OCS ices are ubiquitously detected by infrared observations \citep[e.g.,][]{Boogert2015, McClure2023}. These molecules are also detected in the warm envelopes of IRAS~16293-2422 \citep{Drozdovskaya2018} and the comet 67P/C-G \citep{Calmonte2016}. The SO$_2$/OCS ratios are $\sim$~0.2--0.5 in molecular cloud ice and in IRAS~16293-2422, consistent with the value in the V883 Ori disk, suggesting that the sulfur-bearing ices in the V883 Ori disk have similar origin. On the other hand, the SO$_2$/OCS ratio is higher in comet 67P/C-G ($\sim$~280) than these values by three orders of magnitude, which may indicate the potential chemical evolution from the ISM.


\section{Summary} \label{sec:summary}
We presented the ALMA Band 3 observations of COMs in the V883 Ori disk at an angular resolution of 0\farcs3--0\farcs4. We analyzed the disk-integrated spectra in detail to obtain the spatial distribution of the COM emission and the column densities of COMs. Our major findings are summarized as follows:
\begin{enumerate}
    \item[1.] We robustly identified eleven oxygen-bearing COMs (including isotopologues) in the disk-integrated spectra, where \dimethylether, \propenal, \propanal, $^{13}$\methylformate, and CH$_3$O$^{13}$CHO are the first detection in the V883 Ori disk. We also tentatively identified five COMs (including isotopologues) and detected two sulfur-bearing molecules, OCS and SO$_2$. Nitrogen-bearing COMs are not detected except for a tentative detection of C$_2$H$_3$CN.
    \item[2.] The detailed analyses of the line profiles revealed the inner emission cavity ($\sim40$\,au radius), similar to the previous sub-mm observations. This indicates that the COM emissions are suppressed even in Band 3 where the dust continuum emission is fainter, possibly due to the higher dust temperature in the midplane caused by the viscous accretion heating. In addition, the destruction of COMs in the cavity region by the strong UV/X-ray/cosmic-ray radiations from the central outbursting protostar may explain the inner emission cavity of COMs.  Our ALMA Band 3 observations suggest that the observations in longer wavelengths, where the dust continuum emission should be fainter, with ALMA Band 1, VLA, and future ngVLA, are essential to directly probe the chemistry of the innermost region of the V883 Ori disk.
    
    \item[3.] We found that the column density ratios of complex molecules with respect to \methanol are significantly higher than those in the warm protostellar envelopes of IRAS 16293-2422, and similar to the values measured in comet 67P/C-G. This may indicate that the formation of complex molecules occurs in protoplanetary disks, which can cause the chemical evolution en route to planetary systems.
    
    \item[4.] We characterized the $^{13}$C-fractionation pattern of COMs in protoplanetary disks for the first time. The $^{12}$C/$^{13}$C ratios of \acetaldehyde, \methylformate, and \dimethylether consistently show a lower value ($\sim$\,20--30) compared to the canonical ISM ratios ($\sim$\,69). These COMs could be formed from CO enriched in $^{13}$C due to the exchange reaction with $^{13}$C$^+$ in an environment characterized with a high gas-phase C/O ratios.
    \item[5.]  We also constrained the D/H ratios of multiple COMs in protoplanetary disks for the first time. The D/H ratios of \acetaldehyde and \methylformate shows lower values than those in protostellar envelopes IRAS 16293B, implying the formation of these molecules in warmer dust grain surfaces in protoplanetary disks. 
\end{enumerate}

\bigskip
\textrm{The authors thank the anonymous referee for comments.} We thank Tomohiro~Yoshida and Gianni~Cataldi for the fruitful discussion on the technical aspects of the present work. We also thank Kiyoaki~Doi for the discussion on the dust continuum emission of the V883 Ori disk. 
Y.Y. is financially supported by Grant-in-Aid for the Japan
Society for the Promotion of Science (JSPS) Fellows (KAKENHI Grant Number JP23KJ0636) and International Graduate Program for Excellence in Earth-Space Science (IGPEES) of the University of Tokyo.
S.N. and Y.O. are supported by RIKEN Special Postdoctoral Researcher Program (Fellowships).
S.N.~is also grateful for support from Grants-in-Aid for JSPS (Japan Society for the Promotion of Science) Fellows Grant Number JP23KJ0329, and MEXT/JSPS Grants-in-Aid for Scientific Research (KAKENHI) Grant Numbers JP20K22376, JP23K13155, and JP23H05441.
Y.A. acknowledges support by MEXT/JSPS Grants-in-Aid for Scientific Research (KAKENHI) Grant Numbers JP20H05847 and JP21H04495 and NAOJ ALMA Scientific Research Grant code 2019-13B..
Y.O. is supported by MEXT/JSPS Grants-in-Aid for Scientific Research (KAKENHI) Grant Number JP22K20390.
H.N. is supported by MEXT/JSPS Grants-in-Aid for Scientific Research (KAKENHI) Grant Numbers JP18H05441 and JP19K03910.
N.S. is supported by MEXT/JSPS Grants-in-Aid for Scientific Research (KAKENHI) Grant Numbers JP20H05845 and JP20H00182, and a pioneering project in RIKEN (Evolution of Matter in the Universe) .
This paper makes use of the following ALMA data: ADS/JAO.ALMA\#2021.1.00357.S. 
ALMA is a partnership of ESO (representing its member states), NSF (USA) and NINS (Japan), together with NRC (Canada), MOST and ASIAA (Taiwan), and KASI (Republic of Korea), in cooperation with the Republic of Chile. The Joint ALMA Observatory is operated by ESO, AUI/NRAO and NAOJ. 


%

\vspace{5mm}
\facilities{ALMA}


\software{CASA \citep{CASA}, Numpy \citep{Numpy}, Scipy \citep{SciPy}, Astropy \citep{AstropyI, AstropyII, AstropyIII}, Matplotlib \citep{Matplotlib}, GoFish \citep{GoFish}, bettermoments \citep{bettermoments}, emcee \citep{emcee}}



\appendix
\section{Channel Maps of the COM emission}\label{appendix:channel_maps}
\begin{figure*}
\plotone{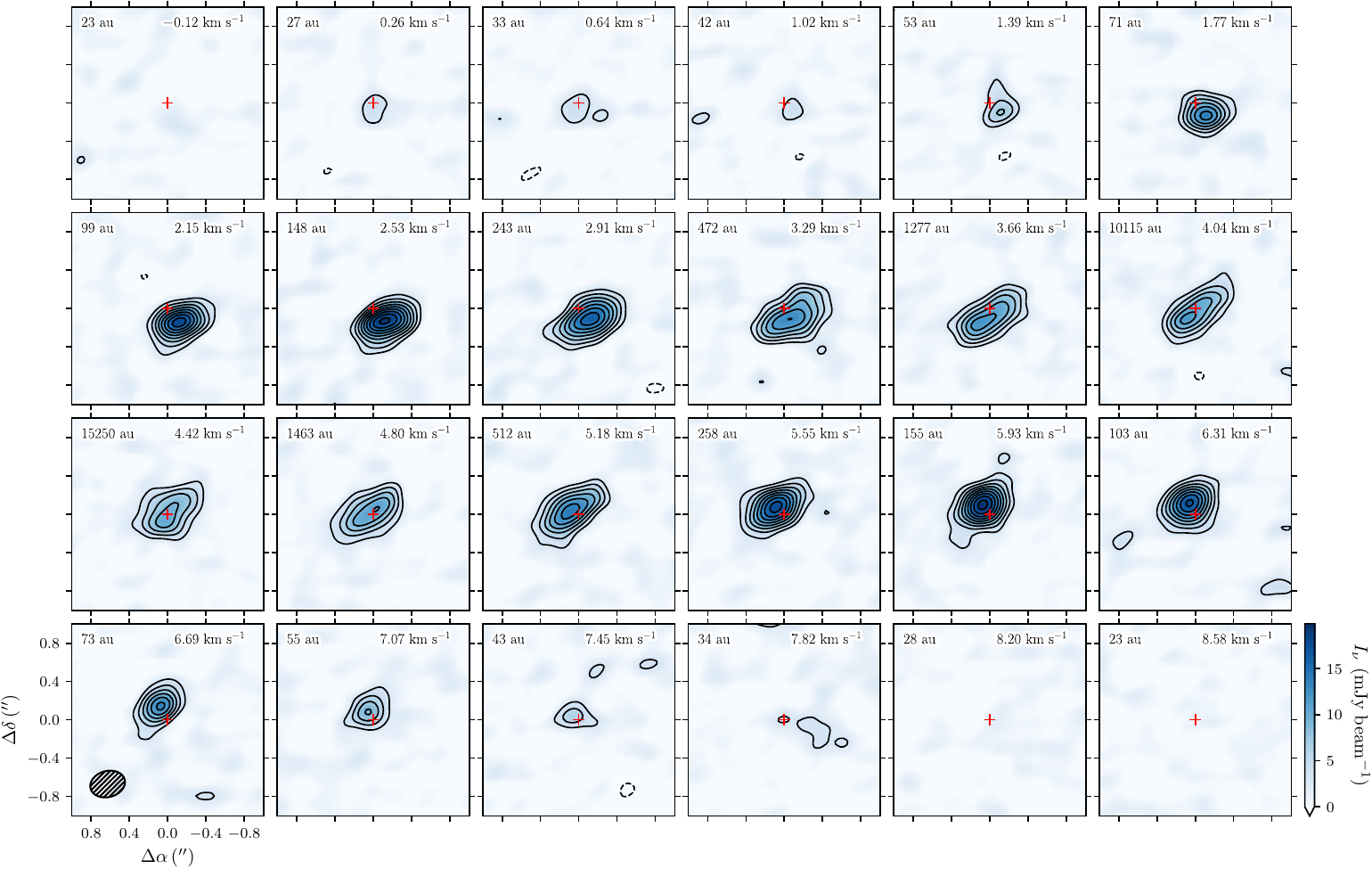}
\caption{Channel maps of the \methanol $2_{-1,1}$ -- $1_{-1,0}$ E $v_t=0$ transition. The black contours mark the [$-$3, 3, 5, 7, ...]$\,\times\,\sigma$ levels. The dashed contours indicate the negative values. The numbers in the upper-right and upper-left corners indicate the velocity of the channel and its corresponding Keplerian radius ($r = GM_\star\sin^2i/(v - v_\mathrm{sys})^2$), respectively. The red cross in each panel indicate the position of the disk center. The beam is shown in the lower-left corner of the lower-left panel.}
\label{fig:channelmap_CH3OH}
\end{figure*}

\begin{figure*}
\plotone{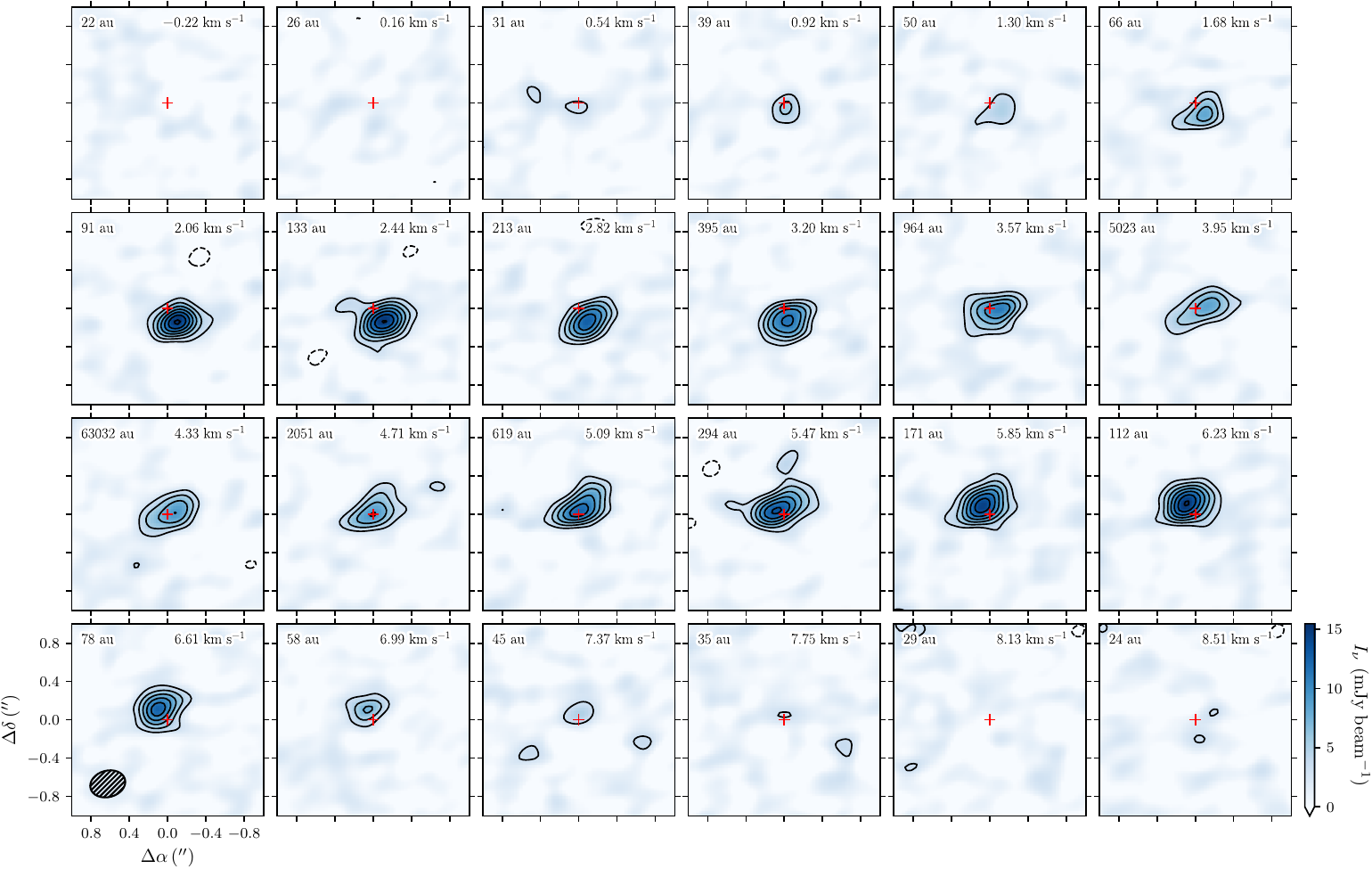}
\caption{Same as Figure \ref{fig:channelmap_CH3OH}, but for \acetaldehyde $5_{2,3}$ -- $4_{2,2}$ E $v_t=0$ transition.}
\label{fig:channelmap_CH3CHO}
\end{figure*}

\begin{figure*}
\plotone{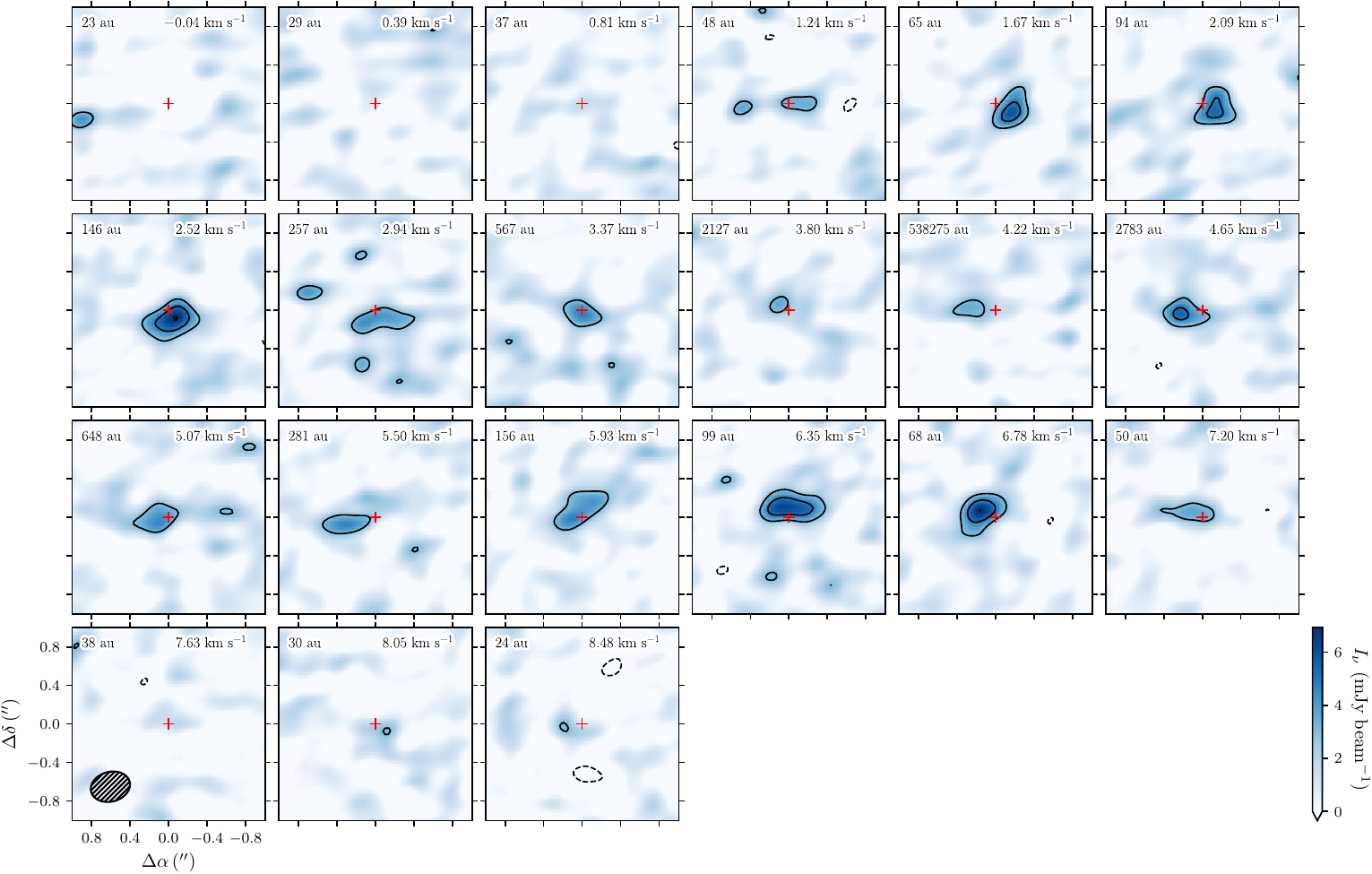}
\caption{Same as Figure \ref{fig:channelmap_CH3OH}, but for \methylformate $8_{5,3}$ -- $7_{5,2}$ A $v_t=0$ transition.}
\label{fig:channelmap_CH3OCHO}
\end{figure*}

Figure \ref{fig:channelmap_CH3OH}, \ref{fig:channelmap_CH3CHO}, and \ref{fig:channelmap_CH3OCHO} show the channel maps of the bright, unblended transitions of \methanol, \acetaldehyde, and \methylformate, respectively, which are used for the line profile analysis in Section \ref{subsubsec:line_profile_analysis}. The comparison to the corresponding Keplerian radius (i.e., $r = GM_\star\sin^2 i / (v - v_\mathrm{sys})^2$) at each velocity channel is made.

\section{JvM effect and continuum subtraction}\label{appendix:JvM_effect}
Here we describe a brief summary of the Jorsater \& van Moorsel (JvM) effect \citep{JvM, Czekala2021}, and subsequently discuss the continuum subtraction methods, which could mitigate the JvM effect.

The JvM effect could result in incorrect measurements of the flux density on the interferometric images produced by the CLEAN deconvolution method, which was originally discussed in \citet{JvM} and extensively investigated in \citet{Czekala2021}. During the imaging process, the CLEAN algorithm iteratively extracts the CLEAN components from the residual image, and this process will be terminated when the peak value of the residual image within the CLEAN mask reaches a user-defined threshold (typically 2--3 $\times$ the RMS level; see Figure 2 in \citet{Czekala2021} for a comprehensive summary of this procedure). The final image product of CLEAN is the sum of the CLEANed model image (convolved with a CLEAN beam determined by a Gaussian fit to the dirty beam) and the residual images, which have inconsistent units (Jansky per CLEAN beam and Jansky per dirty beam). Therefore, if the dirty beam significantly deviates from the CLEAN beam (Gaussian) and the faint emission below the CLEAN threshold remains on the residual image, the intensity scale in the final image can be incorrect (i.e., sum of values with two different units). The impact of this effect depends on the emission strengths, the CLEAN threshold, and the deviation of the dirty beam shape from the CLEAN beam shape: if a significant portion of the emission remains on the residual image and the deviation is large, the intensity scale in the final image will drastically modified from the correct intensity scale \citep{JvM, Czekala2021}.  

To correct for this unit inconsistency, \citet{Czekala2021} applied the ``JvM correction'' to the line emission in protoplanetary disks, where the residual image is scaled before the addition with the CLEANed model image by the ratio of the dirty beam area to the CLEAN beam area, forcing the unit of the residual image to be Jansky per CLEAN beam. However, \citet{Casassus2022} pointed out that the JvM correction can artificially manipulate the noise levels and exaggerate the signal-to-noise ratio of the emission, as the correction will scale the residual image that contains the noise typically by a factor smaller than unity (see also Appendix A in \citealt{Casassus2023}). So far, no concrete solution for this problem has been drawn.

\begin{figure*}
\epsscale{1.15}
\plotone{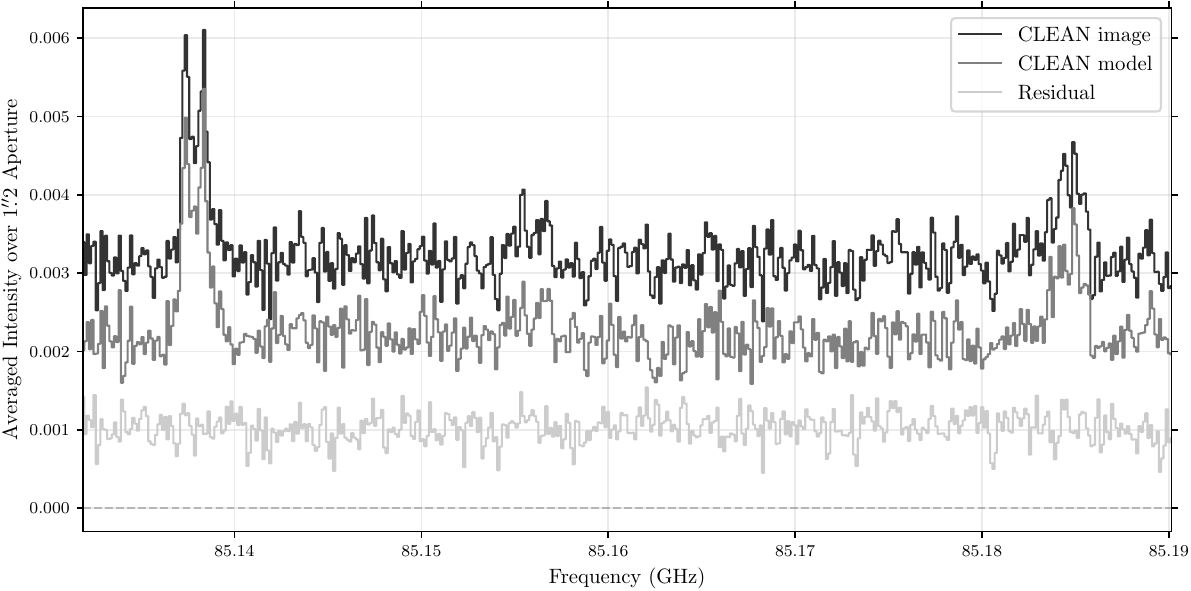}
\caption{Line plus continuum spectra of SPW 2 averaged over 1\farcs2 aperture for the CLEAN image (dark gray), the CLEAN model (gray), and the residual (light gray). The CLEAN model spectrum (in unit of Jy pixel$^{-1}$ by default) is scaled up by a ratio of the CLEAN beam area to the pixel area for visual clarity. The spectral line emission is almost fully recovered into the CLEAN model and none of that remains in the residual spectrum. The horizontal dashed line indicates the zero-flux level.}
\label{fig:JvM_spectrum}
\end{figure*}

We applied the JvM correction to our data following \citet{Czekala2021}, which obtained a JvM $\epsilon$ (the ratio of dirty beam area to the CLEAN beam area) of $\sim0.28$, indicating that the effect is severe. This could result in a drastic underestimation of the uncertainties of physical quantities derived from the fits described in Section \ref{sec:analysis_result}. Another possible solution is CLEANing deeper (e.g., down to 0.5--1 $\times$ the RMS level) so that all the emission components are recovered, but this would be unrealistic as too deep CLEANing will not converge and it is practically unfeasible to CLEAN down to such deep levels particularly for the line image cubes with a lot of channels. Alternatively, we mitigate the JvM effect and recover correct flux scale as much as possible by employing the continuum subtraction on the image plane (with the CASA task \texttt{imcontsub}) rather than on the visibility plane (with the CASA task \texttt{uvcontsub}). The line emission in our data is almost co-spatial to the strong continuum emission, and therefore imaging without continuum subtraction can almost completely include the line emission into the CLEAN model, thanks to the strong line plus continuum intensity well above the CLEAN threshold. While the continuum component is still affected by the JvM effect, the line emission can be safely recovered by subsequent continuum subtraction on the image plane. Figure \ref{fig:JvM_spectrum} shows an example demonstrating that almost no spectral line component remain in the residual spectrum and the line emission is almost fully recovered in the CLEANed image. We note that this method is only applicable to the spectral line data associated with a strong continuum emission. If the continuum emission is faint in the emitting region of the spectral line, the spectral line component will still remain in the residual even with the continuum subtraction on the image plane. 

\section{Detected transitions}\label{appendix:transitions}
Table \ref{tab:transitions} lists the transitions detected in our observations for each molecular species, including the blended ones.

\startlongtable
\begin{deluxetable*}{ccCCCCc}
\tablecaption{Detected Transitions for Each Species\label{tab:transitions}}
\tablehead{\colhead{Species} & \colhead{Transition} & \colhead{$\nu_0$ (GHz)} & \colhead{log$_{10}A_\mathrm{ul}$ (s$^{-1}$)} & \colhead{$g_\mathrm{u}$} & \colhead{$E_\mathrm{u}$ (K)} & \colhead{SPW}}
\decimals
\startdata
\hline 
\multicolumn{7}{c}{Methanol (\methanol)} \\
\hline 
CH$_3$OH$^{\ddagger\ast}$ & $2_{-1,2}$ -- $1_{-1,1}$ E $v_t=1$ & 96.492163 & -5.5963 & 20.0 & 298 & 7 \\
CH$_3$OH$^{\ddagger\ast}$ & $2_{0,2}$ -- $1_{0,1}$ E $v_t=1$ & 96.493551 & -5.4706 & 20.0 & 307 & 7 \\
CH$_3$OH$^\ddagger$ & $2_{1,1}$ -- $1_{1,0}$ E $v_t=1$ & 96.501713 & -5.5967 & 20.0 & 420 & 7 \\
CH$_3$OH & $2_{0,2}$ -- $1_{0,1}$ A $v_t=1$ & 96.513686 & -5.4709 & 20.0 & 430 & 7 \\
CH$_3$OH$^{\dagger\ast}$ & $2_{1,2}$ -- $1_{1,1}$ E $v_t=0$ & 96.739358 & -5.5923 & 20.0 & 12 & 9 \\
CH$_3$OH$^{\dagger\ast}$ & $2_{0,2}$ -- $1_{0,1}$ A $v_t=0$ & 96.741371 & -5.4676 & 20.0 & 6 & 9 \\
CH$_3$OH$^\ast$ & $2_{0,2}$ -- $1_{0,1}$ E $v_t=0$ & 96.744545 & -5.4676 & 20.0 & 20 & 9 \\
CH$_3$OH$^\ast$ & $2_{-1,1}$ -- $1_{-1,0}$ E $v_t=0$ & 96.755501 & -5.581 & 20.0 & 28 & 9 \\
CH$_3$OH$^\ast$ & $2_{1,1}$ -- $1_{1,0}$ A $v_t=0$ & 97.582798 & -5.5807 & 20.0 & 21 & 10 \\
CH$_3$OH$^\dagger$ & $21_{6,16}$ -- $22_{5,17}$ A $v_t=0$ & 97.677684 & -5.8404 & 172.0 & 729 & 10 \\
CH$_3$OH$^\dagger$ & $21_{6,15}$ -- $22_{5,18}$ A $v_t=0$ & 97.678803 & -5.8403 & 172.0 & 729 & 10 \\
\hline 
\multicolumn{7}{c}{Methanol (CH$_2$DOH)} \\
\hline 
CH$_2$DOH$^\ddagger$ & $2_{1,1}$ -- $2_{0,2}$ e0 & 86.668751 & -5.3322 & 5.0 & 10 & 4 \\
CH$_2$DOH$^\ddagger$ & $4_{1,3}$ -- $4_{0,4}$ o1 & 97.870192 & -5.3363 & 9.0 & 44 & 10 \\
CH$_2$DOH & $4_{0,4}$ -- $3_{1,3}$ e0 & 98.031213 & -5.616 & 9.0 & 21 & 10 \\
\hline 
\multicolumn{7}{c}{Methyl Formate (\methylformate)} \\
\hline 
CH$_3$OCHO & $7_{6,1}$ -- $6_{6,0}$ E $v_t=1$ & 85.157135 & -5.6248 & 30.0 & 228 & 3 \\
CH$_3$OCHO$^\dagger$ & $7_{5,3}$ -- $6_{5,2}$ A $v_t=1$ & 85.185466 & -5.3602 & 30.0 & 220 & 3 \\
CH$_3$OCHO$^\dagger$ & $7_{5,2}$ -- $6_{5,1}$ A $v_t=1$ & 85.186063 & -5.3602 & 30.0 & 220 & 3 \\
CH$_3$OCHO & $7_{4,3}$ -- $6_{4,2}$ E $v_t=1$ & 85.506219 & -5.2142 & 30.0 & 214 & 1 \\
CH$_3$OCHO & $7_{5,3}$ -- $6_{5,2}$ E $v_t=1$ & 85.55338 & -5.3527 & 30.0 & 220 & 1 \\
CH$_3$OCHO & $7_{6,1}$ -- $6_{6,0}$ E $v_t=0$ & 85.919209 & -5.6138 & 30.0 & 40 & 2 \\
CH$_3$OCHO$^{\dagger\ast}$ & $7_{6,2}$ -- $6_{6,1}$ E $v_t=0$ & 85.926553 & -5.6137 & 30.0 & 40 & 2 \\
CH$_3$OCHO$^{\dagger\ast}$ & $7_{6,2}$ -- $6_{6,1}$ A $v_t=0$ & 85.927227 & -5.6136 & 30.0 & 40 & 2 \\
CH$_3$OCHO$^{\dagger\ast}$ & $7_{6,1}$ -- $6_{6,0}$ A $v_t=0$ & 85.927227 & -5.6136 & 30.0 & 40 & 2 \\
CH$_3$OCHO & $7_{4,4}$ -- $7_{3,5}$ E $v_t=0$ & 96.507882 & -6.1542 & 30.0 & 27 & 7 \\
CH$_3$OCHO & $8_{4,5}$ -- $8_{3,6}$ A $v_t=0$ & 96.709259 & -5.9454 & 34.0 & 31 & 9 \\
CH$_3$OCHO & $7_{4,3}$ -- $7_{3,5}$ E $v_t=0$ & 96.776715 & -6.4523 & 30.0 & 27 & 9 \\
CH$_3$OCHO & $5_{4,2}$ -- $5_{3,3}$ A $v_t=0$ & 96.794121 & -6.1026 & 22.0 & 19 & 9 \\
CH$_3$OCHO & $8_{5,3}$ -- $7_{5,2}$ E $v_t=1$ & 97.577303 & -5.0823 & 34.0 & 225 & 10 \\
CH$_3$OCHO & $8_{3,6}$ -- $7_{3,5}$ A $v_t=1$ & 97.597161 & -4.9359 & 34.0 & 214 & 10 \\
CH$_3$OCHO$^\dagger$ & $8_{7,2}$ -- $7_{7,1}$ E $v_t=1$ & 97.65127 & -5.4962 & 34.0 & 240 & 10 \\
CH$_3$OCHO$^\dagger$ & $10_{4,7}$ -- $10_{3,8}$ E $v_t=0$ & 97.65127 & -5.9015 & 42.0 & 43 & 10 \\
CH$_3$OCHO & $8_{4,5}$ -- $7_{4,4}$ A $v_t=1$ & 97.661401 & -4.9937 & 34.0 & 219 & 10 \\
CH$_3$OCHO$^\ddagger$ & $10_{4,7}$ -- $10_{3,8}$ A $v_t=0$ & 97.69426 & -5.8957 & 42.0 & 43 & 10 \\
CH$_3$OCHO$^\ddagger$ & $12_{1,11}$ -- $12_{0,12}$ A $v_t=1$ & 97.727054 & -6.2029 & 50.0 & 234 & 10 \\
CH$_3$OCHO & $8_{6,3}$ -- $7_{6,2}$ E $v_t=1$ & 97.738738 & -5.2243 & 34.0 & 232 & 10 \\
CH$_3$OCHO & $8_{4,4}$ -- $7_{4,3}$ A $v_t=1$ & 97.752885 & -4.9924 & 34.0 & 219 & 10 \\
CH$_3$OCHO$^\ddagger$ & $8_{4,4}$ -- $8_{3,5}$ E $v_t=1$ & 97.871147 & -5.967 & 34.0 & 219 & 10 \\
CH$_3$OCHO$^\ddagger$ & $10_{4,7}$ -- $10_{3,8}$ A $v_t=1$ & 97.878933 & -5.8854 & 42.0 & 230 & 10 \\
CH$_3$OCHO & $8_{5,4}$ -- $7_{5,3}$ E $v_t=1$ & 97.885663 & -5.0787 & 34.0 & 224 & 10 \\
CH$_3$OCHO & $8_{4,4}$ -- $7_{4,3}$ E $v_t=1$ & 97.897118 & -4.9875 & 34.0 & 219 & 10 \\
CH$_3$OCHO$^\ddagger$ & $21_{4,17}$ -- $21_{3,18}$ A $v_t=1$ & 98.066305 & -5.7833 & 86.0 & 337 & 10 \\
CH$_3$OCHO$^\ddagger$ & $8_{4,5}$ -- $7_{4,4}$ E $v_t=1$ & 98.176293 & -4.9853 & 34.0 & 218 & 10 \\
CH$_3$OCHO & $8_{7,1}$ -- $7_{7,0}$ E $v_t=0$ & 98.182336 & -5.4902 & 34.0 & 53 & 10 \\
CH$_3$OCHO$^{\dagger\ast}$ & $8_{7,2}$ -- $7_{7,1}$ A $v_t=0$ & 98.190658 & -5.4899 & 34.0 & 53 & 10 \\
CH$_3$OCHO$^{\dagger\ast}$ & $8_{7,1}$ -- $7_{7,0}$ A $v_t=0$ & 98.190658 & -5.4899 & 34.0 & 53 & 10 \\
CH$_3$OCHO$^{\dagger\ast}$ & $8_{7,2}$ -- $7_{7,1}$ E $v_t=0$ & 98.19146 & -5.49 & 34.0 & 53 & 10 \\
CH$_3$OCHO$\ast$ & $8_{6,2}$ -- $7_{6,1}$ E $v_t=0$ & 98.270501 & -5.218 & 34.0 & 45 & 10 \\
CH$_3$OCHO$^{\dagger\ast}$ & $8_{6,3}$ -- $7_{6,2}$ E $v_t=0$ & 98.278921 & -5.2178 & 34.0 & 45 & 10 \\
CH$_3$OCHO$^{\dagger\ast}$ & $8_{6,3}$ -- $7_{6,2}$ A $v_t=0$ & 98.279762 & -5.2178 & 34.0 & 45 & 10 \\
CH$_3$OCHO$^{\dagger\ast}$ & $8_{6,2}$ -- $7_{6,1}$ A $v_t=0$ & 98.279762 & -5.2178 & 34.0 & 45 & 10 \\
CH$_3$OCHO$^{\dagger\ast}$ & $9_{0,9}$ -- $8_{1,8}$ A $v_t=1$ & 98.423165 & -5.701 & 38.0 & 212 & 10 \\
CH$_3$OCHO$^{\dagger\ast}$ & $8_{5,3}$ -- $7_{5,2}$ E $v_t=0$ & 98.424207 & -5.0722 & 34.0 & 37 & 10 \\
CH$_3$OCHO$^{\dagger\ast}$ & $8_{5,4}$ -- $7_{5,3}$ E $v_t=0$ & 98.431803 & -5.072 & 34.0 & 37 & 10 \\
CH$_3$OCHO$^{\dagger\ast}$ & $8_{5,4}$ -- $7_{5,3}$ A $v_t=0$ & 98.43276 & -5.0719 & 34.0 & 37 & 10 \\
CH$_3$OCHO & $8_{5,3}$ -- $7_{5,2}$ A $v_t=0$ & 98.435802 & -5.0719 & 34.0 & 37 & 10 \\
CH$_3$OCHO & $8_{4,5}$ -- $7_{4,3}$ E $v_t=0$ & 98.443186 & -6.4657 & 34.0 & 31 & 10 \\
\hline 
\multicolumn{7}{c}{Methyl Formate (CH$_3$O$^{13}$CHO)} \\
\hline 
CH$_3$O$^{13}$CHO$^\dagger$ & $8_{-6,3}$ -- $7_{-6,2}$ E $v_t=0$ & 97.549352 & -5.2185 & 34.0 & 44 & 10 \\
CH$_3$O$^{13}$CHO$^\dagger$ & $8_{6,3}$ -- $7_{6,2}$ A $v_t=0$ & 97.55014 & -5.2185 & 34.0 & 44 & 10 \\
CH$_3$O$^{13}$CHO$^\dagger$ & $8_{6,2}$ -- $7_{6,1}$ A $v_t=0$ & 97.550183 & -5.2185 & 34.0 & 44 & 10 \\
CH$_3$O$^{13}$CHO$^\ddagger$ & $8_{-5,4}$ -- $7_{-5,3}$ E $v_t=0$ & 97.702479 & -5.0727 & 34.0 & 37 & 10 \\
CH$_3$O$^{13}$CHO$^\ddagger$ & $8_{5,4}$ -- $7_{5,3}$ A $v_t=0$ & 97.703323 & -5.0726 & 34.0 & 37 & 10 \\
CH$_3$O$^{13}$CHO & $8_{-3,6}$ -- $7_{-3,5}$ E $v_t=0$ & 97.874331 & -4.9221 & 34.0 & 27 & 10 \\
CH$_3$O$^{13}$CHO$^\ddagger$ & $8_{3,6}$ -- $7_{3,5}$ A $v_t=0$ & 97.878443 & -4.9215 & 34.0 & 27 & 10 \\
CH$_3$O$^{13}$CHO$^\ddagger$ & $8_{4,4}$ -- $7_{4,3}$ E $v_t=0$ & 98.019644 & -4.9931 & 34.0 & 31 & 10 \\
\hline 
\multicolumn{7}{c}{Methyl Formate ($^{13}$CH$_3$OCHO)} \\
\hline
$^{13}$CH$_3$OCHO$^\ddagger$ & $8_{3,5}$ -- $7_{3,4}$ E $v_t=0$ & 97.816488 & -5.2344 & 34.0 & 22 & 10 \\
$^{13}$CH$_3$OCHO$^\ddagger$ & $8_{-3,5}$ -- $7_{-3,4}$ E $v_t=0$ & 97.816489 & -5.2237 & 34.0 & 26 & 10 \\
$^{13}$CH$_3$OCHO$^\ddagger$ & $8_{3,5}$ -- $7_{3,4}$ A $v_t=0$ & 97.829563 & -5.2228 & 34.0 & 26 & 10 \\
$^{13}$CH$_3$OCHO$^\ddagger$ & $9_{1,9}$ -- $8_{1,8}$ E $v_t=0$ & 97.871736 & -5.1613 & 38.0 & 24 & 10 \\
$^{13}$CH$_3$OCHO$^\ddagger$ & $9_{1,9}$ -- $8_{1,8}$ A $v_t=0$ & 97.873563 & -5.1612 & 38.0 & 24 & 10 \\
$^{13}$CH$_3$OCHO$^\dagger$ & $8_{-1,7}$ -- $7_{-1,6}$ E $v_t=0$ & 98.207248 & -5.1648 & 34.0 & 22 & 10 \\
$^{13}$CH$_3$OCHO$^\dagger$ & $8_{1,7}$ -- $7_{1,6}$ E $v_t=0$ & 98.207251 & -5.1756 & 34.0 & 17 & 10 \\
$^{13}$CH$_3$OCHO$^\dagger$ & $8_{1,7}$ -- $7_{1,6}$ A $v_t=0$ & 98.215266 & -5.1647 & 34.0 & 22 & 10 \\
\hline 
\multicolumn{7}{c}{Acetaldehyde (\acetaldehyde)} \\
\hline 
CH$_3$CHO & $9_{1,8}$ -- $9_{0,9}$ E $v_t=1$ & 85.947624 & -5.559 & 38.0 & 249 & 2 \\
CH$_3$CHO$^\dagger$ & $12_{2,11}$ -- $12_{1,11}$ E $v_t=1$ & 87.303557 & -6.1944 & 50.0 & 285 & 6 \\
CH$_3$CHO$^\dagger$ & $11_{4,7}$ -- $12_{3,9}$ E $v_t=1$ & 87.303714 & -6.2602 & 46.0 & 303 & 6 \\
CH$_3$CHO$\ast$ & $5_{2,3}$ -- $4_{2,2}$ E $v_t=0$ & 96.475524 & -4.6168 & 22.0 & 23 & 7 \\
CH$_3$CHO & $11_{4,8}$ -- $12_{3,10}$ E $v_t=0$ & 96.48895 & -6.1541 & 46.0 & 97 & 7 \\
CH$_3$CHO$^{\dagger\ast}$ & $5_{3,3}$ -- $4_{3,2}$ A $v_t=2$ & 96.716114 & -4.7374 & 22.0 & 420 & 9 \\
CH$_3$CHO$^{\dagger\ast}$ & $5_{3,2}$ -- $4_{3,1}$ A $v_t=2$ & 96.717473 & -4.7374 & 22.0 & 420 & 9 \\
CH$_3$CHO$^{\dagger\ast}$ & $5_{2,3}$ -- $4_{2,2}$ A $v_t=1$ & 96.718409 & -4.6093 & 22.0 & 228 & 9 \\
CH$_3$CHO$^\ast$ & $7_{0,7}$ -- $6_{1,6}$ A $v_t=0$ & 96.765371 & -5.5556 & 30.0 & 25 & 9 \\
CH$_3$CHO$^\ast$ & $5_{2,4}$ -- $4_{2,3}$ E $v_t=1$ & 96.800291 & -4.613 & 22.0 & 226 & 9 \\
CH$_3$CHO & $5_{3,2}$ -- $4_{3,1}$ E $v_t=2$ & 97.612131 & -4.9808 & 22.0 & 419 & 10 \\
CH$_3$CHO & $19_{2,18}$ -- $18_{3,15}$ E $v_t=0$ & 97.796104 & -6.1248 & 78.0 & 183 & 10 \\
CH$_3$CHO & $10_{4,7}$ -- $11_{3,8}$ A $v_t=1$ & 97.941422 & -6.1583 & 42.0 & 290 & 10 \\
CH$_3$CHO & $21_{3,18}$ -- $20_{4,17}$ A $v_t=0$ & 98.20169 & -6.01 & 86.0 & 235 & 10 \\
CH$_3$CHO & $6_{3,3}$ -- $7_{2,5}$ E $v_t=0$ & 98.368631 & -6.245 & 26.0 & 39 & 10 \\
\hline 
\multicolumn{7}{c}{Acetaldehyde (CH$_3^{13}$CHO)} \\
\hline 
CH$_3^{13}$CHO$^{\ddagger\ast}$ & $5_{2,3}$ -- $4_{2,2}$ A $v_t=0$ & 96.494465 & -4.5971 & 11.0 & 22 & 7 \\
\hline 
\multicolumn{7}{c}{Acetaldehyde (CH$_3$CDO)} \\
\hline 
CH$_3$CDO & $5_{1,4}$ -- $4_{1,3}$ E $v_t=0$ & 97.81231 & -4.5235 & 11.0 & 15 & 10 \\
CH$_3$CDO & $5_{1,4}$ -- $4_{1,3}$ A $v_t=0$ & 97.828514 & -4.5234 & 11.0 & 15 & 10 \\
\hline 
\multicolumn{7}{c}{Acetaldehyde (CH$_2$DCHO)} \\
\hline 
CH$_2$DCHO & $11_{2,9}i$ -- $11_{1,10}i$ & 98.412832 & -5.2329 & 23.0 & 66 & 10 \\
\hline 
\multicolumn{7}{c}{Dimethyl Ether (\dimethylether)} \\
\hline 
CH$_3$OCH$_3$$^\dagger$ & $16_{3,14}$ -- $15_{4,11}$ AA & 97.990629 & -5.8984 & 330.0 & 136 & 10 \\
CH$_3$OCH$_3$$^\dagger$ & $16_{3,14}$ -- $15_{4,11}$ EE & 97.993382 & -5.8983 & 528.0 & 136 & 10 \\
CH$_3$OCH$_3$$^\dagger$ & $16_{3,14}$ -- $15_{4,11}$ EA & 97.996098 & -5.8984 & 132.0 & 136 & 10 \\
CH$_3$OCH$_3$$^\dagger$ & $16_{3,14}$ -- $15_{4,11}$ AE & 97.996174 & -5.8984 & 198.0 & 136 & 10 \\
\hline 
\multicolumn{7}{c}{Acetone (\acetone)} \\
\hline 
CH$_3$COCH$_3$$^\dagger$ & $17_{6,11}$ -- $17_{5,12}$ AE & 97.929123 & -4.7694 & 210.0 & 110 & 10 \\
CH$_3$COCH$_3$$^\dagger$ & $17_{6,11}$ -- $17_{5,12}$ & 97.929247 & -4.7694 & 140.0 & 110 & 10 \\
CH$_3$COCH$_3$$^\dagger$ & $17_{7,11}$ -- $17_{6,12}$ AE & 97.930235 & -4.7694 & 70.0 & 110 & 10 \\
CH$_3$COCH$_3$$^\dagger$ & $17_{7,11}$ -- $17_{6,12}$ & 97.930344 & -4.7693 & 140.0 & 110 & 10 \\
CH$_3$COCH$_3$$^\dagger$ & $17_{6,11}$ -- $17_{5,12}$ EE & 98.052399 & -4.7674 & 560.0 & 110 & 10 \\
CH$_3$COCH$_3$$^\dagger$ & $17_{7,11}$ -- $17_{6,12}$ EE & 98.053535 & -4.7674 & 560.0 & 110 & 10 \\
\hline 
\multicolumn{7}{c}{Ethylene Oxide (\ethyleneoxide)} \\
\hline 
$c$-C$_2$H$_4$O$^\ddagger$ & $11_{9,3}$ -- $11_{8,4}$ & 96.501033 & -5.0737 & 115.0 & 146 & 7 \\
$c$-C$_2$H$_4$O$^\ddagger$ & $12_{9,4}$ -- $12_{8,5}$ & 97.728742 & -4.9909 & 75.0 & 169 & 10 \\
\hline 
\multicolumn{7}{c}{Propenal (\propenal)} \\
\hline 
$t$-C$_2$H$_3$CHO$^\ddagger$ & $11_{2,10}$ -- $10_{2,9}$ & 97.815592 & -4.3286 & 23.0 & 36 & 10 \\
$t$-C$_2$H$_3$CHO$^\ddagger$ & $11_{8,3}$ -- $10_{8,2}$ & 97.947054 & -4.6391 & 23.0 & 159 & 10 \\
$t$-C$_2$H$_3$CHO$^\ddagger$ & $11_{8,4}$ -- $10_{8,3}$ & 97.947054 & -4.6391 & 23.0 & 159 & 10 \\
$t$-C$_2$H$_3$CHO$^\ddagger$ & $11_{7,5}$ -- $10_{7,4}$ & 97.947549 & -4.5377 & 23.0 & 129 & 10 \\
$t$-C$_2$H$_3$CHO$^\ddagger$ & $11_{7,4}$ -- $10_{7,3}$ & 97.94755 & -4.5377 & 23.0 & 129 & 10 \\
$t$-C$_2$H$_3$CHO$^\ddagger$ & $11_{9,2}$ -- $10_{9,1}$ & 97.948051 & -4.793 & 23.0 & 194 & 10 \\
$t$-C$_2$H$_3$CHO$^\ddagger$ & $11_{9,3}$ -- $10_{9,2}$ & 97.948051 & -4.793 & 23.0 & 194 & 10 \\
$t$-C$_2$H$_3$CHO$^\ddagger$ & $11_{10,1}$ -- $10_{10,0}$ & 97.950017 & -5.0727 & 23.0 & 233 & 10 \\
$t$-C$_2$H$_3$CHO$^\ddagger$ & $11_{10,2}$ -- $10_{10,1}$ & 97.950017 & -5.0727 & 23.0 & 233 & 10 \\
$t$-C$_2$H$_3$CHO$^\ddagger$ & $11_{6,5}$ -- $10_{6,4}$ & 97.950286 & -4.4656 & 23.0 & 102 & 10 \\
$t$-C$_2$H$_3$CHO$^\ddagger$ & $11_{6,6}$ -- $10_{6,5}$ & 97.950286 & -4.4656 & 23.0 & 102 & 10 \\
$t$-C$_2$H$_3$CHO$^\ddagger$ & $11_{5,7}$ -- $10_{5,6}$ & 97.957003 & -4.4126 & 23.0 & 79 & 10 \\
$t$-C$_2$H$_3$CHO$^\ddagger$ & $11_{5,6}$ -- $10_{5,5}$ & 97.957003 & -4.4126 & 23.0 & 79 & 10 \\
$t$-C$_2$H$_3$CHO$^\dagger$ & $11_{4,8}$ -- $10_{4,7}$ & 97.972119 & -4.3735 & 23.0 & 61 & 10 \\
$t$-C$_2$H$_3$CHO$^\dagger$ & $11_{4,7}$ -- $10_{4,6}$ & 97.972119 & -4.3735 & 23.0 & 61 & 10 \\
$t$-C$_2$H$_3$CHO & $11_{3,9}$ -- $10_{3,8}$ & 98.001321 & -4.3451 & 23.0 & 46 & 10 \\
$t$-C$_2$H$_3$CHO$^\ddagger$ & $11_{3,8}$ -- $10_{3,7}$ & 98.019044 & -4.3448 & 23.0 & 46 & 10 \\
\hline 
\multicolumn{7}{c}{Propanal (\propanal)} \\
\hline 
$s$-C$_2$H$_5$CHO$^\ddagger$ & $14_{4,11}$ -- $14_{3,12}$ E & 87.307042 & -5.1587 & 29.0 & 62 & 6 \\
$s$-C$_2$H$_5$CHO$^\ddagger$ & $14_{4,11}$ -- $14_{3,12}$ A & 87.307594 & -5.1587 & 29.0 & 62 & 6 \\
$s$-C$_2$H$_5$CHO$^\ddagger$ & $29_{6,23}$ -- $29_{5,24}$ E & 96.793208 & -4.8841 & 59.0 & 245 & 9 \\
$s$-C$_2$H$_5$CHO$^\ddagger$ & $29_{6,23}$ -- $29_{5,24}$ A & 96.794186 & -4.8841 & 59.0 & 245 & 9 \\
$s$-C$_2$H$_5$CHO$^\dagger$ & $12_{5,7}$ -- $12_{4,8}$ E & 98.378025 & -5.0192 & 25.0 & 53 & 10 \\
$s$-C$_2$H$_5$CHO$^\dagger$ & $12_{5,7}$ -- $12_{4,8}$ A & 98.378025 & -5.0191 & 25.0 & 53 & 10 \\
$s$-C$_2$H$_5$CHO$^\dagger$ & $29_{7,22}$ -- $29_{6,23}$ A & 98.39443 & -4.847 & 59.0 & 250 & 10 \\
$s$-C$_2$H$_5$CHO$^\dagger$ & $29_{7,22}$ -- $29_{6,23}$ E & 98.394692 & -4.847 & 59.0 & 250 & 10 \\
\hline 
\multicolumn{7}{c}{Others} \\
\hline 
C$_2$H$_3$CN & $9_{1,8}$ -- $8_{1,7}$ & 87.312812 & -4.2778 & 57.0 & 23 & 6 \\
SO$_2$$^\ddagger$ & $7_{3,5}$ -- $8_{2,6}$ & 97.702334 & -5.7413 & 15.0 & 47 & 10 \\
OCS & $7$ -- $6$ & 85.139103 & -5.7658 & 15.0 & 16 & 3
\enddata
\tablenotetext{\dagger}{Blended with the other transitions of the same species.}
\tablenotetext{\ddagger}{Blended with transitions of other species.}
\tablenotetext{\ast}{Transitions excluded from the fit.}
\end{deluxetable*}

\section{Disk-integrated Spectra}\label{appendix:spectra}
Figure \ref{fig:disk-integrated_spectra0}--\ref{fig:disk-integrated_spectra3} show the disk-integrated spectra for all SPWs. The line profile of each transition exhibits the double-peaked feature, which is typical for Keplerian rotation, with some spectral blending. Figure \ref{fig:disk-integrated_spectra_fit0}--\ref{fig:disk-integrated_spectra_fit3} shows the disk-integrated spectra with spectral alignment, overlaid by the best-fit model of the spectral fit (Section \ref{subsec:spectral_fit}). The double-peaked profiles are aligned and appeared to be single-peaked, which are well reproduced with the best-fit model including spectral blending. 

\textrm{We additionally present the zoom-in spectra of CH$_3$O$^{13}$CHO and $^{13}$CH$_3$OCHO transitions detected with no blending from other molecular species in Figure \ref{fig:13C-methylformate_spectra} to confirm the detection of these species, which is critical for the robust measurement of $^{12}$C/$^{13}$C ratios in \methylformate (Section \ref{subsubsec:12C/13C}).}

\begin{figure*}
\plotone{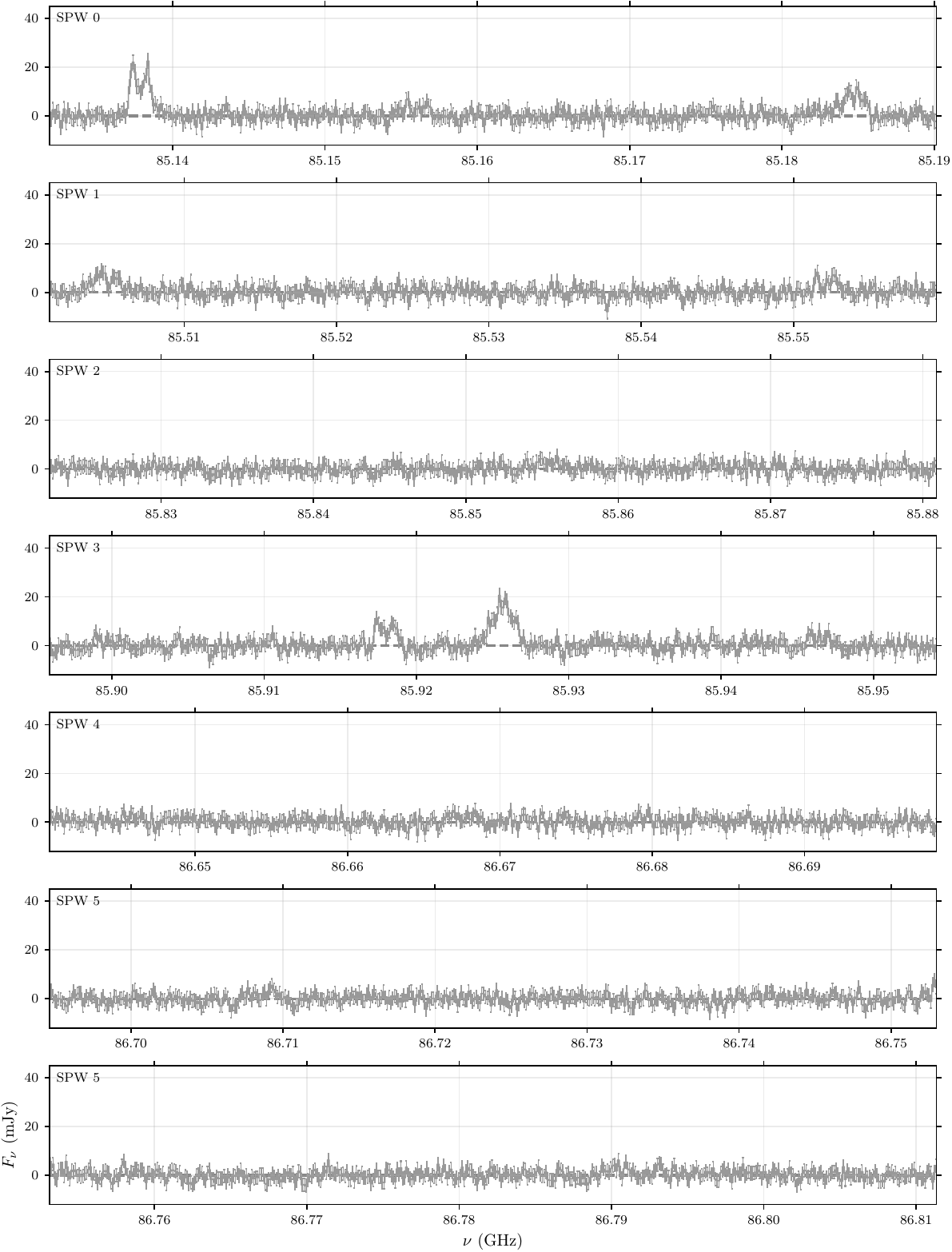}
\caption{Disk-integrated spectra toward V883 Ori without spectral alignment. The emission is integrated over the deprojected disk region with the outer radius of 0\farcs6 (or 240 au).}
\label{fig:disk-integrated_spectra0}
\end{figure*}

\begin{figure*}
\plotone{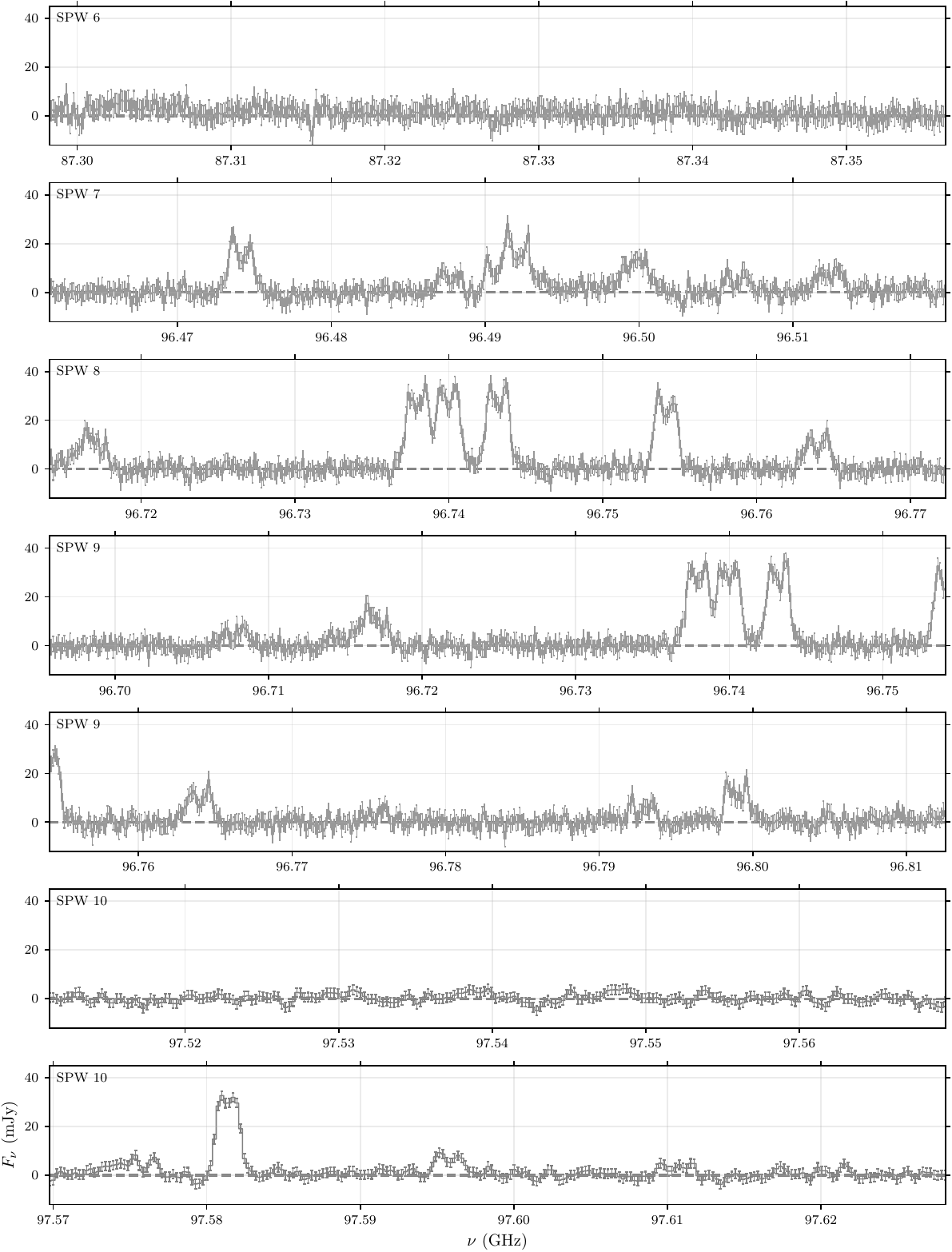}
\caption{Continuation of Figure \ref{fig:disk-integrated_spectra0}.}
\label{fig:disk-integrated_spectra1}
\end{figure*}

\begin{figure*}
\plotone{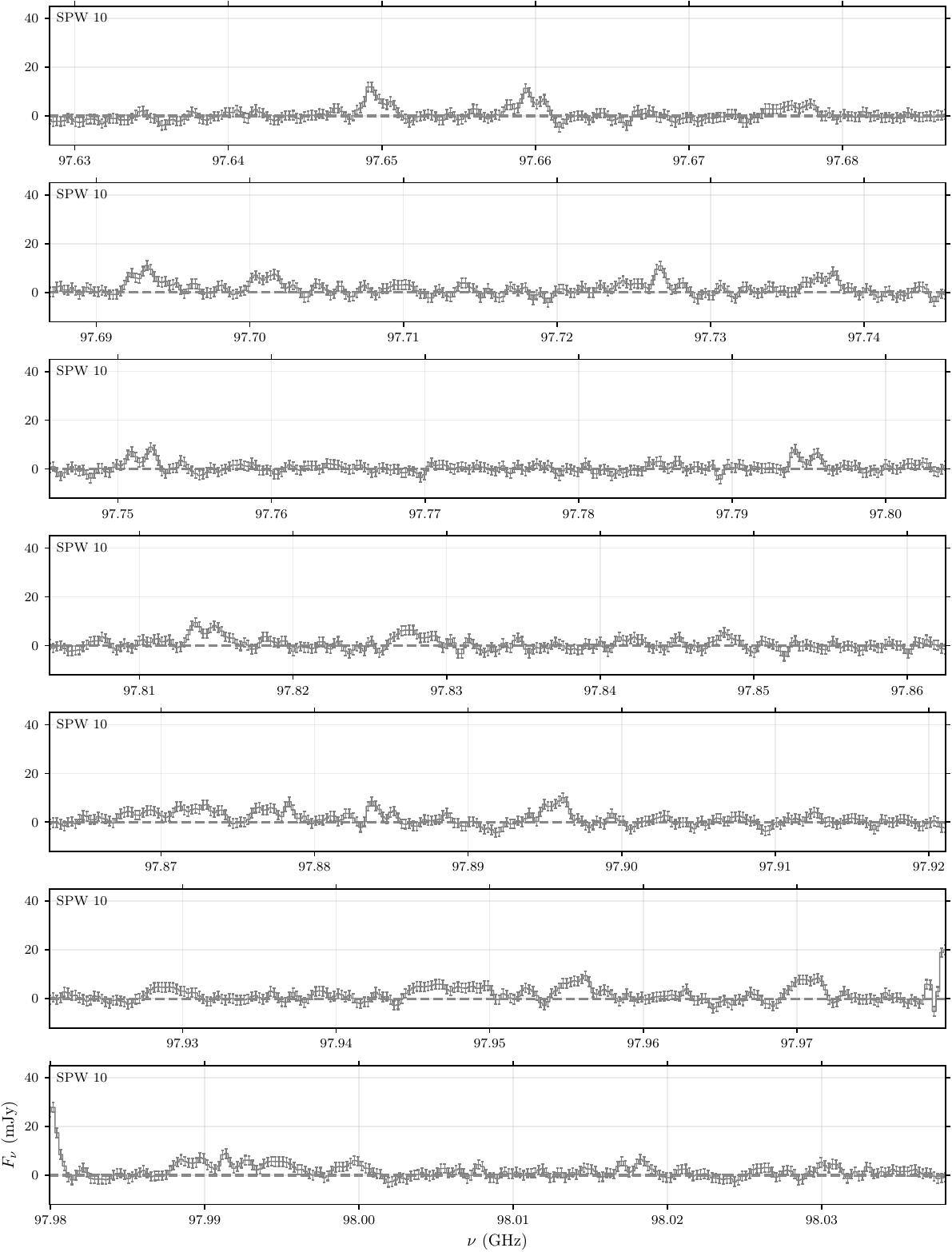}
\caption{Continuation of Figure \ref{fig:disk-integrated_spectra1}.}
\label{fig:disk-integrated_spectra2}
\end{figure*}

\begin{figure*}
\plotone{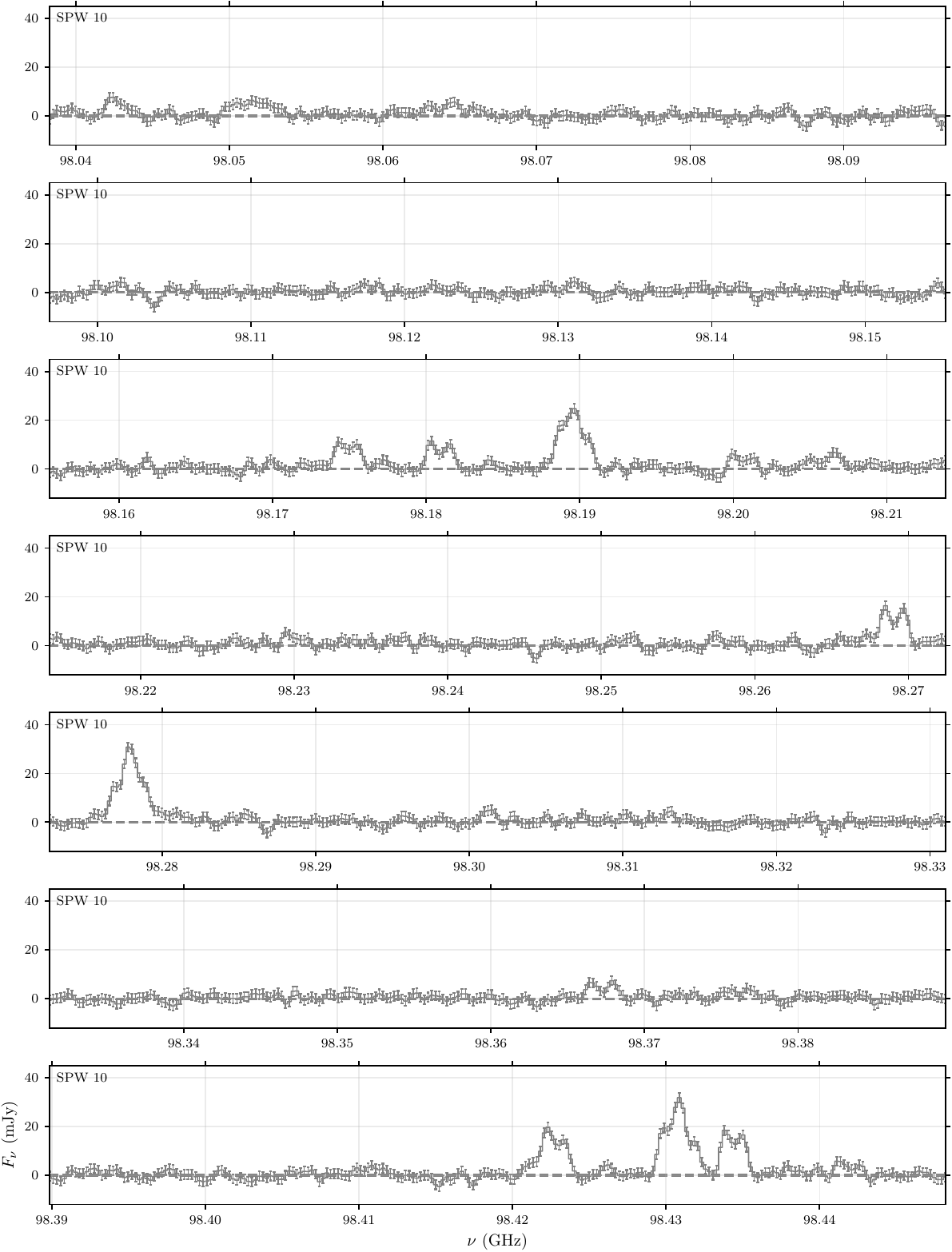}
\caption{Continuation of Figure \ref{fig:disk-integrated_spectra2}.}
\label{fig:disk-integrated_spectra3}
\end{figure*}

\begin{figure*}
\plotone{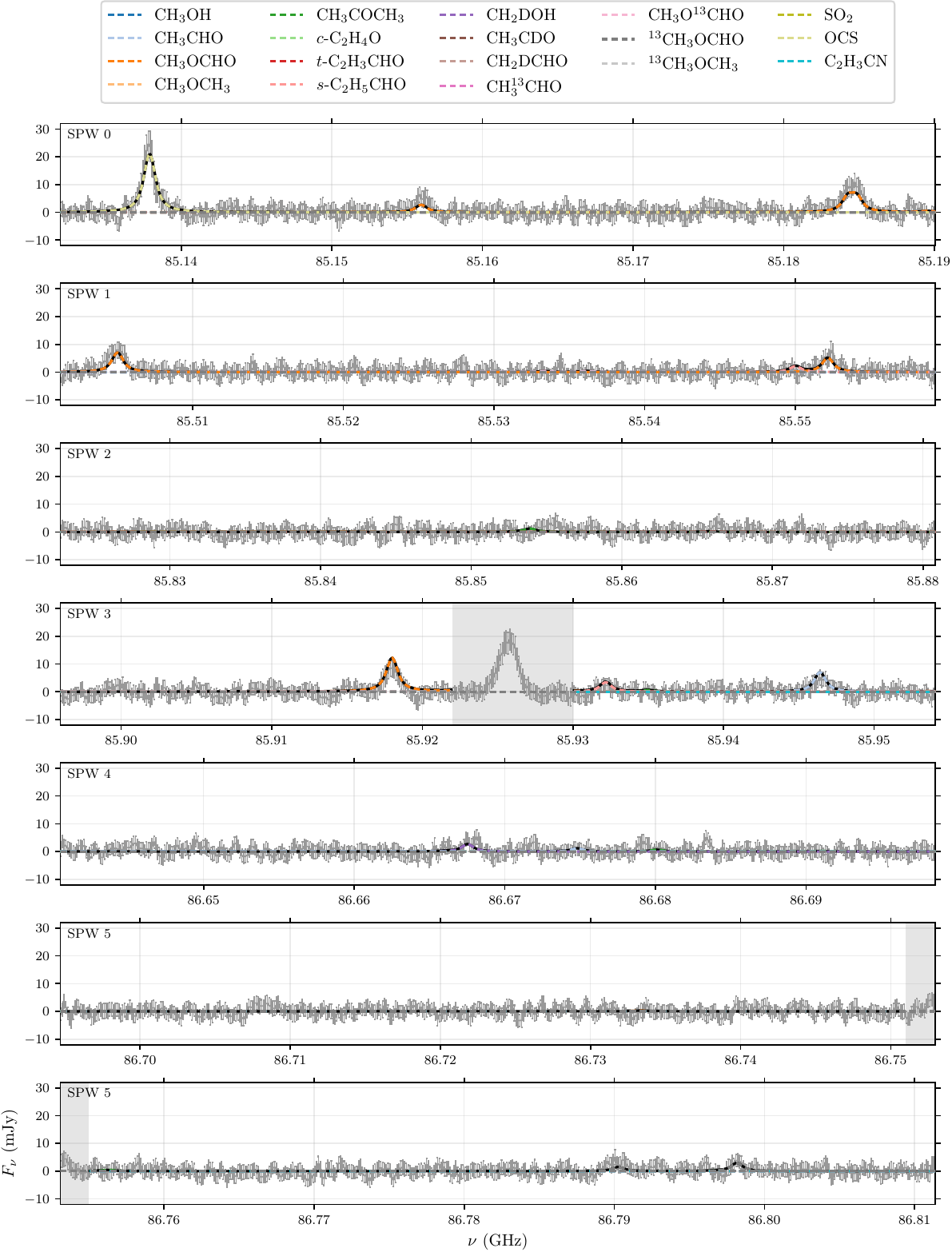}
\caption{Disk-integrated spectra toward V883 Ori after the spectral alignment, overlaid with the best-fit model. The black solid lines indicate the full model composed of all detected species, while dashed colored lines are the model for each species shown in the legend. The gray-shaded region are removed from the fit (see Section \ref{subsec:spectral_fit} for details.)}
\label{fig:disk-integrated_spectra_fit0}
\end{figure*}

\begin{figure*}
\plotone{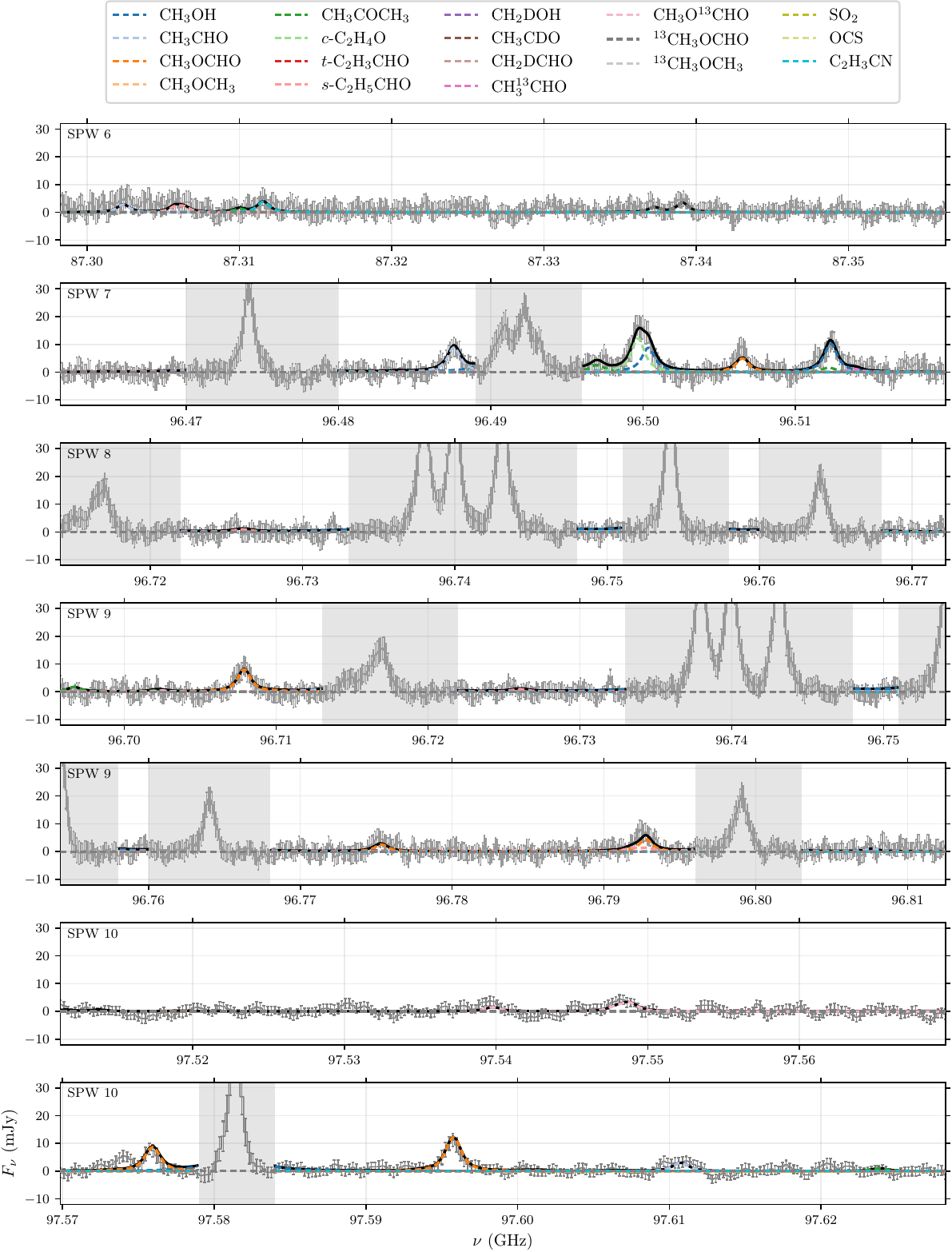}
\caption{Continuation of Figure \ref{fig:disk-integrated_spectra_fit0}.}
\label{fig:disk-integrated_spectra_fit1}
\end{figure*}

\begin{figure*}
\plotone{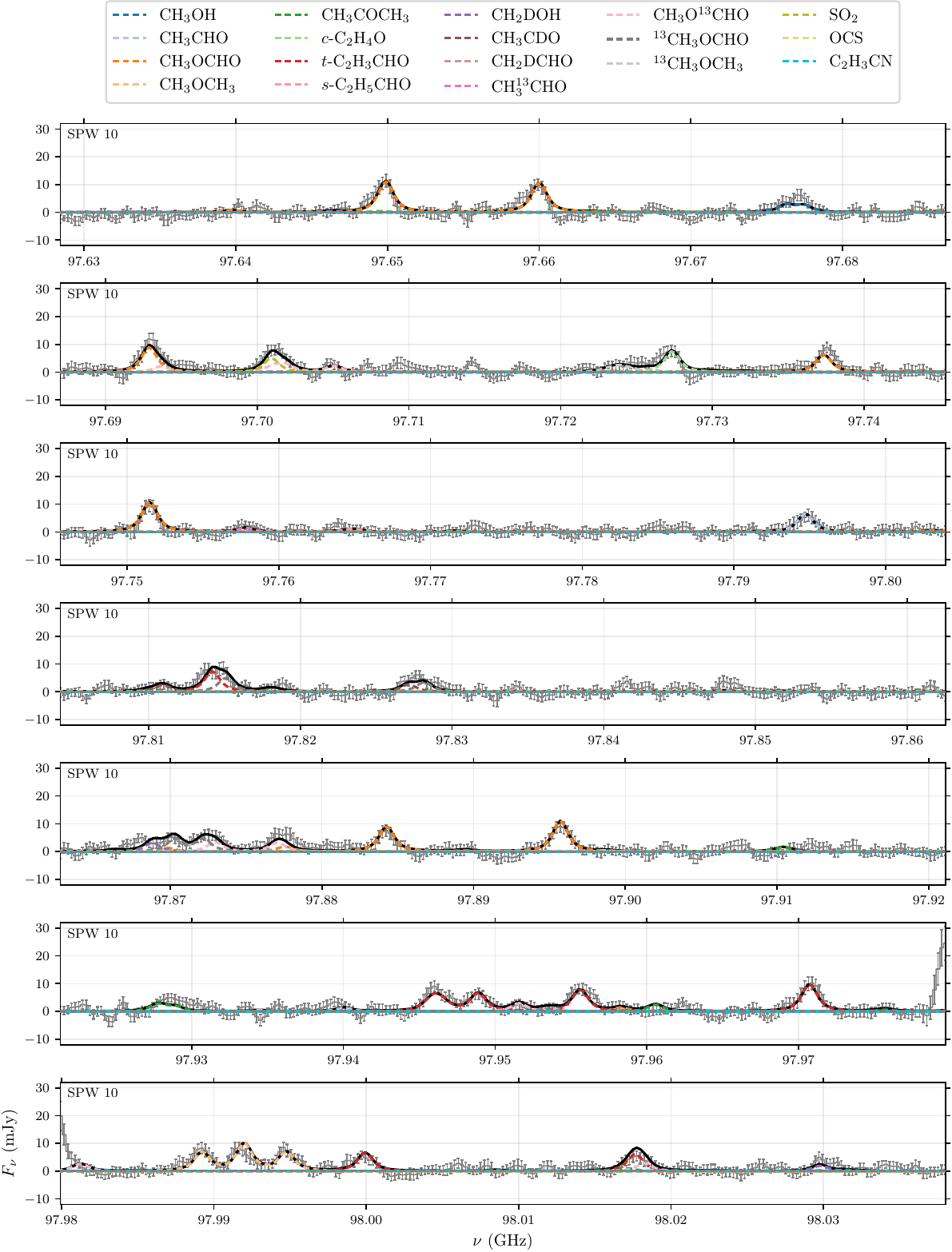}
\caption{Continuation of Figure \ref{fig:disk-integrated_spectra_fit1}.}
\label{fig:disk-integrated_spectra_fit2}
\end{figure*}

\begin{figure*}
\plotone{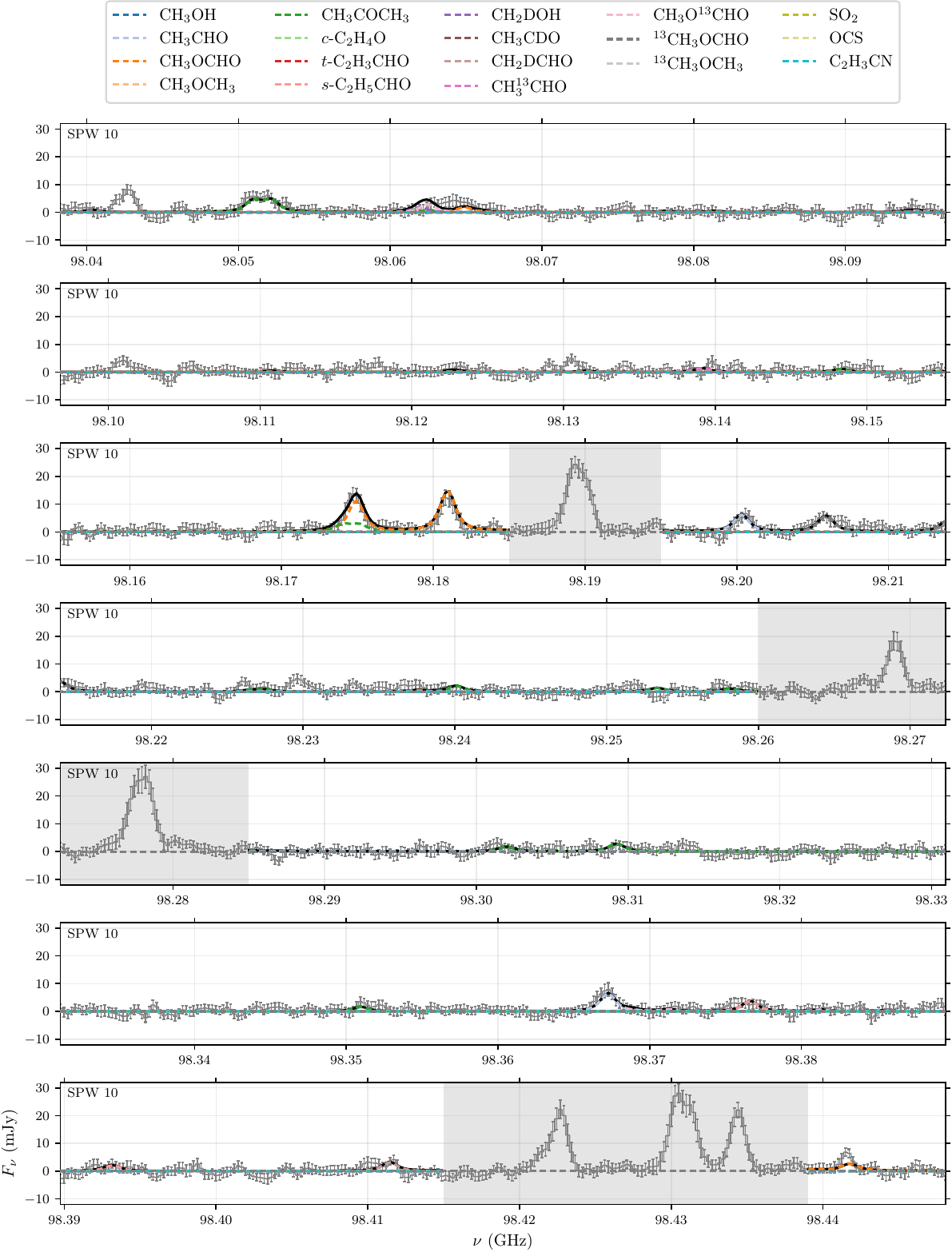}
\caption{Continuation of Figure \ref{fig:disk-integrated_spectra_fit2}.}
\label{fig:disk-integrated_spectra_fit3}
\end{figure*}




\begin{figure*}
\epsscale{0.9}
\plotone{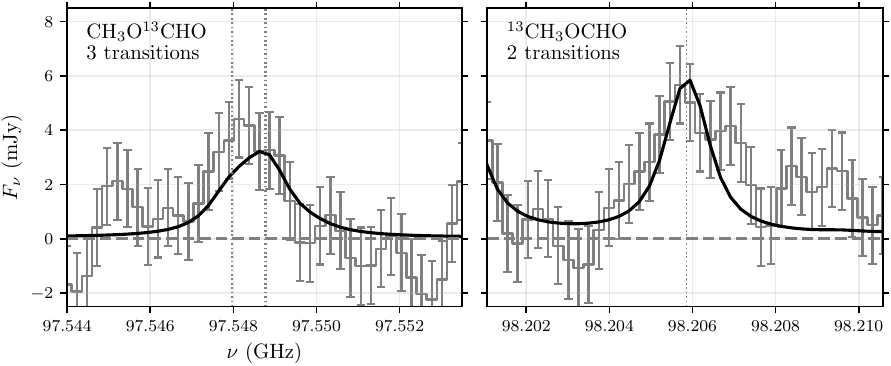}
\caption{Zoom-in spectra of CH$_3$O$^{13}$CHO (left) and $^{13}$CH$_3$OCHO (right) transitions detected with no blending from other molecular species. The gray lines with error bars are the observed spectra, and the black solid lines indicate the best-fit model (see Section \ref{subsec:spectral_fit}). The central frequency of transitions are marked by the vertical dotted lines.}
\label{fig:13C-methylformate_spectra}
\end{figure*}

\section{LTE spectral model}\label{appendix:spectral_model}
Here we describe the details of the spectral model which is fitted to the spectra corrected for Keplerian rotation for column density derivation. We assume that the excitation condition of observed COM emission is well approximated by local thermodynamic equilibrium (LTE), where the gas temperature is equal to the excitation temperature ($T_\mathrm{ex}$) of the emitting molecules. This assumption should be reasonable as the typical gas density in protoplanetary disks ($\gtrsim10^8$\, cm$^{-3}$) is high enough to thermalize the excitation condition. 
We further assume that different species are sufficiently cospatial to share the same emitting region size and excitation temperature. This assumption is also broadly supported by the similar emission extent of different COMs shown in Figure \ref{fig:moment_zero_gallery}. 

The model intensity $I_\nu$ are computed based on a solution of the basic radiative transfer equations for an isothermal, uniform slab:
\begin{equation}
    I_\nu = (B_\nu(T_\mathrm{ex}) - B_\nu(T_\mathrm{CMB})) (1 - e^{-\tau_\nu}),
\end{equation}
where $B_\nu$ is the Planck function for a blackbody radiation, $T_\mathrm{CMB} = 2.73$ K is the temperature of the cosmic microwave background, and $\tau_\nu$ is the optical depth of the line emission at a frequency $\nu$. Following the formulation described in Appendix A of \citet{Yamato2022}, the line optical depth $\tau_\nu$ is computed considering different transitions from various species as
\begin{equation}
    \tau_\nu = \sum_{i}\tau_{0, i} \exp\bigg(-\frac{(\nu - \nu_\mathrm{c})^2}{2 \sigma_\nu^2}\bigg),
\end{equation}
where $i$ is the indices for different transitions, $\tau_{0, i}$ is the optical depth at the line center of $i$th transition, $\nu_\mathrm{c}$ is the central frequency of the transition, and $\sigma_\nu$ is the line width in terms of frequency. The central frequency $\nu_\mathrm{c}$ is calculated as $\nu_\mathrm{c} = \nu_0 (1 - v_\mathrm{sys}/c)$, where $c$ is the speed of light and $v_\mathrm{sys}$ is the systemic velocity of the source, assumed to be $4.25$ km s$^{-1}$ for V883 Ori \citep{Tobin2023}. The frequency line width $\sigma_\nu$ is converted from the velocity line width $\sigma_v$ as $\sigma_\nu = \nu_0 \sigma_v / c$, where $\nu_0$ is the rest frequency of each transition listed in Table \ref{tab:transitions}. In practice, the full width of half maximum (FWHM) of the velocity $\Delta V_\mathrm{FWHM}$ is used as a parameter instead of $\sigma_\nu$ or $\sigma_v$. The optical depth at the line center ($\tau_\mathrm{0, i}$) is calculated as
\begin{equation}
    \tau_{0, i} = \frac{c^2A_\mathrm{u}N_\mathrm{u}}{8\pi\nu_0^2\sqrt{2\pi}\sigma_\nu}\Bigg(\exp\left(\frac{h\nu_0}{k_\mathrm{B}T_\mathrm{ex}}\right) - 1\Bigg),
\end{equation}
\begin{equation}
    \frac{N_\mathrm{u}}{N} = \frac{g_\mathrm{u}}{Q(T_\mathrm{ex})}\exp\left(-\frac{E_\mathrm{u}}{k_\mathrm{B}T_\mathrm{ex}}\right),
\end{equation}
where $A_\mathrm{ul}$ is the Einstein A coefficient for spontaneous emission, $g_\mathrm{u}$ is the upper state degeneracy, $N$ is the molecular column density, $Q$ is the partition function of the molecule, and $E_\mathrm{u}$ is the upper state energy of the transition. While this suit of formulation is similar to that of the eXtended CASA Line Analysis Software Suite (XCLASS; \citealt{Moller2017}), which is used in a previous work on the V883 Ori disk \citep{Lee2019}, we used an independent implementation by ourselves for technical flexibility.

\begin{figure}
\plotone{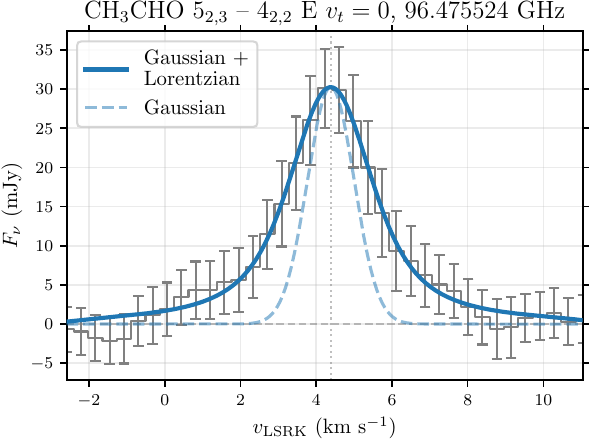}
\caption{Fit of a Gaussian + Lorentzian model (i.e., Voigt profile; solid blue) to the observed spectra of a \acetaldehyde transition at 96.476 GHz (gray). The original Gaussian before the convolution with Lorentzian is indicated by the dashed line. The Gaussian + Lorentzian model well reproduces the wing of the spectra. The horizontal dashed line and vertical dotted line marks the zero-flux level and the systemic velocity ($v_\mathrm{sys} = 4.25$ km s$^{-1}$), respectively.}
\label{fig:Lorentzian_demo}
\end{figure}

The model intensity $I_\nu$ (in unit of Jy sr$^{-1}$) is then integrated over a solid angle of the emitting region $\Omega$, which is common for all transitions and species, to obtain the spectra of flux density (in unit of Jy). 
Finally, the model flux density spectra are convolved with a Lorentz function $f(\nu)$ with a width of $\gamma$,
\begin{equation}
    f(\nu) = \frac{1}{\pi\gamma} \frac{\gamma^2}{\gamma^2 + \nu^2},
\end{equation}
to take line broadening in the spectra corrected for Keplerian rotation. As shown in Figure \ref{fig:Lorentzian_demo}, the spectra corrected for Keplerian rotation still deviates from a simple Gaussian due to (1) the finite spatial resolution or beam smearing, which cannot fully resolve the highest velocity component of Keplerian rotation and (2) potential emission from the elevated disk surface of the back side and front side of the disk, which have different projected velocities. In our data, both (1) and (2) could be dominant causes of the deviation, since the spatial resolution is not so high to fully resolve the disk emission and the COM emission could originate from the warm disk surface in addition to the midplane.
Figure \ref{fig:Lorentzian_demo} demonstrates that the convolution with the Lorentz function well approximates the deviation from Gaussian while conserving the velocity-integrated flux density, which is relevant for column density derivation. Furthermore, to account for the finite spectral resolution, the model spectra is additionally convolved with a Gaussian function with a FWHM of the spectral resolution. This process also accounts for the different spectral resolutions (per channel width) between SPW 10 and others (see Table \ref{tab:cube_properties}). A similar approach has been employed in the spectral line analysis in protoplanetary disks \citep{Cataldi2021}, where they used only a Gaussian convolution to take the line broadening into account (see also \citealt{Bergner2021, Guzman2021}).


\section{Reconstruction of Radial Intensity Profile}\label{appendix:line_profile_analysis_method}
Here we describe details of the forward modeling approach to reconstruct the radial intensity profiles from the line profiles (Section \ref{subsubsec:line_profile_analysis}). The method we employed is similar to the way described in \citet{Bosman2021} but slightly different. First we create the model emission component originating from an annulus at a given radius of the inclined disk. We made a grid of cubes and calculated the line shape of the spatially integrated emission as a function of velocity assuming that the emission is completely originating from the midplane of the Keplerian-rotating disk, i.e.,
\begin{equation}\label{eq:line_profile_model}
    F_{r_i}(v) \propto \int_{\Omega_\mathrm{annulus}}I_{r_i}\exp\left[- \frac{(v - v_\mathrm{kep}(r, \theta) - v_\mathrm{sys})^2}{2\sigma_v^2}\right]d\Omega,
\end{equation}
where $\Omega_\mathrm{annulus}$ is the solid angle of the annulus, $I_{r_i}$ is the surface brightness (in Jy au$^{-2}$ km s$^{-1}$) at the radius of $r_i$, $v_\mathrm{kep}(r, \theta) = \sqrt{GM_\star/r}\,\cos\theta\,\sin i$ is the Keplerian velocity at the deprojected polar disk coordinate $(r, \theta)$, $v_\mathrm{sys}$ is the systemic velocity of the source (fixed to 4.25 km s$^{-1}$), and $\sigma_v$ is the local line width at each position within the disk. We assumed that the local line width is narrower than the spectral resolution (0.5 km s$^{-1}$), where 0.05 km s$^{-1}$ was used practically. The normalization of Equation (\ref{eq:line_profile_model}) is taken as the integration of $F_{r_i}(v)$ along the velocity axis being the spatial integration of $I_{r_i}$. The emission component at each radius $r_i$ is summed over $i$ to construct the disk-integrated spectra, i.e., 
\begin{equation}
    F(v) = \sum_i F_{r_i}(v),
\end{equation}
which is fitted to the observed disk-integrated spectra by varying each $I_{r_i}$ as free parameters. We used a regularly spaced grid in velocity space, which result in the unevenly spaced radial grid due to the non-linear relationship between Keplerian velocity and disk radius (see Equation \ref{eq:Keplerian_rotation}). The spacing of the grid is regulated by the spectral resolution: we used one grid point ($r_i$) per two velocity resolution components to obtain reasonable constraints on each $I_{r_i}$. This treatment resulted in a very sparse radial grid $r_i$ in larger radii, which makes it difficult to infer the radial distribution there. To mitigate this effect, we employed four fits with different radial grid created with different starting radii ($r_0 = 3, 5, 7, 9$\,au) and composed these four fits at the end to construct the final radial intensity profile. The actual fits were conducted by the MCMC algorithm implemented in the \texttt{emcee} package \citep{emcee}. The reconstructed intensity profiles are further convolved with the 1D Gaussian function with a 0\farcs1 FWHM to emulate the beam smearing effect and presented in Figure \ref{fig:radial_profile_from_line_profile}.      

\section{Spectroscopic data for molecules}\label{appendix:spectroscopic_data}
To employ the astronomical modeling of the observed spectra and estimate column density, spectroscopic data such as transition frequencies, intrinsic line strengths, and partition functions are essential information. They are basically retrieved from the Cologne
Database of Molecular Spectroscopy \citep[CDMS;][]{CDMS1, CDMS2, CDMS3} or the Jet Propulsion Laboratory (JPL) database \citep{JPL}. We describe the origin of these data and molecule-specific issues for each species below, with a particular care on the vibratinonal contributions to the partition functions. 

\subsection{Methanol}\label{appendix:specdata_methanol}
Spectroscopic data for \methanol are taken from the CDMS, which are largely based on the theoretical calculation for the ground state ($v_t = 0$) and the first three torsional states ($v_t = 1, 2, 3$) by \citet{Xu2008}. The partition function of \methanol which considers the contributions from these states is also available in the CDMS. 

The data for the deuterated isotopologues, CH$_2$DOH, are taken from the JPL database, which is based on the laboratory experiment for the ground torsional state \citep{Pearson2012}. The partition function available in the JPL database only considers the ground torsional state. \textrm{However, since the spectroscopic calculation of CH$_2$DOH is challenging, the data have a large uncertainty, in particular for the line intensities of $b$-type and $c$-type transitions \citep{Pearson2012, Oyama2023}. The tentatively detected transitions with the present observations are all $b$-type transitions, which could have a factor of $\sim3$ uncertainties at maximum judging from the intensity difference found between the JPL database and the new experiment in ALMA Band 6 frequencies by \citet{Oyama2023}. The derived column density and CH$_2$DOH/CH$_3$OH ratios could thus have a similar level uncertainty, which should be viewed with strong caution.}


\subsection{Acetaldehyde}
The data for \acetaldehyde are taken from the JPL database. This is based on the data listed in \citet{Kleiner1996} for the ground state ($v_t = 0$) and the first two torsional states ($v_t = 1, 2$). The partition function of \acetaldehyde available in the JPL database also considers these three torsional states. 

There are two isomers of $^{13}$C-acetaldehyde: $^{13}$CH$_3$CHO and CH$_3^{13}$CHO. The data of both isomers for the ground state ($v_t = 0$) and the first torsional state ($v_t = 1$) are available at the CDMS based on the laboratory measurement by \citep{Margules2015}. The partition function which considers the contributions from torsional states up to $v_t = 8$ is also available at the CDMS.

There are also two isomers of deuterated acetaldehyde: CH$_2$DCHO and CH$_3$CDO. While the data of the ground states ($v_t = 0$) and the first torsional states ($v_t = 1$) are available for CH$_3$CDO at the CDMS, only the data of the ground states ($v_t = 0$) exists for CH$_2$DCHO. These data are based on the laboratory experiments by \citet{Coudert2019}. The partition functions of these molecules are available at the CDMS. The partition function of CH$_2$CDO are calculated considering the contributions from the first three torsional modes and the approximate contributions from the lowest vibrational mode. The partition function for CH$_2$DCHO are, on the other hand, calculated considering the approximate contributions from the lowest torsional mode and the lowest vibrational mode.

\subsection{Methyl Formate}
We used the spectroscopic data for \methylformate taken from the JPL database. This data is based on the laboratory experiment for the ground states ($v_t = 0$) and the first torsional state ($v_t = 1$) by \citet{Ilyushin2009}. Similar to acetaldehyde, there are two $^{13}$C-isotopologues of methyl formate: $^{13}$CH$_3$OCHO and CH$_3$O$^{13}$CHO. While the data of the ground states ($v_t = 0$) and the first torsional state ($v_t = 1$) for CH$_3$O$^{13}$CHO are available at the CDMS, which is based on the laboratory measurements \citep[][and references therein]{Carvajal2010}, no CDMS entry for $^{13}$CH$_3$OCHO are found. Instead, we compiled the experimental data and theoretical predictions for $^{13}$CH$_3$OCHO transitions from literature \citep{Carvajal2009, Haykal2014, Favre2014}. 

For the partition function of these three molecules, we used the one calculated by \citet{Favre2014}, who fully considered the contributions from torsional (up to $v_t=6$) and vibrational populations. The partition function data exist in the JPL database as well for the normal isotopologues (\methylformate), but only the contributions from $v_t = 0, 1$ states are considered. This resulted in a difference between the one by \citet{Favre2014} and the one in the JPL database by a factor of 1.2 at 150~K and 2.5 at 300~K \citep{Favre2014}. We note that there are difference in the treatment of the degeneracy, where \citet{Favre2014} ignore the nuclear spin degeneracy (i.e., $g_\mathrm{I} = 1$) while the CDMS and JPL entries take $g_\mathrm{I} = 2$. To take this difference into account, we multiply the partition function listed in \citet{Favre2014} by two in practice.

We also examined the data for deuterated methyl formate. Similarly, there are two isomers: CH$_2$DOCHO and CH$_3$OCDO. Spectroscopic data for both molecules are taken from the CDMS, which is based on the experimental measurement by \citet{Coudert2013} and \citet{Margules2009} for CH$_2$DOCHO, and \citet{Margules2010} and \citet{Duan2015} for CH$_3$OCDO. The partition functions of these molecules available in the CDMS only take the ground vibrational state into account. 
\citet{Manigand2019} uses the vibrational correction factor of 1.31 and 1.11 at 115~K for CH$_2$OCHO and CH$_3$OCDO, respectively, assuming that the correction factor is the same as $^{13}$C-isotopologues. We applied these factors to the partition functions of these molecules.

\subsection{Dimethyl Ether}
The data for \dimethylether are taken from the CDMS, which is based on the experimental measurement by \citet{Endres2009} for the ground vibrational states ($v = 0$) with four substates (AA, AE, EA, and EE). The partition function is calculated considering the torsionally excited states ($v_{11} = 1$ and $v_{15} = 1$), C-O-C bending states ($v_7 = 1$), and $v_{11} + v_{15} = 2$, and available at the CDMS. 

The spectroscopic data for $^{13}$C isotopologue ($^{13}$\dimethylether) are also available at the CDMS and based on the laboratory experiments by \citet{Koerber2013}. The partition function in the CDMS is calculated only taking the ground vibrational states into account, and the vibrational correction factor is not yet available.   

\subsection{Acetone}
The data for \acetone are taken from the JPL. The JPL entry compiled the data from \citet{Peter1965}, \citet{Vacherand1986}, \citet{Oldag1992}, and \citet{Groder2002}. The partition function which takes into account the torsional and vibrational state are available at the JPL and used for the spectral fit. 

\subsection{Ethylene Oxide}
The data for \ethyleneoxide are taken from the CDMS. This CDMS entry is based on the experimental data in \citet{Creswell1974} and \citet{Muller2022}. The partition function available at the CDMS takes into account the ground vibrational state only, and the vibrational correction factor is not available.

\subsection{Propenal}
The data for \propenal are taken from the CDMS, which are largely based on the experimental data in \citet{Daly2015} with additional data from \citet{Winnewisser1975} and \citet{Cherniak1966}. The partition function is calculated by considering no vibrational states nor other conformers, and the vibrational and conformational correction factors are not yet available. We thus do not apply any correction factors to the partition function.

\subsection{Propanal}
The data for \propanal are taken from the CDMS, which are largely based on the experimental measurements by \citet{Zingsheim2017} with additional data from \citet{Hardy1982} and \citet{Demaison1987}. The partition function is calculated by considering no vibrational states nor other conformers, and the vibrational and conformational correction factors are not yet available. We thus do not apply any correction factors to the partition function.


\bibliography{sample631}{}

\begin{thebibliography}{}
\expandafter\ifx\csname natexlab\endcsname\relax\def\natexlab#1{#1}\fi
\providecommand{\url}[1]{\href{#1}{#1}}
\providecommand{\dodoi}[1]{doi:~\href{http://doi.org/#1}{\nolinkurl{#1}}}
\providecommand{\doeprint}[1]{\href{http://ascl.net/#1}{\nolinkurl{http://ascl.net/#1}}}
\providecommand{\doarXiv}[1]{\href{https://arxiv.org/abs/#1}{\nolinkurl{https://arxiv.org/abs/#1}}}

\bibitem[{{Ag{\'u}ndez} {et~al.}(2019){Ag{\'u}ndez}, {Marcelino}, {Cernicharo}, {Roueff}, \& {Tafalla}}]{Agundez2019}
{Ag{\'u}ndez}, M., {Marcelino}, N., {Cernicharo}, J., {Roueff}, E., \& {Tafalla}, M. 2019, \aap, 625, A147, \dodoi{10.1051/0004-6361/201935164}

\bibitem[{{Aikawa} {et~al.}(2022){Aikawa}, {Okuzumi}, \& {Pontoppidan}}]{Aikawa2022}
{Aikawa}, Y., {Okuzumi}, S., \& {Pontoppidan}, K. 2022, arXiv e-prints, arXiv:2212.14529, \dodoi{10.48550/arXiv.2212.14529}

\bibitem[{{Aikawa} {et~al.}(2002){Aikawa}, {van Zadelhoff}, {van Dishoeck}, \& {Herbst}}]{Aikawa2002}
{Aikawa}, Y., {van Zadelhoff}, G.~J., {van Dishoeck}, E.~F., \& {Herbst}, E. 2002, \aap, 386, 622, \dodoi{10.1051/0004-6361:20020037}

\bibitem[{{Alarc{\'o}n} {et~al.}(2023){Alarc{\'o}n}, {Casassus}, {lyra}, {P{\'e}rez}, \& {Cieza}}]{Alarcon2023}
{Alarc{\'o}n}, F., {Casassus}, S., {lyra}, W., {P{\'e}rez}, S., \& {Cieza}, L. 2023, arXiv e-prints, arXiv:2311.17195.
\newblock \doarXiv{2311.17195}

\bibitem[{{Altwegg} {et~al.}(2019){Altwegg}, {Balsiger}, \& {Fuselier}}]{Altwegg2019}
{Altwegg}, K., {Balsiger}, H., \& {Fuselier}, S.~A. 2019, \araa, 57, 113, \dodoi{10.1146/annurev-astro-091918-104409}

\bibitem[{{Altwegg} {et~al.}(2017){Altwegg}, {Balsiger}, {Berthelier}, {Bieler}, {Calmonte}, {De Keyser}, {Fiethe}, {Fuselier}, {Gasc}, {Gombosi}, {Owen}, {Le Roy}, {Rubin}, {S{\'e}mon}, \& {Tzou}}]{Altwegg2017}
{Altwegg}, K., {Balsiger}, H., {Berthelier}, J.~J., {et~al.} 2017, Philosophical Transactions of the Royal Society of London Series A, 375, 20160253, \dodoi{10.1098/rsta.2016.0253}

\bibitem[{{Altwegg} {et~al.}(2020{\natexlab{a}}){Altwegg}, {Balsiger}, {Combi}, {De Keyser}, {Drozdovskaya}, {Fuselier}, {Gombosi}, {H{\"a}nni}, {Rubin}, {Schuhmann}, {Schroeder}, \& {Wampfler}}]{Altwegg2020_isotopes}
{Altwegg}, K., {Balsiger}, H., {Combi}, M., {et~al.} 2020{\natexlab{a}}, \mnras, 498, 5855, \dodoi{10.1093/mnras/staa2701}

\bibitem[{{Altwegg} {et~al.}(2020{\natexlab{b}}){Altwegg}, {Balsiger}, {H{\"a}nni}, {Rubin}, {Schuhmann}, {Schroeder}, {S{\'e}mon}, {Wampfler}, {Berthelier}, {Briois}, {Combi}, {Gombosi}, {Cottin}, {De Keyser}, {Dhooghe}, {Fiethe}, \& {Fuselier}}]{Altwegg2020}
{Altwegg}, K., {Balsiger}, H., {H{\"a}nni}, N., {et~al.} 2020{\natexlab{b}}, Nature Astronomy, 4, 533, \dodoi{10.1038/s41550-019-0991-9}

\bibitem[{{Astropy Collaboration} {et~al.}(2013){Astropy Collaboration}, {Robitaille}, {Tollerud}, {Greenfield}, {Droettboom}, {Bray}, {Aldcroft}, {Davis}, {Ginsburg}, {Price-Whelan}, {Kerzendorf}, {Conley}, {Crighton}, {Barbary}, {Muna}, {Ferguson}, {Grollier}, {Parikh}, {Nair}, {Unther}, {Deil}, {Woillez}, {Conseil}, {Kramer}, {Turner}, {Singer}, {Fox}, {Weaver}, {Zabalza}, {Edwards}, {Azalee Bostroem}, {Burke}, {Casey}, {Crawford}, {Dencheva}, {Ely}, {Jenness}, {Labrie}, {Lim}, {Pierfederici}, {Pontzen}, {Ptak}, {Refsdal}, {Servillat}, \& {Streicher}}]{AstropyI}
{Astropy Collaboration}, {Robitaille}, T.~P., {Tollerud}, E.~J., {et~al.} 2013, \aap, 558, A33, \dodoi{10.1051/0004-6361/201322068}

\bibitem[{{Astropy Collaboration} {et~al.}(2018){Astropy Collaboration}, {Price-Whelan}, {Sip{\H{o}}cz}, {G{\"u}nther}, {Lim}, {Crawford}, {Conseil}, {Shupe}, {Craig}, {Dencheva}, {Ginsburg}, {VanderPlas}, {Bradley}, {P{\'e}rez-Su{\'a}rez}, {de Val-Borro}, {Aldcroft}, {Cruz}, {Robitaille}, {Tollerud}, {Ardelean}, {Babej}, {Bach}, {Bachetti}, {Bakanov}, {Bamford}, {Barentsen}, {Barmby}, {Baumbach}, {Berry}, {Biscani}, {Boquien}, {Bostroem}, {Bouma}, {Brammer}, {Bray}, {Breytenbach}, {Buddelmeijer}, {Burke}, {Calderone}, {Cano Rodr{\'\i}guez}, {Cara}, {Cardoso}, {Cheedella}, {Copin}, {Corrales}, {Crichton}, {D'Avella}, {Deil}, {Depagne}, {Dietrich}, {Donath}, {Droettboom}, {Earl}, {Erben}, {Fabbro}, {Ferreira}, {Finethy}, {Fox}, {Garrison}, {Gibbons}, {Goldstein}, {Gommers}, {Greco}, {Greenfield}, {Groener}, {Grollier}, {Hagen}, {Hirst}, {Homeier}, {Horton}, {Hosseinzadeh}, {Hu}, {Hunkeler}, {Ivezi{\'c}}, {Jain}, {Jenness}, {Kanarek}, {Kendrew}, {Kern}, {Kerzendorf}, {Khvalko}, {King}, {Kirkby}, {Kulkarni},
  {Kumar}, {Lee}, {Lenz}, {Littlefair}, {Ma}, {Macleod}, {Mastropietro}, {McCully}, {Montagnac}, {Morris}, {Mueller}, {Mumford}, {Muna}, {Murphy}, {Nelson}, {Nguyen}, {Ninan}, {N{\"o}the}, {Ogaz}, {Oh}, {Parejko}, {Parley}, {Pascual}, {Patil}, {Patil}, {Plunkett}, {Prochaska}, {Rastogi}, {Reddy Janga}, {Sabater}, {Sakurikar}, {Seifert}, {Sherbert}, {Sherwood-Taylor}, {Shih}, {Sick}, {Silbiger}, {Singanamalla}, {Singer}, {Sladen}, {Sooley}, {Sornarajah}, {Streicher}, {Teuben}, {Thomas}, {Tremblay}, {Turner}, {Terr{\'o}n}, {van Kerkwijk}, {de la Vega}, {Watkins}, {Weaver}, {Whitmore}, {Woillez}, {Zabalza}, \& {Astropy Contributors}}]{AstropyII}
{Astropy Collaboration}, {Price-Whelan}, A.~M., {Sip{\H{o}}cz}, B.~M., {et~al.} 2018, \aj, 156, 123, \dodoi{10.3847/1538-3881/aabc4f}

\bibitem[{{Astropy Collaboration} {et~al.}(2022){Astropy Collaboration}, {Price-Whelan}, {Lim}, {Earl}, {Starkman}, {Bradley}, {Shupe}, {Patil}, {Corrales}, {Brasseur}, {N{\"o}the}, {Donath}, {Tollerud}, {Morris}, {Ginsburg}, {Vaher}, {Weaver}, {Tocknell}, {Jamieson}, {van Kerkwijk}, {Robitaille}, {Merry}, {Bachetti}, {G{\"u}nther}, {Aldcroft}, {Alvarado-Montes}, {Archibald}, {B{\'o}di}, {Bapat}, {Barentsen}, {Baz{\'a}n}, {Biswas}, {Boquien}, {Burke}, {Cara}, {Cara}, {Conroy}, {Conseil}, {Craig}, {Cross}, {Cruz}, {D'Eugenio}, {Dencheva}, {Devillepoix}, {Dietrich}, {Eigenbrot}, {Erben}, {Ferreira}, {Foreman-Mackey}, {Fox}, {Freij}, {Garg}, {Geda}, {Glattly}, {Gondhalekar}, {Gordon}, {Grant}, {Greenfield}, {Groener}, {Guest}, {Gurovich}, {Handberg}, {Hart}, {Hatfield-Dodds}, {Homeier}, {Hosseinzadeh}, {Jenness}, {Jones}, {Joseph}, {Kalmbach}, {Karamehmetoglu}, {Ka{\l}uszy{\'n}ski}, {Kelley}, {Kern}, {Kerzendorf}, {Koch}, {Kulumani}, {Lee}, {Ly}, {Ma}, {MacBride}, {Maljaars}, {Muna}, {Murphy}, {Norman},
  {O'Steen}, {Oman}, {Pacifici}, {Pascual}, {Pascual-Granado}, {Patil}, {Perren}, {Pickering}, {Rastogi}, {Roulston}, {Ryan}, {Rykoff}, {Sabater}, {Sakurikar}, {Salgado}, {Sanghi}, {Saunders}, {Savchenko}, {Schwardt}, {Seifert-Eckert}, {Shih}, {Jain}, {Shukla}, {Sick}, {Simpson}, {Singanamalla}, {Singer}, {Singhal}, {Sinha}, {Sip{\H{o}}cz}, {Spitler}, {Stansby}, {Streicher}, {{\v{S}}umak}, {Swinbank}, {Taranu}, {Tewary}, {Tremblay}, {de Val-Borro}, {Van Kooten}, {Vasovi{\'c}}, {Verma}, {de Miranda Cardoso}, {Williams}, {Wilson}, {Winkel}, {Wood-Vasey}, {Xue}, {Yoachim}, {Zhang}, {Zonca}, \& {Astropy Project Contributors}}]{AstropyIII}
{Astropy Collaboration}, {Price-Whelan}, A.~M., {Lim}, P.~L., {et~al.} 2022, \apj, 935, 167, \dodoi{10.3847/1538-4357/ac7c74}

\bibitem[{{Belloche} {et~al.}(2020){Belloche}, {Maury}, {Maret}, {Anderl}, {Bacmann}, {Andr{\'e}}, {Bontemps}, {Cabrit}, {Codella}, {Gaudel}, {Gueth}, {Lef{\`e}vre}, {Lefloch}, {Podio}, \& {Testi}}]{Belloche2020}
{Belloche}, A., {Maury}, A.~J., {Maret}, S., {et~al.} 2020, \aap, 635, A198, \dodoi{10.1051/0004-6361/201937352}

\bibitem[{{Bergner} {et~al.}(2018){Bergner}, {Guzm{\'a}n}, {{\"O}berg}, {Loomis}, \& {Pegues}}]{Bergner2018}
{Bergner}, J.~B., {Guzm{\'a}n}, V.~G., {{\"O}berg}, K.~I., {Loomis}, R.~A., \& {Pegues}, J. 2018, \apj, 857, 69, \dodoi{10.3847/1538-4357/aab664}

\bibitem[{{Bergner} {et~al.}(2021){Bergner}, {{\"O}berg}, {Guzm{\'a}n}, {Law}, {Loomis}, {Cataldi}, {Bosman}, {Aikawa}, {Andrews}, {Bergin}, {Booth}, {Cleeves}, {Czekala}, {Huang}, {Ilee}, {Le Gal}, {Long}, {Nomura}, {M{\'e}nard}, {Qi}, {Schwarz}, {Teague}, {Tsukagoshi}, {Walsh}, {Wilner}, \& {Yamato}}]{Bergner2021}
{Bergner}, J.~B., {{\"O}berg}, K.~I., {Guzm{\'a}n}, V.~V., {et~al.} 2021, \apjs, 257, 11, \dodoi{10.3847/1538-4365/ac143a}

\bibitem[{{Boogert} {et~al.}(2015){Boogert}, {Gerakines}, \& {Whittet}}]{Boogert2015}
{Boogert}, A.~C.~A., {Gerakines}, P.~A., \& {Whittet}, D. C.~B. 2015, \araa, 53, 541, \dodoi{10.1146/annurev-astro-082214-122348}

\bibitem[{{Booth} {et~al.}(2023{\natexlab{a}}){Booth}, {Ilee}, {Walsh}, {Kama}, {Keyte}, {van Dishoeck}, \& {Nomura}}]{Booth2023}
{Booth}, A.~S., {Ilee}, J.~D., {Walsh}, C., {et~al.} 2023{\natexlab{a}}, \aap, 669, A53, \dodoi{10.1051/0004-6361/202244472}

\bibitem[{{Booth} {et~al.}(2023{\natexlab{b}}){Booth}, {Law}, {Temmink}, {Leemker}, \& {Macias}}]{Booth2023_HD169142}
{Booth}, A.~S., {Law}, C.~J., {Temmink}, M., {Leemker}, M., \& {Macias}, E. 2023{\natexlab{b}}, arXiv e-prints, arXiv:2308.07910, \dodoi{10.48550/arXiv.2308.07910}

\bibitem[{{Booth} {et~al.}(2021{\natexlab{a}}){Booth}, {van der Marel}, {Leemker}, {van Dishoeck}, \& {Ohashi}}]{Booth2021}
{Booth}, A.~S., {van der Marel}, N., {Leemker}, M., {van Dishoeck}, E.~F., \& {Ohashi}, S. 2021{\natexlab{a}}, \aap, 651, L6, \dodoi{10.1051/0004-6361/202141057}

\bibitem[{{Booth} {et~al.}(2021{\natexlab{b}}){Booth}, {Walsh}, {Terwisscha van Scheltinga}, {van Dishoeck}, {Ilee}, {Hogerheijde}, {Kama}, \& {Nomura}}]{Booth2021_CH3OH}
{Booth}, A.~S., {Walsh}, C., {Terwisscha van Scheltinga}, J., {et~al.} 2021{\natexlab{b}}, Nature Astronomy, 5, 684, \dodoi{10.1038/s41550-021-01352-w}

\bibitem[{{Bosman} {et~al.}(2021){Bosman}, {Bergin}, {Loomis}, {Andrews}, {van't Hoff}, {Teague}, {{\"O}berg}, {Guzm{\'a}n}, {Walsh}, {Aikawa}, {Alarc{\'o}n}, {Bae}, {Bergner}, {Booth}, {Cataldi}, {Cleeves}, {Czekala}, {Huang}, {Ilee}, {Law}, {Le Gal}, {Liu}, {Long}, {M{\'e}nard}, {Nomura}, {P{\'e}rez}, {Qi}, {Schwarz}, {Sierra}, {Tsukagoshi}, {Yamato}, {Wilner}, \& {Zhang}}]{Bosman2021}
{Bosman}, A.~D., {Bergin}, E.~A., {Loomis}, R.~A., {et~al.} 2021, \apjs, 257, 15, \dodoi{10.3847/1538-4365/ac1433}

\bibitem[{{Brunken} {et~al.}(2022){Brunken}, {Booth}, {Leemker}, {Nazari}, {van der Marel}, \& {van Dishoeck}}]{Brunken2022}
{Brunken}, N. G.~C., {Booth}, A.~S., {Leemker}, M., {et~al.} 2022, \aap, 659, A29, \dodoi{10.1051/0004-6361/202142981}

\bibitem[{{Cabedo} {et~al.}(2023){Cabedo}, {Maury}, {Girart}, {Padovani}, {Hennebelle}, {Houde}, \& {Zhang}}]{Cabedo2023}
{Cabedo}, V., {Maury}, A., {Girart}, J.~M., {et~al.} 2023, \aap, 669, A90, \dodoi{10.1051/0004-6361/202243813}

\bibitem[{{Calmonte} {et~al.}(2016){Calmonte}, {Altwegg}, {Balsiger}, {Berthelier}, {Bieler}, {Cessateur}, {Dhooghe}, {van Dishoeck}, {Fiethe}, {Fuselier}, {Gasc}, {Gombosi}, {H{\"a}ssig}, {Le Roy}, {Rubin}, {S{\'e}mon}, {Tzou}, \& {Wampfler}}]{Calmonte2016}
{Calmonte}, U., {Altwegg}, K., {Balsiger}, H., {et~al.} 2016, \mnras, 462, S253, \dodoi{10.1093/mnras/stw2601}

\bibitem[{{Carvajal} {et~al.}(2010){Carvajal}, {Kleiner}, \& {Demaison}}]{Carvajal2010}
{Carvajal}, M., {Kleiner}, I., \& {Demaison}, J. 2010, \apjs, 190, 315, \dodoi{10.1088/0067-0049/190/2/315}

\bibitem[{{Carvajal} {et~al.}(2009){Carvajal}, {Margul{\`e}s}, {Tercero}, {Demyk}, {Kleiner}, {Guillemin}, {Lattanzi}, {Walters}, {Demaison}, {Wlodarczak}, {Huet}, {M{\o}llendal}, {Ilyushin}, \& {Cernicharo}}]{Carvajal2009}
{Carvajal}, M., {Margul{\`e}s}, L., {Tercero}, B., {et~al.} 2009, \aap, 500, 1109, \dodoi{10.1051/0004-6361/200811456}

\bibitem[{{CASA Team} {et~al.}(2022){CASA Team}, {Bean}, {Bhatnagar}, {Castro}, {Donovan Meyer}, {Emonts}, {Garcia}, {Garwood}, {Golap}, {Gonzalez Villalba}, {Harris}, {Hayashi}, {Hoskins}, {Hsieh}, {Jagannathan}, {Kawasaki}, {Keimpema}, {Kettenis}, {Lopez}, {Marvil}, {Masters}, {McNichols}, {Mehringer}, {Miel}, {Moellenbrock}, {Montesino}, {Nakazato}, {Ott}, {Petry}, {Pokorny}, {Raba}, {Rau}, {Schiebel}, {Schweighart}, {Sekhar}, {Shimada}, {Small}, {Steeb}, {Sugimoto}, {Suoranta}, {Tsutsumi}, {van Bemmel}, {Verkouter}, {Wells}, {Xiong}, {Szomoru}, {Griffith}, {Glendenning}, \& {Kern}}]{CASA}
{CASA Team}, {Bean}, B., {Bhatnagar}, S., {et~al.} 2022, \pasp, 134, 114501, \dodoi{10.1088/1538-3873/ac9642}

\bibitem[{{Casassus} \& {C{\'a}rcamo}(2022)}]{Casassus2022}
{Casassus}, S., \& {C{\'a}rcamo}, M. 2022, \mnras, 513, 5790, \dodoi{10.1093/mnras/stac1285}

\bibitem[{{Casassus} {et~al.}(2023){Casassus}, {Cieza}, {C{\'a}rcamo}, {Ribas}, {Christiaens}, {Rodr{\'\i}guez-Jim{\'e}nez}, {Arce-Tord}, {Bhowmik}, {Chavan}, {Gonz{\'a}lez-Ruilova}, {Mart{\'\i}nez-Brunner}, {Guidotti}, \& {Leiva}}]{Casassus2023}
{Casassus}, S., {Cieza}, L., {C{\'a}rcamo}, M., {et~al.} 2023, arXiv e-prints, arXiv:2307.07416, \dodoi{10.48550/arXiv.2307.07416}

\bibitem[{{Cataldi} {et~al.}(2021){Cataldi}, {Yamato}, {Aikawa}, {Bergner}, {Furuya}, {Guzm{\'a}n}, {Huang}, {Loomis}, {Qi}, {Andrews}, {Bergin}, {Booth}, {Bosman}, {Cleeves}, {Czekala}, {Ilee}, {Law}, {Le Gal}, {Liu}, {Long}, {M{\'e}nard}, {Nomura}, {{\"O}berg}, {Schwarz}, {Teague}, {Tsukagoshi}, {Walsh}, {Wilner}, \& {Zhang}}]{Cataldi2021}
{Cataldi}, G., {Yamato}, Y., {Aikawa}, Y., {et~al.} 2021, \apjs, 257, 10, \dodoi{10.3847/1538-4365/ac143d}

\bibitem[{{Ceccarelli} {et~al.}(2014){Ceccarelli}, {Caselli}, {Bockel{\'e}e-Morvan}, {Mousis}, {Pizzarello}, {Robert}, \& {Semenov}}]{Ceccarelli2014}
{Ceccarelli}, C., {Caselli}, P., {Bockel{\'e}e-Morvan}, D., {et~al.} 2014, in Protostars and Planets VI, ed. H.~{Beuther}, R.~S. {Klessen}, C.~P. {Dullemond}, \& T.~{Henning}, 859--882, \dodoi{10.2458/azu_uapress_9780816531240-ch037}

\bibitem[{{Ceccarelli} {et~al.}(2023){Ceccarelli}, {Codella}, {Balucani}, {Bockelee-Morvan}, {Herbst}, {Vastel}, {Caselli}, {Favre}, {Lefloch}, {Oberg}, \& {Yamamoto}}]{Ceccarelli2023}
{Ceccarelli}, C., {Codella}, C., {Balucani}, N., {et~al.} 2023, in Astronomical Society of the Pacific Conference Series, Vol. 534, Astronomical Society of the Pacific Conference Series, ed. S.~{Inutsuka}, Y.~{Aikawa}, T.~{Muto}, K.~{Tomida}, \& M.~{Tamura}, 379

\bibitem[{{Cherniak} \& {Costain}(1966)}]{Cherniak1966}
{Cherniak}, E.~A., \& {Costain}, C.~C. 1966, \jcp, 45, 104, \dodoi{10.1063/1.1727291}

\bibitem[{{Chuang} {et~al.}(2016){Chuang}, {Fedoseev}, {Ioppolo}, {van Dishoeck}, \& {Linnartz}}]{Chuang2016}
{Chuang}, K.~J., {Fedoseev}, G., {Ioppolo}, S., {van Dishoeck}, E.~F., \& {Linnartz}, H. 2016, \mnras, 455, 1702, \dodoi{10.1093/mnras/stv2288}

\bibitem[{{Cieza} {et~al.}(2016){Cieza}, {Casassus}, {Tobin}, {Bos}, {Williams}, {Perez}, {Zhu}, {Caceres}, {Canovas}, {Dunham}, {Hales}, {Prieto}, {Principe}, {Schreiber}, {Ruiz-Rodriguez}, \& {Zurlo}}]{Cieza2016}
{Cieza}, L.~A., {Casassus}, S., {Tobin}, J., {et~al.} 2016, \nat, 535, 258, \dodoi{10.1038/nature18612}

\bibitem[{{Coudert} {et~al.}(2013){Coudert}, {Drouin}, {Tercero}, {Cernicharo}, {Guillemin}, {Motiyenko}, \& {Margul{\`e}s}}]{Coudert2013}
{Coudert}, L.~H., {Drouin}, B.~J., {Tercero}, B., {et~al.} 2013, \apj, 779, 119, \dodoi{10.1088/0004-637X/779/2/119}

\bibitem[{{Coudert} {et~al.}(2019){Coudert}, {Margul{\`e}s}, {Vastel}, {Motiyenko}, {Caux}, \& {Guillemin}}]{Coudert2019}
{Coudert}, L.~H., {Margul{\`e}s}, L., {Vastel}, C., {et~al.} 2019, \aap, 624, A70, \dodoi{10.1051/0004-6361/201834827}

\bibitem[{{Coutens} {et~al.}(2014){Coutens}, {J{\o}rgensen}, {Persson}, {van Dishoeck}, {Vastel}, \& {Taquet}}]{Coutens2014}
{Coutens}, A., {J{\o}rgensen}, J.~K., {Persson}, M.~V., {et~al.} 2014, \apjl, 792, L5, \dodoi{10.1088/2041-8205/792/1/L5}

\bibitem[{{Creswell} \& {Schwendeman}(1974)}]{Creswell1974}
{Creswell}, R.~A., \& {Schwendeman}, R.~H. 1974, Chemical Physics Letters, 27, 521, \dodoi{10.1016/0009-2614(74)80295-2}

\bibitem[{{Csengeri} {et~al.}(2019){Csengeri}, {Belloche}, {Bontemps}, {Wyrowski}, {Menten}, \& {Bouscasse}}]{Csengeri2019}
{Csengeri}, T., {Belloche}, A., {Bontemps}, S., {et~al.} 2019, \aap, 632, A57, \dodoi{10.1051/0004-6361/201935226}

\bibitem[{{Czekala} {et~al.}(2021){Czekala}, {Loomis}, {Teague}, {Booth}, {Huang}, {Cataldi}, {Ilee}, {Law}, {Walsh}, {Bosman}, {Guzm{\'a}n}, {Gal}, {{\"O}berg}, {Yamato}, {Aikawa}, {Andrews}, {Bae}, {Bergin}, {Bergner}, {Cleeves}, {Kurtovic}, {M{\'e}nard}, {Nomura}, {P{\'e}rez}, {Qi}, {Schwarz}, {Tsukagoshi}, {Waggoner}, {Wilner}, \& {Zhang}}]{Czekala2021}
{Czekala}, I., {Loomis}, R.~A., {Teague}, R., {et~al.} 2021, \apjs, 257, 2, \dodoi{10.3847/1538-4365/ac1430}

\bibitem[{{Daly} {et~al.}(2015){Daly}, {Berm{\'u}dez}, {Kolesnikov{\'a}}, \& {Alonso}}]{Daly2015}
{Daly}, A.~M., {Berm{\'u}dez}, C., {Kolesnikov{\'a}}, L., \& {Alonso}, J.~L. 2015, \apjs, 218, 30, \dodoi{10.1088/0067-0049/218/2/30}

\bibitem[{{Demaison} {et~al.}(1987){Demaison}, {Maes}, {Van Eijck}, {Wlodarczak}, \& {Lasne}}]{Demaison1987}
{Demaison}, J., {Maes}, H., {Van Eijck}, B.~P., {Wlodarczak}, G., \& {Lasne}, M.~C. 1987, Journal of Molecular Spectroscopy, 125, 214, \dodoi{10.1016/0022-2852(87)90208-6}

\bibitem[{{Drozdovskaya} {et~al.}(2022){Drozdovskaya}, {Coudert}, {Margul{\`e}s}, {Coutens}, {J{\o}rgensen}, \& {Manigand}}]{Drozdovskaya2022}
{Drozdovskaya}, M.~N., {Coudert}, L.~H., {Margul{\`e}s}, L., {et~al.} 2022, \aap, 659, A69, \dodoi{10.1051/0004-6361/202142863}

\bibitem[{{Drozdovskaya} {et~al.}(2019){Drozdovskaya}, {van Dishoeck}, {Rubin}, {J{\o}rgensen}, \& {Altwegg}}]{Drozdovskaya2019}
{Drozdovskaya}, M.~N., {van Dishoeck}, E.~F., {Rubin}, M., {J{\o}rgensen}, J.~K., \& {Altwegg}, K. 2019, \mnras, 490, 50, \dodoi{10.1093/mnras/stz2430}

\bibitem[{{Drozdovskaya} {et~al.}(2018){Drozdovskaya}, {van Dishoeck}, {J{\o}rgensen}, {Calmonte}, {van der Wiel}, {Coutens}, {Calcutt}, {M{\"u}ller}, {Bjerkeli}, {Persson}, {Wampfler}, \& {Altwegg}}]{Drozdovskaya2018}
{Drozdovskaya}, M.~N., {van Dishoeck}, E.~F., {J{\o}rgensen}, J.~K., {et~al.} 2018, \mnras, 476, 4949, \dodoi{10.1093/mnras/sty462}

\bibitem[{{Drozdovskaya} {et~al.}(2021){Drozdovskaya}, {Schroeder I}, {Rubin}, {Altwegg}, {van Dishoeck}, {Kulterer}, {De Keyser}, {Fuselier}, \& {Combi}}]{Drozdovskaya2021}
{Drozdovskaya}, M.~N., {Schroeder I}, I. R.~H.~G., {Rubin}, M., {et~al.} 2021, \mnras, 500, 4901, \dodoi{10.1093/mnras/staa3387}

\bibitem[{{Duan} {et~al.}(2015){Duan}, {Carvajal}, {Yu}, {Pearson}, {Drouin}, \& {Kleiner}}]{Duan2015}
{Duan}, C., {Carvajal}, M., {Yu}, S., {et~al.} 2015, \aap, 576, A39, \dodoi{10.1051/0004-6361/201425328}

\bibitem[{{Eistrup} {et~al.}(2016){Eistrup}, {Walsh}, \& {van Dishoeck}}]{Eistrup2016}
{Eistrup}, C., {Walsh}, C., \& {van Dishoeck}, E.~F. 2016, \aap, 595, A83, \dodoi{10.1051/0004-6361/201628509}

\bibitem[{{Endres} {et~al.}(2009){Endres}, {Drouin}, {Pearson}, {M{\"u}ller}, {Lewen}, {Schlemmer}, \& {Giesen}}]{Endres2009}
{Endres}, C.~P., {Drouin}, B.~J., {Pearson}, J.~C., {et~al.} 2009, \aap, 504, 635, \dodoi{10.1051/0004-6361/200912409}

\bibitem[{{Endres} {et~al.}(2016){Endres}, {Schlemmer}, {Schilke}, {Stutzki}, \& {M{\"u}ller}}]{CDMS3}
{Endres}, C.~P., {Schlemmer}, S., {Schilke}, P., {Stutzki}, J., \& {M{\"u}ller}, H. S.~P. 2016, Journal of Molecular Spectroscopy, 327, 95, \dodoi{10.1016/j.jms.2016.03.005}

\bibitem[{{Favre} {et~al.}(2014){Favre}, {Carvajal}, {Field}, {J{\o}rgensen}, {Bisschop}, {Brouillet}, {Despois}, {Baudry}, {Kleiner}, {Bergin}, {Crockett}, {Neill}, {Margul{\`e}s}, {Huet}, \& {Demaison}}]{Favre2014}
{Favre}, C., {Carvajal}, M., {Field}, D., {et~al.} 2014, \apjs, 215, 25, \dodoi{10.1088/0067-0049/215/2/25}

\bibitem[{{Favre} {et~al.}(2018){Favre}, {Fedele}, {Semenov}, {Parfenov}, {Codella}, {Ceccarelli}, {Bergin}, {Chapillon}, {Testi}, {Hersant}, {Lefloch}, {Fontani}, {Blake}, {Cleeves}, {Qi}, {Schwarz}, \& {Taquet}}]{Favre2018}
{Favre}, C., {Fedele}, D., {Semenov}, D., {et~al.} 2018, \apjl, 862, L2, \dodoi{10.3847/2041-8213/aad046}

\bibitem[{{Foreman-Mackey} {et~al.}(2013){Foreman-Mackey}, {Hogg}, {Lang}, \& {Goodman}}]{emcee}
{Foreman-Mackey}, D., {Hogg}, D.~W., {Lang}, D., \& {Goodman}, J. 2013, \pasp, 125, 306, \dodoi{10.1086/670067}

\bibitem[{{Furlan} {et~al.}(2016){Furlan}, {Fischer}, {Ali}, {Stutz}, {Stanke}, {Tobin}, {Megeath}, {Osorio}, {Hartmann}, {Calvet}, {Poteet}, {Booker}, {Manoj}, {Watson}, \& {Allen}}]{Furlan2016}
{Furlan}, E., {Fischer}, W.~J., {Ali}, B., {et~al.} 2016, \apjs, 224, 5, \dodoi{10.3847/0067-0049/224/1/5}

\bibitem[{{Furuya} \& {Aikawa}(2014)}]{Furuya2014}
{Furuya}, K., \& {Aikawa}, Y. 2014, \apj, 790, 97, \dodoi{10.1088/0004-637X/790/2/97}

\bibitem[{{Furuya} {et~al.}(2011){Furuya}, {Aikawa}, {Sakai}, \& {Yamamoto}}]{Furuya2011}
{Furuya}, K., {Aikawa}, Y., {Sakai}, N., \& {Yamamoto}, S. 2011, \apj, 731, 38, \dodoi{10.1088/0004-637X/731/1/38}

\bibitem[{{Furuya} {et~al.}(2022){Furuya}, {Lee}, \& {Nomura}}]{Furuya2022}
{Furuya}, K., {Lee}, S., \& {Nomura}, H. 2022, \apj, 938, 29, \dodoi{10.3847/1538-4357/ac9233}

\bibitem[{{Garrod} \& {Herbst}(2006)}]{Garrod2006}
{Garrod}, R.~T., \& {Herbst}, E. 2006, \aap, 457, 927, \dodoi{10.1051/0004-6361:20065560}

\bibitem[{{Groner} {et~al.}(2002){Groner}, {Albert}, {Herbst}, {De Lucia}, {Lovas}, {Drouin}, \& {Pearson}}]{Groder2002}
{Groner}, P., {Albert}, S., {Herbst}, E., {et~al.} 2002, \apjs, 142, 145, \dodoi{10.1086/341221}

\bibitem[{{Guzm{\'a}n} {et~al.}(2021){Guzm{\'a}n}, {Bergner}, {Law}, {{\"O}berg}, {Walsh}, {Cataldi}, {Aikawa}, {Bergin}, {Czekala}, {Huang}, {Andrews}, {Loomis}, {Zhang}, {Le Gal}, {Alarc{\'o}n}, {Ilee}, {Teague}, {Cleeves}, {Wilner}, {Long}, {Schwarz}, {Bosman}, {P{\'e}rez}, {M{\'e}nard}, \& {Liu}}]{Guzman2021}
{Guzm{\'a}n}, V.~V., {Bergner}, J.~B., {Law}, C.~J., {et~al.} 2021, \apjs, 257, 6, \dodoi{10.3847/1538-4365/ac1440}

\bibitem[{{Hardy} {et~al.}(1982){Hardy}, {Cox}, {Fliege}, \& {Dreizler}}]{Hardy1982}
{Hardy}, J.~A., {Cox}, A.~P., {Fliege}, E., \& {Dreizler}, H. 1982, Zeitschrift Naturforschung Teil A, 37, 1035, \dodoi{10.1515/zna-1982-0910}

\bibitem[{Harris {et~al.}(2020)Harris, Millman, van~der Walt, Gommers, Virtanen, Cournapeau, Wieser, Taylor, Berg, Smith, Kern, Picus, Hoyer, van Kerkwijk, Brett, Haldane, del R{\'{i}}o, Wiebe, Peterson, G{\'{e}}rard-Marchant, Sheppard, Reddy, Weckesser, Abbasi, Gohlke, \& Oliphant}]{Numpy}
Harris, C.~R., Millman, K.~J., van~der Walt, S.~J., {et~al.} 2020, Nature, 585, 357, \dodoi{10.1038/s41586-020-2649-2}

\bibitem[{{H{\"a}ssig} {et~al.}(2017){H{\"a}ssig}, {Altwegg}, {Balsiger}, {Berthelier}, {Bieler}, {Calmonte}, {Dhooghe}, {Fiethe}, {Fuselier}, {Gasc}, {Gombosi}, {Le Roy}, {Luspay-Kuti}, {Mandt}, {Rubin}, {Tzou}, {Wampfler}, \& {Wurz}}]{Hassig2017}
{H{\"a}ssig}, M., {Altwegg}, K., {Balsiger}, H., {et~al.} 2017, \aap, 605, A50, \dodoi{10.1051/0004-6361/201630140}

\bibitem[{{Haykal} {et~al.}(2014){Haykal}, {Carvajal}, {Tercero}, {Kleiner}, {L{\'o}pez}, {Cernicharo}, {Motiyenko}, {Huet}, {Guillemin}, \& {Margul{\`e}s}}]{Haykal2014}
{Haykal}, I., {Carvajal}, M., {Tercero}, B., {et~al.} 2014, \aap, 568, A58, \dodoi{10.1051/0004-6361/201322937}

\bibitem[{{Herbst} \& {van Dishoeck}(2009)}]{Herbst2009}
{Herbst}, E., \& {van Dishoeck}, E.~F. 2009, \araa, 47, 427, \dodoi{10.1146/annurev-astro-082708-101654}

\bibitem[{{H{\"o}gbom}(1974)}]{Hogbom1974}
{H{\"o}gbom}, J.~A. 1974, \aaps, 15, 417

\bibitem[{Hunter(2007)}]{Matplotlib}
Hunter, J.~D. 2007, Computing in Science \& Engineering, 9, 90, \dodoi{10.1109/MCSE.2007.55}

\bibitem[{{Ilee} {et~al.}(2021){Ilee}, {Walsh}, {Booth}, {Aikawa}, {Andrews}, {Bae}, {Bergin}, {Bergner}, {Bosman}, {Cataldi}, {Cleeves}, {Czekala}, {Guzm{\'a}n}, {Huang}, {Law}, {Le Gal}, {Loomis}, {M{\'e}nard}, {Nomura}, {{\"O}berg}, {Qi}, {Schwarz}, {Teague}, {Tsukagoshi}, {Wilner}, {Yamato}, \& {Zhang}}]{Ilee2021}
{Ilee}, J.~D., {Walsh}, C., {Booth}, A.~S., {et~al.} 2021, \apjs, 257, 9, \dodoi{10.3847/1538-4365/ac1441}

\bibitem[{{Ilyushin} {et~al.}(2009){Ilyushin}, {Kryvda}, \& {Alekseev}}]{Ilyushin2009}
{Ilyushin}, V., {Kryvda}, A., \& {Alekseev}, E. 2009, Journal of Molecular Spectroscopy, 255, 32, \dodoi{10.1016/j.jms.2009.01.016}

\bibitem[{{Jensen} {et~al.}(2021){Jensen}, {J{\o}rgensen}, {Kristensen}, {Coutens}, {van Dishoeck}, {Furuya}, {Harsono}, \& {Persson}}]{Jensen2021}
{Jensen}, S.~S., {J{\o}rgensen}, J.~K., {Kristensen}, L.~E., {et~al.} 2021, \aap, 650, A172, \dodoi{10.1051/0004-6361/202140560}

\bibitem[{{J{\o}rgensen} {et~al.}(2020){J{\o}rgensen}, {Belloche}, \& {Garrod}}]{Jorgensen2020}
{J{\o}rgensen}, J.~K., {Belloche}, A., \& {Garrod}, R.~T. 2020, \araa, 58, 727, \dodoi{10.1146/annurev-astro-032620-021927}

\bibitem[{{J{\o}rgensen} {et~al.}(2016){J{\o}rgensen}, {van der Wiel}, {Coutens}, {Lykke}, {M{\"u}ller}, {van Dishoeck}, {Calcutt}, {Bjerkeli}, {Bourke}, {Drozdovskaya}, {Favre}, {Fayolle}, {Garrod}, {Jacobsen}, {{\"O}berg}, {Persson}, \& {Wampfler}}]{Jorgensen2016}
{J{\o}rgensen}, J.~K., {van der Wiel}, M.~H.~D., {Coutens}, A., {et~al.} 2016, \aap, 595, A117, \dodoi{10.1051/0004-6361/201628648}

\bibitem[{{J{\o}rgensen} {et~al.}(2018){J{\o}rgensen}, {M{\"u}ller}, {Calcutt}, {Coutens}, {Drozdovskaya}, {{\"O}berg}, {Persson}, {Taquet}, {van Dishoeck}, \& {Wampfler}}]{Jorgensen2018}
{J{\o}rgensen}, J.~K., {M{\"u}ller}, H.~S.~P., {Calcutt}, H., {et~al.} 2018, \aap, 620, A170, \dodoi{10.1051/0004-6361/201731667}

\bibitem[{{Jorsater} \& {van Moorsel}(1995)}]{JvM}
{Jorsater}, S., \& {van Moorsel}, G.~A. 1995, \aj, 110, 2037, \dodoi{10.1086/117668}

\bibitem[{{Kleiner} {et~al.}(1996){Kleiner}, {Lovas}, \& {Godefroid}}]{Kleiner1996}
{Kleiner}, I., {Lovas}, F.~J., \& {Godefroid}, M. 1996, Journal of Physical and Chemical Reference Data, 25, 1113, \dodoi{10.1063/1.555983}

\bibitem[{{Koerber} {et~al.}(2013){Koerber}, {Bisschop}, {Endres}, {Kleshcheva}, {Pohl}, {Klein}, {Lewen}, \& {Schlemmer}}]{Koerber2013}
{Koerber}, M., {Bisschop}, S.~E., {Endres}, C.~P., {et~al.} 2013, \aap, 558, A112, \dodoi{10.1051/0004-6361/201321898}

\bibitem[{{Koga} {et~al.}(2022){Koga}, {Kawasaki}, \& {Machida}}]{Koga2022}
{Koga}, S., {Kawasaki}, Y., \& {Machida}, M.~N. 2022, \mnras, 515, 6073, \dodoi{10.1093/mnras/stac2115}

\bibitem[{{Kuhn} \& {Hillenbrand}(2019)}]{Kuhn2019}
{Kuhn}, M.~A., \& {Hillenbrand}, L.~A. 2019, \apj, 883, 117, \dodoi{10.3847/1538-4357/ab3a3f}

\bibitem[{{Langer} {et~al.}(1984){Langer}, {Graedel}, {Frerking}, \& {Armentrout}}]{Langer1984}
{Langer}, W.~D., {Graedel}, T.~E., {Frerking}, M.~A., \& {Armentrout}, P.~B. 1984, \apj, 277, 581, \dodoi{10.1086/161730}

\bibitem[{{Law} {et~al.}(2023{\natexlab{a}}){Law}, {Booth}, \& {{\"O}berg}}]{Law2023_SiS}
{Law}, C.~J., {Booth}, A.~S., \& {{\"O}berg}, K.~I. 2023{\natexlab{a}}, \apjl, 952, L19, \dodoi{10.3847/2041-8213/acdfd0}

\bibitem[{{Law} {et~al.}(2021){Law}, {Teague}, {Loomis}, {Bae}, {{\"O}berg}, {Czekala}, {Andrews}, {Aikawa}, {Alarc{\'o}n}, {Bergin}, {Bergner}, {Booth}, {Bosman}, {Calahan}, {Cataldi}, {Cleeves}, {Furuya}, {Guzm{\'a}n}, {Huang}, {Ilee}, {Le Gal}, {Liu}, {Long}, {M{\'e}nard}, {Nomura}, {P{\'e}rez}, {Qi}, {Schwarz}, {Soto}, {Tsukagoshi}, {Yamato}, {van't Hoff}, {Walsh}, {Wilner}, \& {Zhang}}]{Law2021}
{Law}, C.~J., {Teague}, R., {Loomis}, R.~A., {et~al.} 2021, \apjs, 257, 4, \dodoi{10.3847/1538-4365/ac1439}

\bibitem[{{Law} {et~al.}(2022){Law}, {Crystian}, {Teague}, {{\"O}berg}, {Rich}, {Andrews}, {Bae}, {Flaherty}, {Guzm{\'a}n}, {Huang}, {Ilee}, {Kastner}, {Loomis}, {Long}, {P{\'e}rez}, {P{\'e}rez}, {Qi}, {Rosotti}, {Ru{\'\i}z-Rodr{\'\i}guez}, {Tsukagoshi}, \& {Wilner}}]{Law2022}
{Law}, C.~J., {Crystian}, S., {Teague}, R., {et~al.} 2022, \apj, 932, 114, \dodoi{10.3847/1538-4357/ac6c02}

\bibitem[{{Law} {et~al.}(2023{\natexlab{b}}){Law}, {Teague}, {{\"O}berg}, {Rich}, {Andrews}, {Bae}, {Benisty}, {Facchini}, {Flaherty}, {Isella}, {Jin}, {Hashimoto}, {Huang}, {Loomis}, {Long}, {Mu{\~n}oz-Romero}, {Paneque-Carre{\~n}o}, {P{\'e}rez}, {Qi}, {Schwarz}, {Stadler}, {Tsukagoshi}, {Wilner}, \& {van der Plas}}]{Law2023}
{Law}, C.~J., {Teague}, R., {{\"O}berg}, K.~I., {et~al.} 2023{\natexlab{b}}, \apj, 948, 60, \dodoi{10.3847/1538-4357/acb3c4}

\bibitem[{{Le Gal} {et~al.}(2021){Le Gal}, {{\"O}berg}, {Teague}, {Loomis}, {Law}, {Walsh}, {Bergin}, {M{\'e}nard}, {Wilner}, {Andrews}, {Aikawa}, {Booth}, {Cataldi}, {Bergner}, {Bosman}, {Cleeves}, {Czekala}, {Furuya}, {Guzm{\'a}n}, {Huang}, {Ilee}, {Nomura}, {Qi}, {Schwarz}, {Tsukagoshi}, {Yamato}, \& {Zhang}}]{LeGal2021}
{Le Gal}, R., {{\"O}berg}, K.~I., {Teague}, R., {et~al.} 2021, \apjs, 257, 12, \dodoi{10.3847/1538-4365/ac2583}

\bibitem[{{Lee} {et~al.}(2022){Lee}, {Codella}, {Ceccarelli}, \& {L{\'o}pez-Sepulcre}}]{Lee2022}
{Lee}, C.-F., {Codella}, C., {Ceccarelli}, C., \& {L{\'o}pez-Sepulcre}, A. 2022, \apj, 937, 10, \dodoi{10.3847/1538-4357/ac8c28}

\bibitem[{{Lee} {et~al.}(2019){Lee}, {Lee}, {Baek}, {Aikawa}, {Cieza}, {Yoon}, {Herczeg}, {Johnstone}, \& {Casassus}}]{Lee2019}
{Lee}, J.-E., {Lee}, S., {Baek}, G., {et~al.} 2019, Nature Astronomy, 3, 314, \dodoi{10.1038/s41550-018-0680-0}

\bibitem[{{Loomis} {et~al.}(2018){Loomis}, {Cleeves}, {{\"O}berg}, {Aikawa}, {Bergner}, {Furuya}, {Guzman}, \& {Walsh}}]{Loomis2018}
{Loomis}, R.~A., {Cleeves}, L.~I., {{\"O}berg}, K.~I., {et~al.} 2018, \apj, 859, 131, \dodoi{10.3847/1538-4357/aac169}

\bibitem[{{Lykke} {et~al.}(2017){Lykke}, {Coutens}, {J{\o}rgensen}, {van der Wiel}, {Garrod}, {M{\"u}ller}, {Bjerkeli}, {Bourke}, {Calcutt}, {Drozdovskaya}, {Favre}, {Fayolle}, {Jacobsen}, {{\"O}berg}, {Persson}, {van Dishoeck}, \& {Wampfler}}]{Lykke2017}
{Lykke}, J.~M., {Coutens}, A., {J{\o}rgensen}, J.~K., {et~al.} 2017, \aap, 597, A53, \dodoi{10.1051/0004-6361/201629180}

\bibitem[{{Manigand} {et~al.}(2019){Manigand}, {Calcutt}, {J{\o}rgensen}, {Taquet}, {M{\"u}ller}, {Coutens}, {Wampfler}, {Ligterink}, {Drozdovskaya}, {Kristensen}, {van der Wiel}, \& {Bourke}}]{Manigand2019}
{Manigand}, S., {Calcutt}, H., {J{\o}rgensen}, J.~K., {et~al.} 2019, \aap, 623, A69, \dodoi{10.1051/0004-6361/201832844}

\bibitem[{{Manigand} {et~al.}(2020){Manigand}, {J{\o}rgensen}, {Calcutt}, {M{\"u}ller}, {Ligterink}, {Coutens}, {Drozdovskaya}, {van Dishoeck}, \& {Wampfler}}]{Manigand2020}
{Manigand}, S., {J{\o}rgensen}, J.~K., {Calcutt}, H., {et~al.} 2020, \aap, 635, A48, \dodoi{10.1051/0004-6361/201936299}

\bibitem[{{Manigand} {et~al.}(2021){Manigand}, {Coutens}, {Loison}, {Wakelam}, {Calcutt}, {M{\"u}ller}, {J{\o}rgensen}, {Taquet}, {Wampfler}, {Bourke}, {Kulterer}, {van Dishoeck}, {Drozdovskaya}, \& {Ligterink}}]{Manigand2021}
{Manigand}, S., {Coutens}, A., {Loison}, J.~C., {et~al.} 2021, \aap, 645, A53, \dodoi{10.1051/0004-6361/202038113}

\bibitem[{{Margul{\`e}s} {et~al.}(2009){Margul{\`e}s}, {Coudert}, {M{\o}llendal}, {Guillemin}, {Huet}, \& {Jane{\v{c}}kov{\`a}}}]{Margules2009}
{Margul{\`e}s}, L., {Coudert}, L.~H., {M{\o}llendal}, H., {et~al.} 2009, Journal of Molecular Spectroscopy, 254, 55, \dodoi{10.1016/j.jms.2008.12.007}

\bibitem[{{Margul{\`e}s} {et~al.}(2015){Margul{\`e}s}, {Motiyenko}, {Ilyushin}, \& {Guillemin}}]{Margules2015}
{Margul{\`e}s}, L., {Motiyenko}, R.~A., {Ilyushin}, V.~V., \& {Guillemin}, J.~C. 2015, \aap, 579, A46, \dodoi{10.1051/0004-6361/201425478}

\bibitem[{{Margul{\`e}s} {et~al.}(2010){Margul{\`e}s}, {Huet}, {Demaison}, {Carvajal}, {Kleiner}, {M{\o}llendal}, {Tercero}, {Marcelino}, \& {Cernicharo}}]{Margules2010}
{Margul{\`e}s}, L., {Huet}, T.~R., {Demaison}, J., {et~al.} 2010, \apj, 714, 1120, \dodoi{10.1088/0004-637X/714/2/1120}

\bibitem[{{McClure} {et~al.}(2023){McClure}, {Rocha}, {Pontoppidan}, {Crouzet}, {Chu}, {Dartois}, {Lamberts}, {Noble}, {Pendleton}, {Perotti}, {Qasim}, {Rachid}, {Smith}, {Sun}, {Beck}, {Boogert}, {Brown}, {Caselli}, {Charnley}, {Cuppen}, {Dickinson}, {Drozdovskaya}, {Egami}, {Erkal}, {Fraser}, {Garrod}, {Harsono}, {Ioppolo}, {Jim{\'e}nez-Serra}, {Jin}, {J{\o}rgensen}, {Kristensen}, {Lis}, {McCoustra}, {McGuire}, {Melnick}, {{\~A}-berg}, {Palumbo}, {Shimonishi}, {Sturm}, {van Dishoeck}, \& {Linnartz}}]{McClure2023}
{McClure}, M.~K., {Rocha}, W.~R.~M., {Pontoppidan}, K.~M., {et~al.} 2023, Nature Astronomy, 7, 431, \dodoi{10.1038/s41550-022-01875-w}

\bibitem[{{Millar} {et~al.}(1989){Millar}, {Bennett}, \& {Herbst}}]{Millar1989}
{Millar}, T.~J., {Bennett}, A., \& {Herbst}, E. 1989, \apj, 340, 906, \dodoi{10.1086/167444}

\bibitem[{{M{\"o}ller} {et~al.}(2017){M{\"o}ller}, {Endres}, \& {Schilke}}]{Moller2017}
{M{\"o}ller}, T., {Endres}, C., \& {Schilke}, P. 2017, \aap, 598, A7, \dodoi{10.1051/0004-6361/201527203}

\bibitem[{{M{\"u}ller} {et~al.}(2022){M{\"u}ller}, {Guillemin}, {Lewen}, \& {Schlemmer}}]{Muller2022}
{M{\"u}ller}, H. S.~P., {Guillemin}, J.-C., {Lewen}, F., \& {Schlemmer}, S. 2022, Journal of Molecular Spectroscopy, 384, 111584, \dodoi{10.1016/j.jms.2022.111584}

\bibitem[{{M{\"u}ller} {et~al.}(2005){M{\"u}ller}, {Schl{\"o}der}, {Stutzki}, \& {Winnewisser}}]{CDMS2}
{M{\"u}ller}, H. S.~P., {Schl{\"o}der}, F., {Stutzki}, J., \& {Winnewisser}, G. 2005, Journal of Molecular Structure, 742, 215, \dodoi{10.1016/j.molstruc.2005.01.027}

\bibitem[{{M{\"u}ller} {et~al.}(2001){M{\"u}ller}, {Thorwirth}, {Roth}, \& {Winnewisser}}]{CDMS1}
{M{\"u}ller}, H.~S.~P., {Thorwirth}, S., {Roth}, D.~A., \& {Winnewisser}, G. 2001, \aap, 370, L49, \dodoi{10.1051/0004-6361:20010367}

\bibitem[{{Mumma} \& {Charnley}(2011)}]{Mumma2011}
{Mumma}, M.~J., \& {Charnley}, S.~B. 2011, \araa, 49, 471, \dodoi{10.1146/annurev-astro-081309-130811}

\bibitem[{{Nazari} {et~al.}(2023){Nazari}, {Tabone}, {van't Hoff}, {J{\o}rgensen}, \& {van Dishoeck}}]{Nazari2023}
{Nazari}, P., {Tabone}, B., {van't Hoff}, M. L.~R., {J{\o}rgensen}, J.~K., \& {van Dishoeck}, E.~F. 2023, \apjl, 951, L38, \dodoi{10.3847/2041-8213/acdde4}

\bibitem[{{Nomura} {et~al.}(2009){Nomura}, {Aikawa}, {Nakagawa}, \& {Millar}}]{Nomura2009}
{Nomura}, H., {Aikawa}, Y., {Nakagawa}, Y., \& {Millar}, T.~J. 2009, \aap, 495, 183, \dodoi{10.1051/0004-6361:200810206}

\bibitem[{{Nomura} {et~al.}(2023){Nomura}, {Furuya}, {Cordiner}, {Charnley}, {Alexander}, {Nixon}, {Guzman}, {Yurimoto}, {Tsukagoshi}, \& {Iino}}]{Nomura2023_PPVII}
{Nomura}, H., {Furuya}, K., {Cordiner}, M.~A., {et~al.} 2023, in Astronomical Society of the Pacific Conference Series, Vol. 534, Astronomical Society of the Pacific Conference Series, ed. S.~{Inutsuka}, Y.~{Aikawa}, T.~{Muto}, K.~{Tomida}, \& M.~{Tamura}, 1075

\bibitem[{{Notsu} {et~al.}(2020){Notsu}, {Eistrup}, {Walsh}, \& {Nomura}}]{Notsu2020}
{Notsu}, S., {Eistrup}, C., {Walsh}, C., \& {Nomura}, H. 2020, \mnras, 499, 2229, \dodoi{10.1093/mnras/staa2944}

\bibitem[{{Notsu} {et~al.}(2021){Notsu}, {van Dishoeck}, {Walsh}, {Bosman}, \& {Nomura}}]{Notsu2021}
{Notsu}, S., {van Dishoeck}, E.~F., {Walsh}, C., {Bosman}, A.~D., \& {Nomura}, H. 2021, \aap, 650, A180, \dodoi{10.1051/0004-6361/202140667}

\bibitem[{{{\"O}berg} {et~al.}(2009){{\"O}berg}, {Garrod}, {van Dishoeck}, \& {Linnartz}}]{Oberg2009}
{{\"O}berg}, K.~I., {Garrod}, R.~T., {van Dishoeck}, E.~F., \& {Linnartz}, H. 2009, \aap, 504, 891, \dodoi{10.1051/0004-6361/200912559}

\bibitem[{{{\"O}berg} {et~al.}(2015){{\"O}berg}, {Guzm{\'a}n}, {Furuya}, {Qi}, {Aikawa}, {Andrews}, {Loomis}, \& {Wilner}}]{Oberg2015}
{{\"O}berg}, K.~I., {Guzm{\'a}n}, V.~V., {Furuya}, K., {et~al.} 2015, \nat, 520, 198, \dodoi{10.1038/nature14276}

\bibitem[{{Okoda} {et~al.}(2021){Okoda}, {Oya}, {Abe}, {Komaki}, {Watanabe}, \& {Yamamoto}}]{Okoda2021}
{Okoda}, Y., {Oya}, Y., {Abe}, S., {et~al.} 2021, \apj, 923, 168, \dodoi{10.3847/1538-4357/ac2c6c}

\bibitem[{{Okoda} {et~al.}(2022){Okoda}, {Oya}, {Imai}, {Sakai}, {Watanabe}, {L{\'o}pez-Sepulcre}, {Saigo}, \& {Yamamoto}}]{Okoda2022}
{Okoda}, Y., {Oya}, Y., {Imai}, M., {et~al.} 2022, \apj, 935, 136, \dodoi{10.3847/1538-4357/ac7ff4}

\bibitem[{{Oldag} \& {Sutter}(1992)}]{Oldag1992}
{Oldag}, F., \& {Sutter}, D.~H. 1992, Zeitschrift Naturforschung Teil A, 47, 527, \dodoi{10.1515/zna-1992-0315}

\bibitem[{{Oyama} {et~al.}(2023){Oyama}, {Ohno}, {Tamanai}, {Watanabe}, {Yamamoto}, {Sakai}, {Zeng}, {Nakatani}, \& {Sakai}}]{Oyama2023}
{Oyama}, T., {Ohno}, Y., {Tamanai}, A., {et~al.} 2023, \apj, 957, 4, \dodoi{10.3847/1538-4357/acf320}

\bibitem[{{Padovani} {et~al.}(2020){Padovani}, {Ivlev}, {Galli}, {Offner}, {Indriolo}, {Rodgers-Lee}, {Marcowith}, {Girichidis}, {Bykov}, \& {Kruijssen}}]{Padovani2020}
{Padovani}, M., {Ivlev}, A.~V., {Galli}, D., {et~al.} 2020, \ssr, 216, 29, \dodoi{10.1007/s11214-020-00654-1}

\bibitem[{{Paneque-Carre{\~n}o} {et~al.}(2023){Paneque-Carre{\~n}o}, {Miotello}, {van Dishoeck}, {Tabone}, {Izquierdo}, \& {Facchini}}]{PanequeCarreno2023}
{Paneque-Carre{\~n}o}, T., {Miotello}, A., {van Dishoeck}, E.~F., {et~al.} 2023, \aap, 669, A126, \dodoi{10.1051/0004-6361/202244428}

\bibitem[{{Pearson} {et~al.}(2012){Pearson}, {Yu}, \& {Drouin}}]{Pearson2012}
{Pearson}, J.~C., {Yu}, S., \& {Drouin}, B.~J. 2012, Journal of Molecular Spectroscopy, 280, 119, \dodoi{10.1016/j.jms.2012.06.012}

\bibitem[{{Peter} \& {Dreizler}(1965)}]{Peter1965}
{Peter}, R., \& {Dreizler}, H. 1965, Zeitschrift Naturforschung Teil A, 20, 301, \dodoi{10.1515/zna-1965-0224}

\bibitem[{{Pickett} {et~al.}(1998){Pickett}, {Poynter}, {Cohen}, {Delitsky}, {Pearson}, \& {M{\"u}ller}}]{JPL}
{Pickett}, H.~M., {Poynter}, R.~L., {Cohen}, E.~A., {et~al.} 1998, \jqsrt, 60, 883, \dodoi{10.1016/S0022-4073(98)00091-0}

\bibitem[{{Poch} {et~al.}(2020){Poch}, {Istiqomah}, {Quirico}, {Beck}, {Schmitt}, {Theul{\'e}}, {Faure}, {Hily-Blant}, {Bonal}, {Raponi}, {Ciarniello}, {Rousseau}, {Potin}, {Brissaud}, {Flandinet}, {Filacchione}, {Pommerol}, {Thomas}, {Kappel}, {Mennella}, {Moroz}, {Vinogradoff}, {Arnold}, {Erard}, {Bockel{\'e}e-Morvan}, {Leyrat}, {Capaccioni}, {De Sanctis}, {Longobardo}, {Mancarella}, {Palomba}, \& {Tosi}}]{Poch2020}
{Poch}, O., {Istiqomah}, I., {Quirico}, E., {et~al.} 2020, Science, 367, aaw7462, \dodoi{10.1126/science.aaw7462}

\bibitem[{{Richard} {et~al.}(2021){Richard}, {J{\o}rgensen}, {Margul{\`e}s}, {Motiyenko}, {Guillemin}, \& {Groner}}]{Richard2021}
{Richard}, C., {J{\o}rgensen}, J.~K., {Margul{\`e}s}, L., {et~al.} 2021, \aap, 651, A120, \dodoi{10.1051/0004-6361/202141282}

\bibitem[{{Roberts} \& {Millar}(2000)}]{Roberts2000}
{Roberts}, H., \& {Millar}, T.~J. 2000, \aap, 361, 388

\bibitem[{{Rubin} {et~al.}(2017){Rubin}, {Altwegg}, {Balsiger}, {Berthelier}, {Bieler}, {Calmonte}, {Combi}, {De Keyser}, {Engrand}, {Fiethe}, {Fuselier}, {Gasc}, {Gombosi}, {Hansen}, {H{\"a}ssig}, {Le Roy}, {Mezger}, {Tzou}, {Wampfler}, \& {Wurz}}]{Rubin2017}
{Rubin}, M., {Altwegg}, K., {Balsiger}, H., {et~al.} 2017, \aap, 601, A123, \dodoi{10.1051/0004-6361/201730584}

\bibitem[{{Rubin} {et~al.}(2019){Rubin}, {Altwegg}, {Balsiger}, {Berthelier}, {Combi}, {De Keyser}, {Drozdovskaya}, {Fiethe}, {Fuselier}, {Gasc}, {Gombosi}, {H{\"a}nni}, {Hansen}, {Mall}, {R{\`e}me}, {Schroeder}, {Schuhmann}, {S{\'e}mon}, {Waite}, {Wampfler}, \& {Wurz}}]{Rubin2019}
---. 2019, \mnras, 489, 594, \dodoi{10.1093/mnras/stz2086}

\bibitem[{{Schuhmann} {et~al.}(2019){Schuhmann}, {Altwegg}, {Balsiger}, {Berthelier}, {De Keyser}, {Fuselier}, {Gasc}, {Gombosi}, {H{\"a}nni}, {Rubin}, {S{\'e}mon}, {Tzou}, \& {Wampfler}}]{Schuhmann2019}
{Schuhmann}, M., {Altwegg}, K., {Balsiger}, H., {et~al.} 2019, ACS Earth and Space Chemistry, 3, 1854, \dodoi{10.1021/acsearthspacechem.9b00094}

\bibitem[{{Schwarz} {et~al.}(2018){Schwarz}, {Bergin}, {Cleeves}, {Zhang}, {{\"O}berg}, {Blake}, \& {Anderson}}]{Schwarz2018}
{Schwarz}, K.~R., {Bergin}, E.~A., {Cleeves}, L.~I., {et~al.} 2018, \apj, 856, 85, \dodoi{10.3847/1538-4357/aaae08}

\bibitem[{{Smith} {et~al.}(2021){Smith}, {Gudipati}, {Smith}, \& {Lewis}}]{Smith2021}
{Smith}, L.~R., {Gudipati}, M.~S., {Smith}, R.~L., \& {Lewis}, R.~D. 2021, \aap, 656, A82, \dodoi{10.1051/0004-6361/202141529}

\bibitem[{{Smith} {et~al.}(2015){Smith}, {Pontoppidan}, {Young}, \& {Morris}}]{Smith2015}
{Smith}, R.~L., {Pontoppidan}, K.~M., {Young}, E.~D., \& {Morris}, M.~R. 2015, \apj, 813, 120, \dodoi{10.1088/0004-637X/813/2/120}

\bibitem[{{Strom} \& {Strom}(1993)}]{Strom1993}
{Strom}, K.~M., \& {Strom}, S.~E. 1993, \apjl, 412, L63, \dodoi{10.1086/186941}

\bibitem[{{Taquet} {et~al.}(2016){Taquet}, {Wirstr{\"o}m}, \& {Charnley}}]{Taquet2016}
{Taquet}, V., {Wirstr{\"o}m}, E.~S., \& {Charnley}, S.~B. 2016, \apj, 821, 46, \dodoi{10.3847/0004-637X/821/1/46}

\bibitem[{Teague(2019)}]{GoFish}
Teague, R. 2019, The Journal of Open Source Software, 4, 1632, \dodoi{10.21105/joss.01632}

\bibitem[{{Teague} \& {Foreman-Mackey}(2018)}]{bettermoments}
{Teague}, R., \& {Foreman-Mackey}, D. 2018, Research Notes of the American Astronomical Society, 2, 173, \dodoi{10.3847/2515-5172/aae265}

\bibitem[{{Tielens} \& {Hagen}(1982)}]{Tielens1982}
{Tielens}, A.~G.~G.~M., \& {Hagen}, W. 1982, \aap, 114, 245

\bibitem[{{Tobin} {et~al.}(2023){Tobin}, {van't Hoff}, {Leemker}, {van Dishoeck}, {Paneque-Carre{\~n}o}, {Furuya}, {Harsono}, {Persson}, {Cleeves}, {Sheehan}, \& {Cieza}}]{Tobin2023}
{Tobin}, J.~J., {van't Hoff}, M. L.~R., {Leemker}, M., {et~al.} 2023, \nat, 615, 227, \dodoi{10.1038/s41586-022-05676-z}

\bibitem[{{Vacherand} {et~al.}(1986){Vacherand}, {Van Eijck}, {Burie}, \& {Demaison}}]{Vacherand1986}
{Vacherand}, J.~M., {Van Eijck}, B.~P., {Burie}, J., \& {Demaison}, J. 1986, Journal of Molecular Spectroscopy, 118, 355, \dodoi{10.1016/0022-2852(86)90175-X}

\bibitem[{{van der Marel} {et~al.}(2021){van der Marel}, {Booth}, {Leemker}, {van Dishoeck}, \& {Ohashi}}]{vanderMarel2021}
{van der Marel}, N., {Booth}, A.~S., {Leemker}, M., {van Dishoeck}, E.~F., \& {Ohashi}, S. 2021, \aap, 651, L5, \dodoi{10.1051/0004-6361/202141051}

\bibitem[{{van 't Hoff} {et~al.}(2020){van 't Hoff}, {Bergin}, {J{\o}rgensen}, \& {Blake}}]{vantHoff2020}
{van 't Hoff}, M. L.~R., {Bergin}, E.~A., {J{\o}rgensen}, J.~K., \& {Blake}, G.~A. 2020, \apjl, 897, L38, \dodoi{10.3847/2041-8213/ab9f97}

\bibitem[{{van 't Hoff} {et~al.}(2018){van 't Hoff}, {Tobin}, {Trapman}, {Harsono}, {Sheehan}, {Fischer}, {Megeath}, \& {van Dishoeck}}]{vantHoff2018}
{van 't Hoff}, M. L.~R., {Tobin}, J.~J., {Trapman}, L., {et~al.} 2018, \apjl, 864, L23, \dodoi{10.3847/2041-8213/aadb8a}

\bibitem[{Virtanen {et~al.}(2020)Virtanen, Gommers, Oliphant, Haberland, Reddy, Cournapeau, Burovski, Peterson, Weckesser, Bright, {van der Walt}, Brett, Wilson, Millman, Mayorov, Nelson, Jones, Kern, Larson, Carey, Polat, Feng, Moore, {VanderPlas}, Laxalde, Perktold, Cimrman, Henriksen, Quintero, Harris, Archibald, Ribeiro, Pedregosa, {van Mulbregt}, \& {SciPy 1.0 Contributors}}]{SciPy}
Virtanen, P., Gommers, R., Oliphant, T.~E., {et~al.} 2020, Nature Methods, 17, 261, \dodoi{10.1038/s41592-019-0686-2}

\bibitem[{{Walsh} {et~al.}(2014){Walsh}, {Millar}, {Nomura}, {Herbst}, {Widicus Weaver}, {Aikawa}, {Laas}, \& {Vasyunin}}]{Walsh2014}
{Walsh}, C., {Millar}, T.~J., {Nomura}, H., {et~al.} 2014, \aap, 563, A33, \dodoi{10.1051/0004-6361/201322446}

\bibitem[{{Walsh} {et~al.}(2016){Walsh}, {Loomis}, {{\"O}berg}, {Kama}, {van 't Hoff}, {Millar}, {Aikawa}, {Herbst}, {Widicus Weaver}, \& {Nomura}}]{Walsh2016}
{Walsh}, C., {Loomis}, R.~A., {{\"O}berg}, K.~I., {et~al.} 2016, \apjl, 823, L10, \dodoi{10.3847/2041-8205/823/1/L10}

\bibitem[{{Watanabe} \& {Kouchi}(2002)}]{Watanabe2002}
{Watanabe}, N., \& {Kouchi}, A. 2002, \apjl, 571, L173, \dodoi{10.1086/341412}

\bibitem[{{Watanabe} {et~al.}(2003){Watanabe}, {Shiraki}, \& {Kouchi}}]{Watanabe2003}
{Watanabe}, N., {Shiraki}, T., \& {Kouchi}, A. 2003, \apjl, 588, L121, \dodoi{10.1086/375634}

\bibitem[{{Wei} {et~al.}(2019){Wei}, {Nomura}, {Lee}, {Ip}, {Walsh}, \& {Millar}}]{Wei2019}
{Wei}, C.-E., {Nomura}, H., {Lee}, J.-E., {et~al.} 2019, \apj, 870, 129, \dodoi{10.3847/1538-4357/aaf390}

\bibitem[{{Wilson}(1999)}]{Wilson1999}
{Wilson}, T.~L. 1999, Reports on Progress in Physics, 62, 143, \dodoi{10.1088/0034-4885/62/2/002}

\bibitem[{{Winnewisser} {et~al.}(1975){Winnewisser}, {Winnewisser}, {Honda}, \& {Hirota}}]{Winnewisser1975}
{Winnewisser}, M., {Winnewisser}, G., {Honda}, T., \& {Hirota}, E. 1975, Zeitschrift Naturforschung Teil A, 30, 1001, \dodoi{10.1515/zna-1975-0814}

\bibitem[{{Woods} \& {Willacy}(2009)}]{Woods2009}
{Woods}, P.~M., \& {Willacy}, K. 2009, \apj, 693, 1360, \dodoi{10.1088/0004-637X/693/2/1360}

\bibitem[{{Xu} {et~al.}(2008){Xu}, {Fisher}, {Lees}, {Shi}, {Hougen}, {Pearson}, {Drouin}, {Blake}, \& {Braakman}}]{Xu2008}
{Xu}, L.-H., {Fisher}, J., {Lees}, R.~M., {et~al.} 2008, Journal of Molecular Spectroscopy, 251, 305, \dodoi{10.1016/j.jms.2008.03.017}

\bibitem[{{Yamato} {et~al.}(2022){Yamato}, {Furuya}, {Aikawa}, {Persson}, {Tobin}, {J{\o}rgensen}, \& {Kama}}]{Yamato2022}
{Yamato}, Y., {Furuya}, K., {Aikawa}, Y., {et~al.} 2022, \apj, 941, 75, \dodoi{10.3847/1538-4357/ac9ea5}

\bibitem[{{Yang} {et~al.}(2021){Yang}, {Sakai}, {Zhang}, {Murillo}, {Zhang}, {Higuchi}, {Zeng}, {L{\'o}pez-Sepulcre}, {Yamamoto}, {Lefloch}, {Bouvier}, {Ceccarelli}, {Hirota}, {Imai}, {Oya}, {Sakai}, \& {Watanabe}}]{Yang2021}
{Yang}, Y.-L., {Sakai}, N., {Zhang}, Y., {et~al.} 2021, \apj, 910, 20, \dodoi{10.3847/1538-4357/abdfd6}

\bibitem[{{Yen} {et~al.}(2016){Yen}, {Koch}, {Liu}, {Puspitaningrum}, {Hirano}, {Lee}, \& {Takakuwa}}]{Yen2016}
{Yen}, H.-W., {Koch}, P.~M., {Liu}, H.~B., {et~al.} 2016, \apj, 832, 204, \dodoi{10.3847/0004-637X/832/2/204}

\bibitem[{{Yoshida} {et~al.}(2022){Yoshida}, {Nomura}, {Furuya}, {Tsukagoshi}, \& {Lee}}]{Yoshida2022_12CO13CO}
{Yoshida}, T.~C., {Nomura}, H., {Furuya}, K., {Tsukagoshi}, T., \& {Lee}, S. 2022, \apj, 932, 126, \dodoi{10.3847/1538-4357/ac6efb}

\bibitem[{{Zhang} {et~al.}(2017){Zhang}, {Bergin}, {Blake}, {Cleeves}, \& {Schwarz}}]{Zhang2017}
{Zhang}, K., {Bergin}, E.~A., {Blake}, G.~A., {Cleeves}, L.~I., \& {Schwarz}, K.~R. 2017, Nature Astronomy, 1, 0130, \dodoi{10.1038/s41550-017-0130}

\bibitem[{{Zingsheim} {et~al.}(2017){Zingsheim}, {M{\"u}ller}, {Lewen}, {J{\o}rgensen}, \& {Schlemmer}}]{Zingsheim2017}
{Zingsheim}, O., {M{\"u}ller}, H. S.~P., {Lewen}, F., {J{\o}rgensen}, J.~K., \& {Schlemmer}, S. 2017, Journal of Molecular Spectroscopy, 342, 125, \dodoi{10.1016/j.jms.2017.07.008}

\end{thebibliography}
\bibliographystyle{aasjournal}



\end{document}